\documentclass{aa}
\usepackage{rotating}
\usepackage{txfonts}
\usepackage{booktabs}
\usepackage{graphicx}
\usepackage{graphics}
\usepackage{natbib}
\usepackage{array}
\usepackage{multirow}
\usepackage{lscape}
\usepackage{pdflscape}
\usepackage[utf8]{inputenc}
\usepackage[T1]{fontenc}
\usepackage{longtable}
\usepackage{supertabular,booktabs}
\bibpunct{(}{)}{;}{a}{}{,} 

\newcommand{\msun}{\mbox{M$_{\odot}$}}

\begin{document}
\title{The chemical structure of the Class~0 protostellar envelope NGC~1333 IRAS~4A\thanks{Based on Herschel observations. Herschel is an ESA space observatory with science instruments provided by European-led Principal Investigator consortia and with important participation from NASA.}}

\author{E.~Koumpia\inst{\ref{inst1},\ref{inst2},\ref{inst9}}, D.A.~Semenov\inst{\ref{inst3}}, F.F.S.~van der Tak\inst{\ref{inst1},\ref{inst2}}, A.C.A.~Boogert\inst{\ref{inst4}}, E.~Caux\inst{\ref{inst5},\ref{inst6}}}

\institute{SRON Netherlands Institute for Space Research, Landleven 12, 9747 AD Groningen, The Netherlands\label{inst1}
\and
Kapteyn Institute, University of Groningen, Landleven 12, 9747 AD Groningen, The Netherlands\label{inst2}
\and 
Max Planck Institute for Astronomy, Königstuhl 17, 69117 Heidelberg, Germany\label{inst3}
\and
Universities Space Research Association, Stratospheric Observatory for Infrared Astronomy, NASA Ames Research Center, MS 232-11, Moffett Field,
CA 94035, USA\label{inst4}
\and
Université de Toulouse, UPS-OMP, IRAP, Toulouse, France\label{inst5}\and 
CNRS, IRAP, 9 Av. Colonel Roche, BP 44346, F-31028 Toulouse Cedex 4, France\label{inst6} 
\and \email{ev.koumpia@gmail.com}\label{inst9}
}
\date{Received date/Accepted date}

\abstract
{It is not well known what drives the chemistry of a protostellar envelope, in particular the role of the stellar mass and the outflows on its chemical enrichment.}{We study the chemical structure of NGC 1333 IRAS 4A in order to (i) investigate the influence of the outflows on the chemistry, (ii) constrain the age of our object, (iii) compare it with a typical high-mass protostellar envelope.}{In our analysis we use JCMT line mapping and HIFI pointed spectra. To study the influence of the outflow on the degree of deuteration, we compare JCMT maps of HCO$^{+}$ and DCO$^{+}$ with non-LTE (RADEX) models in a region that spatially covers the outflow activity of IRAS 4A. To study the envelope chemistry, we derive empirical molecular abundance profiles for the observed species using the radiative transfer code (RATRAN) and adopting a 1D dust density/temperature profile from the literature. We compare our best-fit observed abundance profiles with the predictions from the time dependent gas grain chemical code (ALCHEMIC).} {The CO, HCN, HNC and CN abundance require an enhanced UV field which points towards an outflow cavity. The abundances (wrt H$_{2}$) are 1 to 2 orders of magnitude lower than those observed in the high mass protostellar envelope (AFGL 2591), while they are found to be similar within factors of a few with respect to CO. Differences in UV radiation may be responsible for such chemical differentiation, but temperature differences seem a more plausible explanation. The CH$_{3}$OH modeled abundance profile points towards an age of $>$ 4$\times$10$^{4}$ yrs for IRAS 4A. The spatial distribution of H$_{2}$D$^{+}$ differs from that of other deuterated species, indicating an origin from a foreground colder layer ($<$ 20 K).}{The observed abundances can be explained by passive heating towards the high mass protostellar envelope, while the presence of UV cavity channels become more important toward the low-mass protostellar envelope.}


\titlerunning{The chemical structure of the Class~0 protostellar envelope NGC~1333 IRAS~4A} 
\authorrunning{Koumpia et al.} 
 \maketitle


\section{Introduction} 

During low-mass ($<$ 2 \msun) star formation a rotating cloud of gas and dust collapses under gravitational forces. The central protostar increases in mass through the accretion disk that surrounds it. The main mechanisms that retard the gravitational collapse are the thermal pressure, magnetic fields, and turbulence \citep{Hennebelle2009,Evans2011,Luhman2012,Tan2015}. Turbulence can be enriched by energetic outflows from young stellar objects (YSOs) which may further trigger star formation in nearby gas \citep{Quillen2005}. 

Molecular outflows are prominent during the earliest stages of star formation, especially when collimated jets are driven in the youngest 
(10$^{3}$--10$^{4}$ years) embedded protostars \citep{Arce2007}. Class~0 protostars are 
still in their main accretion phase and they also drive the most powerful outflows. The impact of the ejected material on the surrounding cloud 
causes shock fronts. These lead to changes in the chemical composition and the enhancement of the abundance of several species in the surroundings. 
\citet{Fontani2014} have found enhancement of HDCO/H$_{2}$CO towards the shock location of a Class~0 object, L1157~mm (d$=$250~pc), 
reporting a deuterated molecule as a shock tracer 
for the first time. 

The strong outflow activity and winds that YSOs produce result in high-velocity gas, but also in the evacuation of regions near the protostar. Such cavities have been previously seen as a ``hole” in the 1.3~mm continuum emission \citep[e.g., near NGC~1333~IRAS4 and SVS13;][]{Lefloch1998}. 
UV radiation from the protostellar system (mainly due to accretion) is expected to play a crucial role in such environments \citep{Stauber2004, Visser2012}.

During the cold and dense pre-collapse phase, molecular complexity increases by rapid ion-molecule gas-phase reactions followed by gradual freeze-out and build up of ices (H$_{2}$O, CO, NH$_{3}$) and surface processes. While collapsing, the radiation that comes from the forming protostar heats the inner parts of 
the envelope making surface radicals mobile and highly reactive. Later, these freshly formed complex ices thermally desorb, further boosting rich chemical processes in the gas and creating a ``hot corino'' \citep[e.g.,][]{Ceccarelli2004,Garrod2006,Ceccarelli2008}. The hot corinos refer to inner regions ($<$200~au) with an increase of the temperature above 100~K, as a result of passive heating from the protostar. An analog characterizes the high-mass protostars. Previous independent studies on high- and low-mass protostellar envelopes suggest that there is an abundance difference of a few orders of magnitude in several species \citep[e.g., H$_2$CO, CH$_3$OH][]{vanderTak2000,Maret2004,Maret2005}. In addition, \citet{Herbst2009} present complex organic molecule abundances relative to CH$_{3}$OH that are similar (within factors of a few) for low- and high-mass YSOs. 

In this work we are interested in a direct abundance comparison among high-mass and low-mass protostellar envelopes with the use of datasets from the same instruments and similar methodology. In particular, we aim to answer a) what the chemical structure of low mass protostellar envelopes is and how it compares to high mass protostellar envelopes, b) how does the temperature profile affect the abundance profile of several species in the inner envelope (``hot corino''), the freeze--out zones, and the outer parts of protostellar envelopes, c) how do the outflows influence the chemistry of the surroundings of a protostellar envelope and what is the role of outflow cavities in the observed abundances and finally, d) if the deuteration can be used as an outflow/shock tracer. 
For this purpose we use the low-mass protostellar envelope IRAS~4A, which appears 
as the brightest (sub)--mm continuum object in NGC~1333~IRAS~4 region and is classified as a Class~0 object \citep{Andre1993}. IRAS~4A is a prototypical well studied Class~0 object and of great interest as it is among the first \citep{Mathieu1994} proto--binary systems ever detected. NGC~1333 is one of the nearest \citep[D=235~pc;][]{Hirota2008} and youngest \citep[$<$~1~Myr;][]{Gutermuth2008} star forming 
regions. IRAS~4A is a binary system, consisting of two deeply 
embedded Class~0 YSOs with a separation of 1.$\arcsec$8 (420~au at a distance of 235~pc). 
The binary nature of IRAS~4A was first observed in 0.84~mm CSO--JCMT interferometric 
high--resolution submillimeter continuum observations \citep{Lay1995} and resolved at millimeter wavelengths using the BIMA array by \citet{Looney2000}. 
They were also found to share a common circumbinary envelope \citep{Looney2003}. 

In addition, a spectral line and continuum survey using the SMA interferometer was performed by \citet{Jorgensen2007} where inverse P--Cygni $^{13}$CO 2--1 line profiles have been found. 
These profiles indicate infall motions, which are also characteristic of the Class~0 stage. \citet{DiFrancesco2001} reported inverse P--Cygni profiles in 
CS and H$_{2}$CO, tracing high-density gas as observed with IRAM Plateau de Bure.

IRAS~4A has been suggested to have a ‘‘hot corino’’ \citep{Maret2004}. Multitransition observations of species such as H$_{2}$CO and CH$_{3}$OH towards 4A revealed abundance enhancements in the warmest inner regions ($>$100~K) by up to 2 orders of magnitude \citep{Maret2004, Maret2005}. The same abundance enhancement 
can also occur in outflows on larger scales as a result of ice mantle sputtering in shocks \citep{Bachiller1997,Tafalla2000}. 
Mantle sputtering is thought to play a role when outflow speeds reach about 10~km~s$^{-1}$ and is independent of gas density. In faster shocks with speeds as high as 20--25~km~s$^{-1}$ the mantles vaporize completely \citep{Guillet2011} while grain core sputtering is still inefficient. H$_{2}$CO and CH$_{3}$OH have been found to trace outflow activity of IRAS~4A \citep{Jorgensen2007,Koumpia2016}, which makes it difficult to distinguish between a ‘‘hot corino’’ chemistry and the enhancement due to shocks caused by protostellar outflows. 

The highly collimated outflows from IRAS~4A have been mapped in several CO transitions \citep{Knee2000,Jorgensen2007,Yildiz2012}. IRAS~4A shows two bipolar outflows, one with a N--S orientation and the other with P.A. $\sim$45$^{\circ}$. The usual interpretation of the CO and SiO outflow has been that a N--S component arises in A$_{2}$ which becomes bent to a P.A. $\sim$45$^{\circ}$ angle at short distance to the North. There is no evidence for a flow from A$_{1}$, the brighter of the millimeter sources. \citet{Marvel2008} have used water masers to trace the small-scale motions of the IRAS 4A outflows, with somewhat puzzling results. One interpretation has been that there may be a third component of the system quite close to A$_{2}$, which drives the outflow. 

Our study mainly aims to a) constrain the chemical structure of the protostellar envelope of IRAS~4A and compare it with a high-mass protostellar envelope, b) investigate the presence of deuteration towards the outflow, and c) investigate the influence of the temperature over outflow activity to the observed abundance profiles.  
In this article we present HIFI and JCMT observations of a range of chemically diverse species towards IRAS~4A. 
To study the outflow chemistry, we estimate the excitation temperature and column density of H$_{2}$CO in the envelope and outflow using 
population diagrams. We proceed with modeling our observed maps with RADEX in order to determine the DCO$^{+}$/HCO$^{+}$ abundance ratio and test the enhancement observed by \citet{Fontani2014}. To study the envelope chemistry, we apply the 1D physical model determined by \citet{Krist2012} that takes into account the  
temperature and density gradients of the envelope and we run RATRAN in order to fit our observations. We apply constant and jump-like abundance profiles. In addition, we run chemical models that similarly take into account the physical structure of the protostellar envelope and apply the abundance profiles we determine to our RATRAN models. 

We compare this low--mass case with the high--mass case of AFGL~2591 for which a similar analysis was performed by \citet{Kazmierczak2015}. Lastly, 
we try to constrain the chemical age of the IRAS~4A from our time-dependent chemical models.

\section{Observations and data reduction} 

\subsection{HIFI: observations and reduction}

Observations of the YSO NGC~1333~IRAS~4A (RA=03$^{h}$29$^{m}$10.3$^{s}$, Dec=$+$31$^{o}$13\arcmin31\arcsec [J2000])
were made with the HIFI instrument at the Herschel Space Observatory, as part of the CHESS  guaranteed  time  key  programme \citep[Chemical Herschel Survey of Star Forming Regions;][]{Ceccarelli2010}. Full spectral
scans were made in bands 2a and 2b, covering the spectral ranges 626.01--721.48 GHz 
(415.81--479.23~$\mu$m) and 714.02--800.90 GHz (374.58--420.16~$\mu$m), respectively, 
at a resolution of 1.1 MHz (0.41--0.53~km~s$^{-1}$). In this frequency range, the Herschel/HIFI Half Power Beam Width (HPBW) is 26.5--33.9\arcsec
\citep{Roelfsema2012}. The spectral scan was performed in Dual Beam Switch (DBS) mode with a normal
chop frequency of 0.17~Hz. Each sky frequency was covered four times to facilitate the double
sideband deconvolution process. The instrument stability settings were calculated in HSPOT by
fixing the minimum and maximum goal resolutions to 1.1~MHz and setting the 1~GHz reference
option without continuum optimization. 


The observations were processed with the pipeline at the Herschel Science Center with HIPE 7.1.0 and
retrieved from the Herschel Science Archive. Further post--pipeline level 2 processing was done in
HIPE 8.0. Since we are interested in full spectral coverage, only the WBS spectra were considered, although many HRS spectra in narrower ranges 
were also available. 
Spectral regions affected by ‘spurs’ and not automatically detected by
the HIFI pipeline were flagged out and ignored. 
Polynomial baselines were subtracted using the FitBaseline task by masking
the lines interactively. The overlapping sidebands were deconvolved with the
doDeconvolution task in HIPE by applying the default settings. 
The observed line intensities in units of antenna temperature were corrected for loss 
in the sidelobes by converting them to main beam temperatures using the 
main beam efficiency of $\eta_{mb}$ (B$_{eff}$/F$_{eff}$) of 0.75 \citep{Roelfsema2012}.

\subsection{James Clerk Maxwell Telescope (JCMT): observations and reduction}

These observations are a part of the 
JCMT Spectral Legacy Survey \citep[SLS;][]{Plume2007}. The 
Auto--Correlation Spectral Imaging System (ACSIS) was used at
the James Clerk Maxwell Telescope (JCMT\footnote{The James Clerk Maxwell Telescope is operated by the Joint Astronomy Centre on behalf of the Science and Technology Facilities Council of the United Kingdom, the Netherlands Organization for Scientific Research, and the National Research Council of Canada.}) on Mauna Kea, Hawaii. 

We have 2\arcmin$\times$2\arcmin~maps from the HARP--B instrument, which provides high velocity resolution (1~MHz, $\sim$ 1~km~s$^{-1}$). These observations cover 
the frequency window between 330 and 373 GHz. The angular resolution of JCMT is 
$\sim 15$$\arcsec$ at 345~GHz, which is equivalent to $\sim 3500$ au at the distance of NGC~1333~IRAS~4 \citep{Choi2004}. The beam efficiency is 0.63 \citep{Buckle2009}. 

Details regarding the reduction and line detections of this dataset can be found in \citet{Koumpia2016}.

\section{Observational results} 

\subsection{Line detections}

The Single Side Band (SSB) H and V--polarization HIFI spectra were searched independently for line detections. 
We consider as safe detections 
the signals detected in both polarizations with $>$ 3$\times$RMS (RMS $\sim$ 0.01~K--0.04~K) after averaging, and with widths of at least two channels $>$ 0.9~km~s$^{-1}$ (single channel; 1.1 MHz, $\sim$0.47~km~s$^{-1}$). 
Table~\ref{HIFI_lines} presents the line list with secure detections. The detected lines were 
identified by producing single temperature LTE models in CASSIS\footnote{CASSIS has been developed by IRAP-UPS/CNRS (http://cassis.irap.omp.eu)} of the species detected in the JCMT Spectral Legacy Survey by \citet{Koumpia2016}. 
The species that were identified following this process are: CO, $^{13}$CO, C$^{18}$O, CS, HCN, HCO$^{+}$, N$_{2}$H$^{+}$, H$_{2}$CO, CH$_{3}$OH, H$_{2}$S and 
H$_{2}$O. We inspected our HIFI spectra for lines of SO, SO$_{2}$, SiO, HNC, and H$_{2}$O isotopologs, but found nothing. This is not surprising considering that the predicted intensity of the transitions in the HIFI range, assuming a beam filling of unity, is comparable with the measured RMS for HNC and a few orders of magnitude lower than the measured RMS for SO, SO$_{2}$ and SiO (RMS$<$0.03~K). The transitions of these latter species are expected to show such weak lines in the 626--801 GHz regime mainly because of their a) high critical densities ($>$10$^{8}$~cm$^{-3}$) and/or b) low Einstein coefficients ($<$10$^{-5}$~s$^{-1}$) compared to the observed transition in the JCMT regime (330--373 GHz).

The JCMT observations have been described in more detail in \citet{Koumpia2016}, where the detected species towards IRAS~4A are also presented. 
In addition to the species detected in the HIFI survey and excluding H$_{2}$S and H$_{2}$O, we detect the deuterated species DCO$^{+}$, HDCO, D$_{2}$CO, the S--bearing species SO, SO$_{2}$, OCS and finally SiO, HNC, CN, C$_{2}$H. Finally, we detect o--H$_{2}$D$^{+}$ emission in the vicinity of IRAS~4A, but not directly towards the source. The overall rms noise level in the JCMT data ranges between 0.005 and 0.05~K at a velocity resolution of 0.9~km~s$^{-1}$.

\subsection{Line profiles}

\begin{figure}[h]
\begin{center}  
\includegraphics[scale=0.3, angle=270,trim={0 0 0 -150},clip=true]{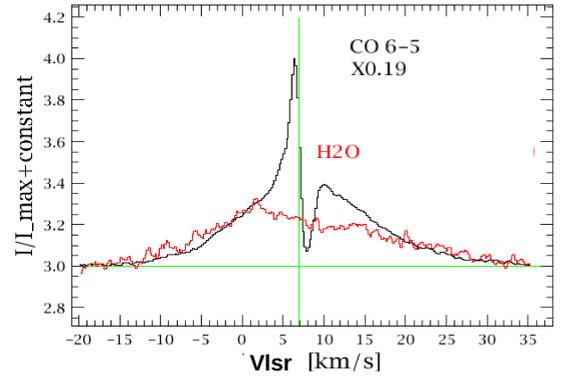} 
\end{center}
\caption{Normalized line profiles of H$_{2}$O 2$_{11}$--2$_{02}$ (red) and CO 6--5 (black) 
scaled by 0.19 for easier comparison. The H$_{2}$O profile is very similar to the CO broad wing emission. The vertical green line 
represents the central velocity of the source at 6.7~km$s^{-1}$.}
\label{fig:prof1}
\end{figure}

\begin{figure}[h]
\includegraphics[scale=0.3, angle=270,trim={0 0 0 -150},clip=true]{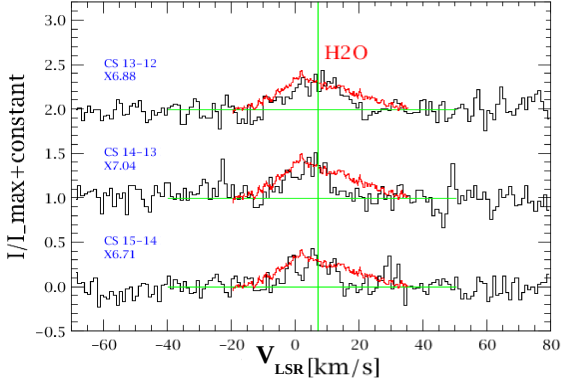} 
\caption{Normalized CS line profiles (black) scaled by the indicated factors
 compared to the broader profile of H$_{2}$O 2$_{11}$--2$_{02}$ (red). The vertical green line 
represents the central velocity of the source.}
\label{fig:prof2}
\end{figure}

\begin{figure}[h]
\includegraphics[scale=0.3,trim={0 0 -150 150},clip=true]{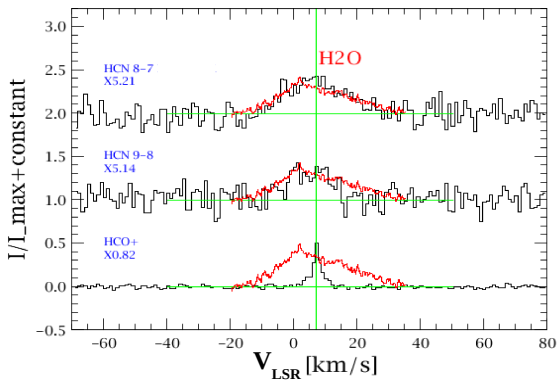} 
\caption{Normalized HCN and HCO$^{+}$ line profiles (black) scaled by the indicated factors compared to the H$_{2}$O 2$_{11}$--2$_{02}$ (red). The shape of the HCN line profile is comparable to the H$_{2}$O but HCO$^{+}$ is narrower. The vertical green line represents the central velocity of the source at 6.7~km$s^{-1}$.}
\label{fig:prof3}
\end{figure}

Examples of the line profiles from the species detected with HIFI and JCMT are plotted in 
Figures~\ref{fig:prof1}--\ref{fig:prof3}. Large line profile variations are observed. The CO 6--5 line shows narrow emission (FWHM$\sim$0.8~km~s$^{-1}$) and 
absorption 
peaks accompanied by prominent wings extending over 25~km~s$^{-1}$ towards blue- and red-shifted velocities. 
The absorption is absent in the other CO isotopologs, and the wings are much weaker.
On the other hand, the H$_{2}$O 2$_{11}$--2$_{02}$ line shows only very broad emission, very similar to the CO 6--5
wings (Figure~\ref{fig:prof1}). 

\vspace{0.3cm}

Visually, the detected lines can be divided into three groups. These groups are characterized by line profiles that are dominated
by a) absorption features such as CO 6--5 (Figure~\ref{fig:prof1}) and H$_{2}$S 2$_{12}$--1$_{01}$, b) broad emission (FWHM$=$4--11 km~s$^{-1}$; Figure~\ref{fig:prof2}) such as $^{13}$CO 6--5, CS 13--12, H$_{2}$O 2$_{11}$--2$_{02}$, HCN 8--7, CH$_{3}$OH 5$_{32}$--4$_{22}$ and H$_{2}$CO 9$_{18}$--8$_{17}$, and c) narrow emission (FWHM$<$4~km~s$^{-1}$; Figure~\ref{fig:prof3}) such as C$^{18}$O 6--5, H$_{2}$S 2$_{12}$--1$_{01}$, HCO$^{+}$ 8--7 and N$_{2}$H$^{+}$ 8--7.


To quantify the difference between the line profiles, they were decomposed with a simultaneous
fit of three Gaussians: broad emission, narrow emission, and narrow absorption components (Table \ref{HIFI_lines}). Our observations show that there is considerable variation in the widths of the broad emission line components. With
median FWHM values of 4.6 and 5.1 km~s$^{-1}$, the H$_{2}$CO and CH$_{3}$OH lines are narrower than the
H$_{2}$O line (11~km~s$^{-1}$). The H$_{2}$O line itself is in fact asymmetric, and can be fitted with an additional,
narrower Gaussian (V$_{LSR}$ $=$ 0.55~km~s$^{-1}$, FWHM$=$3.86~km~s$^{-1}$). Finally, it is worth mentioning
that the peak positions of the narrow N$_{2}$H$^{+}$ lines are similar to the absorption component seen in CO 6--5 and H$_{2}$S ($\sim$7.5~km~s$^{-1}$). 
The existence of such variation in the shape of the lines can be a result of 
the different regions that these lines trace. A broad component is indicative 
of outflow activity, a narrow component arises from dynamically quiescent gas (i.e., envelope) and the absorption is due to infall motions or the presence of foreground material. This foreground material can be         a foreground cloud not eventually bended to the source itself, but it can also be the external and colder part of the envelope which absorbs the emission of the inner and warmer parts of this envelope.

\vspace{1cm}

\section{Chemistry of the IRAS~4A outflow}

\subsection{LTE column densities}

The protostellar envelope of NGC~1333~IRAS~4A has previously been 
studied by various authors 
\citep[e.g.,][]{Maret2004, Maret2005}, which is not the case 
for its associated outflow. In this section we aim to estimate the 
column densities and excitation temperatures from
the observed emission in the envelope and outflow of IRAS~4A and compare them. In our approach we also add the observed HIFI 
transitions that were not available in previous studies. 

A widely used method to estimate the column density of a molecule is the population diagram \citep{Goldsmith99}. When LTE applies, the T$_{ex}$ equals the gas kinetic temperature, otherwise it provides only a lower limit. 

The column density of the upper state N$_{u}$ and the rotational temperature T$_{rot}$ ($=$T$_{ex}$) are determined by:

\begin{equation}
\frac{N_{u}}{g_{u}} = x \frac{\int T_{mb}dV}{v\mu^{2}S} = \frac{N_{T}}{Q(T_{rot})}e^{-\frac{E_{u}}{T_{rot}}}
,\end{equation} 
 
where x $=$ 8.591$\times$10$^{37}$ 8$\pi$k/hc$^{3}$, N$_{u}$ the column density of the upper energy level (cm$^{-2}$), g$_{u}$ the degeneracy of the upper energy level, T$_{mb}$ the main beam temperature (K), dV the velocity range (km~s$^{-1}$), $\nu$ the frequency (Hz), $\mu$ the dipole moment, S the line strength, N$_{T}$ the total column density (cm$^{-2}$), T$_{rot}$ the rotational temperature (K), Q the partition function and E$_{u}$ the upper energy level. Plotting ln(N$_{u}$/g$_{u}$) versus E$_{u}$/k results in a straight line with a slope of 1/T$_{rot}$.

In non-LTE excitation the population of each level may be characterized by a different excitation temperature T$_{ex}$ ($\neq$T$_{rot}$). The population diagrams can also take into account optical depth and beam effects due to different angular resolutions among the lines by using the modified equation: 

\begin{equation}
ln\frac{N_{u}}{g_{u}} = ln\frac{N_{T,thin}}{Q(T_{rot})} - \frac{E_{u}}{kT_{ex}} - ln(C_{\tau}) + ln(f)
,\end{equation} 

where T$_{ex}$ is the excitation temperature, \emph{C$_{\tau}$}$=$ $\tau$/(1-e$^{-\tau}$) is the optical depth correction factor, and \emph{f} 
the beam dilution which is defined as the size of the telescope beam 
over the size of the emitting region and which is assumed equal for all lines. A more detailed description of the method and formulas used can be found in 
\citet{Goldsmith99}.

The velocity information of our lines allows us to fit multiple Gaussians and separate, at first order, the outflow activity from the emission that comes from the dynamically quiescent envelope. For a more accurate determination of the temperature, species with many observed transitions are preferred. 
We perform this analysis for H$_{2}$CO for which we have observed 12 transitions with JCMT \citep{Koumpia2016} and HIFI (current study). In this approach we also include four extra transitions as observed using IRAM and presented in \citet{Maret2004}.

\begin{figure}[h]
\includegraphics[scale=0.37, angle=270]{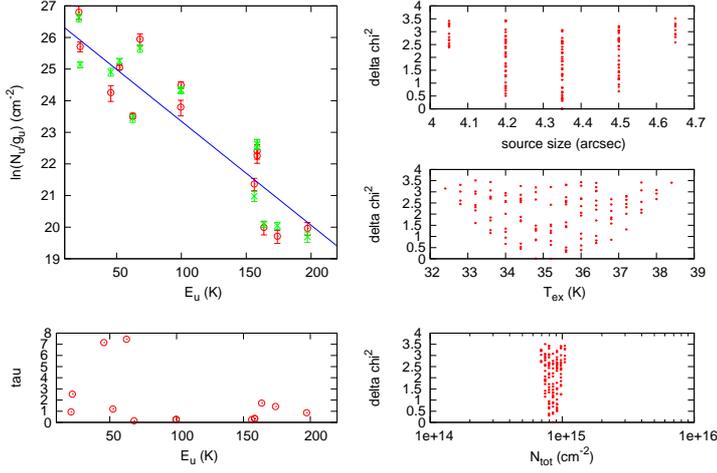} 
\caption{Results of population diagram based on the narrow component (envelope) of H$_{2}$CO as observed with JCMT and HIFI (150 K <E$_{up}$< 200 K) and the additional IRAM transitions as taken by \citet{Maret2004}. The plot presents the 
resulting excitation temperature (T$_{ex}$$\sim$$T_{rot}$), column density (N$_{tot}$), optical depth (tau) and size of the emitting area. The red symbols represent the observed data, the green symbols represent the best fitted data, and the blue line represents the rotation diagram fit. The three right panels show the resulting values after using the $\chi^2$ method to converge to the solution.}
\label{fig:H2CO_pop}
\end{figure}

\begin{figure}[h]
\includegraphics[scale=0.37, angle=270]{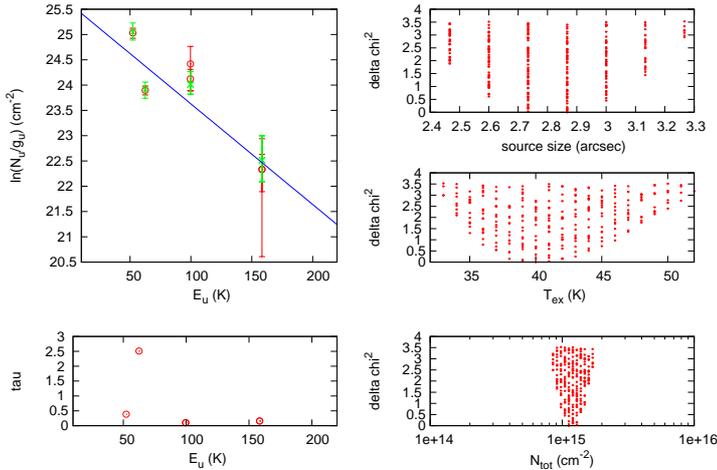} 
\caption{Same as in Figure~\ref{fig:H2CO_pop} for the broad component (8.5$<$FWHM$<$12~km~s$^{-1}$;outflow) of H$_{2}$CO. The H$_{2}$CO transitions observed with HIFI were fitted by a significantly more narrow single component (FWHM$<$4.7~km~s$^{-1}$), and therefore are not included in this plot.}
\label{fig:H2CO_pop_out}
\end{figure}

Figures~\ref{fig:H2CO_pop} and \ref{fig:H2CO_pop_out} present the population diagrams for the envelope and the outflow respectively, and the 
resulting excitation temperature ($\sim$rotational temperature), column density, optical depth and size of the emitting area. We find T$_{ex}$$\sim35$$\pm$0.5~K and N(H$_{2}$CO) of 8.5$^{+1.2}_{-0.8}$$\times10$$^{14}$cm$^{-2}$ for the envelope and T$_{ex}$$\sim41.5$$\pm$4~K and N(H$_{2}$CO) of 1.2$^{+0.5}_{-0.2}$$\times10$$^{15}$cm$^{-2}$ for the outflow. The error estimates have been computed during a $\chi^2$ minimization procedure. \citet{Maret2004} modeled only the envelope and found a factor of 2 lower column density and a factor of 1.2 lower T$_{ex}$. We attribute these differences to the use of our HIFI transitions, which probe denser/warmer regions. 

The excitation temperature provides only a lower limit for the 
gas kinetic temperature when the source is not in LTE. Our results point towards $\sim$20\% lower T$_{ex}$ and $\sim$30\% lower H$_{2}$CO column 
density for the envelope compared to the outflow, which is significant given our error estimates ($\sim$10\% and $\sim$14\% respectively).

\subsection{Deuterated species and outflows}

\label{deut}

When it comes to deuterated species in the field of star formation, the pre-stellar cores have been characterized as ``deuterium fractionation factories'' \citep{Ceccarelli2014}, due to the high deuteration degrees that have been observed towards them in various studies \citep[$>$10\%][]{Bacmann2003,Crapsi05}. Several studies have found a lower degree of deuteration towards Class~0 objects \citet[$\sim$1\%--10\%; e.g.][]{Crapsi2005a,Shah2001}.

More recently, deuterated species have also been employed as shock tracers by \citet{Fontani2014}. They found an enhancement of HDCO/H$_{2}$CO ($\sim$10\%) towards the eastern wall of the cavity excavated by the shock associated with the Class~0 object, L1157~mm. This is at least an order of magnitude larger than the HDCO/H$_{2}$CO ratio of the surrounding material. We aim to study the distribution of the deuterated species in the region and examine the deuteration towards the outflow of NGC~1333 IRAS~4A, which is also a Class~0 object. 

\subsubsection{In search of ions and deuterated species towards the outflow}

\begin{figure}[h]
\includegraphics[scale=0.4]{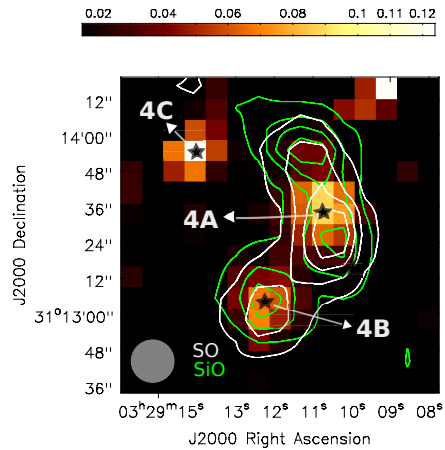}
\caption{C$_{2}$H map overplotted with SiO in green contours (0.013, 0.03, 0.04, and 0.05 K; rms$\sim$0.005~K) and S in white contours (0.07, 0.02, 0.03 and 0.04 K; rms$\sim$0.005~K).}
\label{fig:maps_sio}
\end{figure}

The outflow activity of IRAS~4A has been traced by many species. SiO is expected to be a tracer of outflow shocks 
due to sputtering of Si off dust grains \citep{Schilke1997}. \citet{Choi2005} found it to trace the jet in the case of NGC~1333 IRAS~4A.
They have presented a map of SiO 1--0 that traces the outflow activity of IRAS~4A 
and the spatial distribution of a narrow line component offset at $\sim$7.6~km~s$^{-1}$. Our dataset shows that SiO 8--7 has its primary peak at the shock position north of IRAS~4A \citep[R1;][]{Santangelo2014}, while SO 
also emits significantly at that position (Figure~\ref{fig:maps_sio}). The same figure shows the integrated intensity map of C$_{2}$H 
(core; from $+$5 to $+$9~km~s$^{-1}$) 
which was found to trace the envelopes of the three protostars. Its three peaks 
are in alignment with the continuum peaks as observed by \citet{Sandell2001}.

\begin{figure}[h]
\includegraphics[scale=0.35, angle=270]{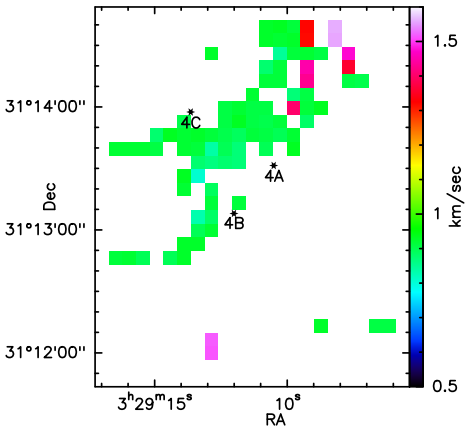}
\caption{FWHM map of H$_{2}$D$^{+}$. H$_{2}$D$^{+}$ shows narrow line profile ($\sim$1~km~s$^{-1}$) at most positions.}
\label{fig:width_h2dplus}
\end{figure}

\begin{figure}[h]
\includegraphics[scale=0.37]{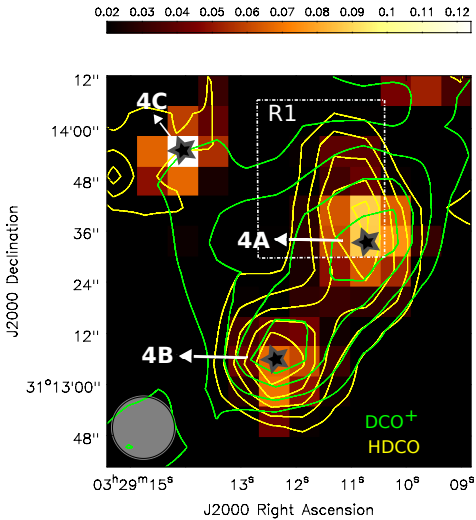}
\caption{C$_{2}$H map overplotted with DCO$^{+}$ in green contours (0.04, 0.06, 0.09, and 0.15 K; rms$\sim$0.005~K) and HDCO in yellow contours (0.013, 0.024, 0.034, 0.045 and 0.06 K; rms$\sim$0.005~K).}
\label{fig:maps_deu_2}
\end{figure}

\begin{figure}[h]
\includegraphics[scale=0.37]{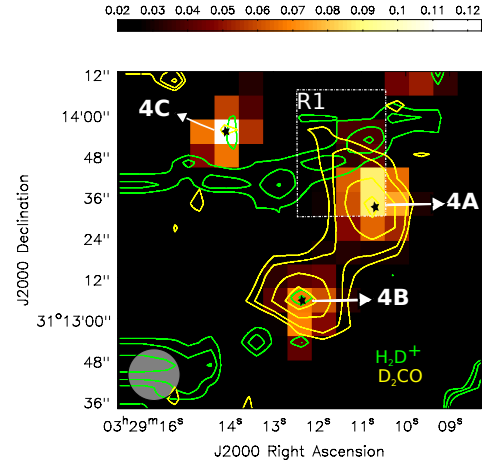} 
\caption{The spatial distribution of deuterated species in IRAS~4 region. Integrated intensity map (core; from $+$5 to $+$9~km~s$^{-1}$) of C$_{2}$H which traces the envelope in colors, overplotted with H$_{2}$D$^{+}$ in green contours and D$_{2}$CO in yellow contours. The contour levels are set to 0.014, 0.016, 0.023 and 0.03 K (rms$\sim$0.005~K).}
\label{fig:maps_deu}
\end{figure}

Figures~\ref{fig:maps_deu_2} and \ref{fig:maps_deu} present the spatial distribution of the 
observed deuterated species D$_{2}$CO, HDCO, DCO$^{+}$ and H$_{2}$D$^{+}$ as overplotted with C$_{2}$H. HDCO and D$_{2}$CO trace mainly the protostellar envelopes and their distribution 
covers part of the outflow-shock area (white rectangle) north-northeast of IRAS~4A \citep[R1;][]{Santangelo2014}. DCO$^{+}$ traces the envelopes 
but it also emits in a more extended area between the sources in the NW--SE direction. Interestingly, H$_{2}$D$^{+}$ shows a very different 
spatial distribution compared to the other deuterated species. It does not follow the distribution of the envelopes and it mainly emits in the NW-SE direction in 
the space between the three sources. The spatial distribution of o-H$_{2}$D$^{+}$ 
as presented in Figure~\ref{fig:maps_deu} appears to be parallel with the narrow component 
of SiO as presented in Figure~1 by \citet{Choi2005}. Lastly, \citet{vanderTak2002} detected the rare ND$_{3}$ species towards the DCO$^{+}$ peak (+23,-06 $\arcsec$ offset from IRAS~4A) which is about 15$\arcsec$ to the South of the H$_{2}$D$^{+}$ peak (+38,+09 $\arcsec$ offset from IRAS~4A). Both ND$_{3}$ and H$_{2}$D$^{+}$ (Fig.~\ref{fig:width_h2dplus}) are characterized by narrow emission (1.0$<$FWHM$<$1.7~km~s$^{-1}$), indicative of quiescent gas or an outflow component perpendicular to the line of sight.

\subsubsection{Origin of o-H$_{2}$D$^{+}$ emission}

\label{cold}

H$_{2}$D$^{+}$ is expected to arise from very cold gas. At very low temperatures ($<$20-30~K) the reaction H$_{3}^{+}$ $+$ HD $\longleftrightarrow$ H$_{2}$D$^{+}$ $+$ H$_{2}$ $+$ $\Delta$E is not balanced by the backward process, increasing the abundance of H$_{2}$D$^{+}$. In addition, the freeze out of CO and N$_{2}$ normally boosts H$_{3}^{+}$ \citep{Roberts2000}, and increases the H$_{2}$D$^{+}$ production rate \citep[e.g.,][]{Bacmann2003,Caselli2003,Caselli2008,Albertsson2013}. 
 
Given the ``peculiar'' spatial distribution of H$_{2}$D$^{+}$, the logical follow-up is to investigate the spatial distribution of HCO$^{+}$ and 
N$_{2}$H$^{+}$ and their deuterated isotopologs. HCO$^{+}$ and N$_{2}$H$^{+}$ are produced through the reactions: H$_{3}^{+}$ $+$ CO $\Longrightarrow$ HCO$^{+}$ $+$ H$_{2}$ and H$_{3}^{+}$ $+$ N$_{2}$ $\Longrightarrow$ N$_{2}$H$^{+}$ $+$ H$_{2}$. In dense and very cold environments both CO and N$_{2}$ are expected to be depleted and thus so do HCO$^{+}$ and N$_{2}$H$^{+}$. Given the fact that N$_{2}$H$^{+}$ is also destroyed by CO, the emission from the two species is usually spatially anti-correlated. Previous observations of N-bearing species (e.g. N$_{2}$H$^{+}$) towards prestellar cores have shown depletion resistance compared to CO, and that N$_{2}$ depletes at later times compared to CO \citep{Bergin2007,Pagani2012}. This contradicts the expectation of a similar behavior of the two species in terms of freezing out and desorption mechanisms, which is based on the ratio of their binding energies \citep[$\sim$1 under astrophysical conditions;][]{Bisschop2006}. More recent calculations and experiments on the topic are presented in \citet{Boogert2015}.

\begin{figure}[h]
\includegraphics[scale=0.42]{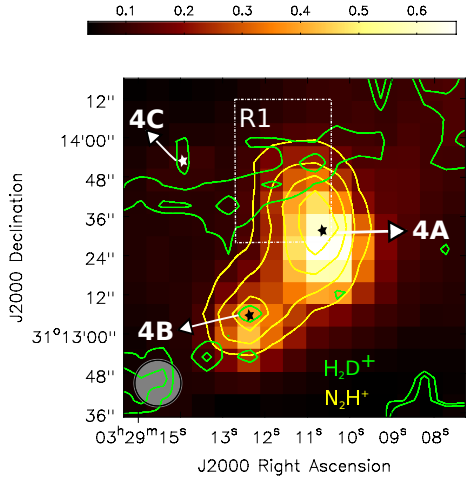}
\caption{HCO$^{+}$ integrated intensity map overplotted with H$_{2}$D$^{+}$ in green contours (0.017, 0.026 K; rms$\sim$0.005~K) and N$_{2}$H$^{+}$ in yellow contours (0.14, 0.27, 0.41 and 0.54 K; rms$\sim$0.005~K).}
\label{fig:maps_deu_3}
\end{figure}

Figure~\ref{fig:maps_deu_3} presents the integrated intensity map of HCO$^{+}$ overplotted with H$_{2}$D$^{+}$ and N$_{2}$H$^{+}$ in contours. 
This figure shows that N$_{2}$H$^{+}$ and HCO$^{+}$ 
have a similar spatial distribution, tracing mostly the protostellar envelopes and they emit significantly in almost half of the 
H$_{2}$D$^{+}$ slab towards the N-NW axis.

\begin{figure}[h]
\includegraphics[scale=0.42]{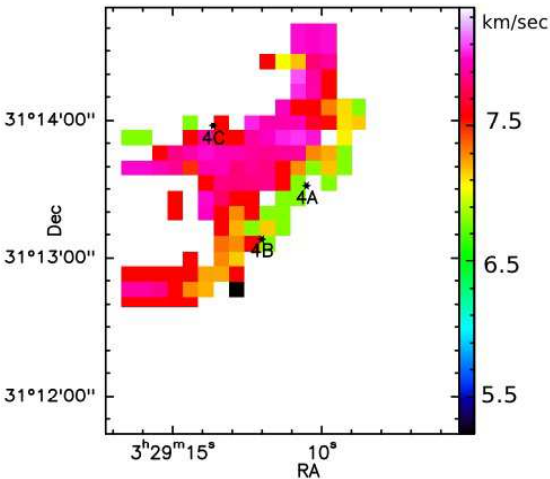}
\caption{Central velocity map of H$_{2}$D$^{+}$. The main layer of H$_{2}$D$^{+}$ is characterized by velocities of $\sim$8~km~s$^{-1}$. The emission in the S--E lobe is $>$ 3~RMS and it seems to be real.}
\label{fig:velo_h2dplus}
\end{figure}

The presence of N$_{2}$H$^{+}$ and HCO$^{+}$ in part of the H$_{2}$D$^{+}$ slab could be a result of the outflow-shock activity in the region 
that removes ices such as CO and N$_{2}$ from grains. If we assume a single layer of gas, such activity should also make H$_{2}$D$^{+}$ less abundant, 
but this is not what we observe. Previous studies including \citet{Choi2004} and \citet{Koumpia2016} provided evidence for 
a foreground cloud at $\sim$8~km~s$^{-1}$ while IRAS~4A and IRAS~4B 
are part of a smaller embedded cloud at 6.7~km~s$^{-1}$. The H$_{2}$D$^{+}$ line is shifted by $\sim$1.5 km~s$^{-1}$ relative to the rest velocity of this cloud (Fig.~\ref{fig:velo_h2dplus}). This could be an indication that the narrow component of SiO 1--0 ($\sim$7.6~km~s$^{-1}$) and the H$_{2}$D$^{+}$ ($\sim$8~km~s$^{-1}$) 
originate from the same foreground layer at the offset velocity. Such narrow SiO emission has been discovered in 
more regions (e.g., G035.39--00.33) where it has been suggested that it originates 
from cold gas associated with a low-velocity shock \citep{Duarte-Cabral2014}. H$_{2}$D$^{+}$ 
emission requires very cold conditions though and is unlikely to be associated with shock activity due to the presence of N$_{2}$H$^{+}$ and 
HCO$^{+}$. This emission probably originates from a colder layer in the foreground and the co-existence of the other two species is rather a projection effect.

\subsubsection{Modeling the deuteration}

Although high deuteration is usually connected with very cold environments, deuterated species have also been associated with shocks in a few studies \citep{Lis2002,Fontani2014,Lis2016}. We aim to model the distribution of the [DCO$^{+}$]/[HCO$^{+}$] ratio in the area covered by our JCMT maps, and examine the deuteration towards the outflow of NGC~1333 IRAS~4A.
 
In order to explore spatial variations in molecular D/H abundance ratios of the region, especially the [DCO$^{+}$]/[HCO$^{+}$] ratio, we use the kinetic temperature map as derived by \citet{Koumpia2016} in the non-LTE radiative transfer program RADEX \citep{vanderTak07} after adopting a constant H$_{2}$ density 
of 3$\times$10$^{5}$ cm$^{-3}$ as suggested in the same work. In addition to the optically thick HCO$^{+}$ \citep{Koumpia2016} we use its optically thin isotopolog H$^{13}$CO$^{+}$ which helps us derive an accurate [DCO$^{+}$]/[HCO$^{+}$]. 

\begin{figure}[h]
\includegraphics[scale=0.37, angle=270]{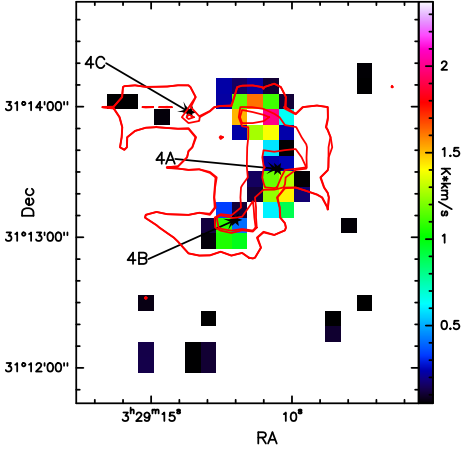}
\caption{SiO 8--7 map overplotted with the contours of [DCO$^{+}$]/[HCO$^{+}$] abundance ratio. The contours are set to 1.3$\%$ and 5.4$\%$. The most outer contour represents the area covered during the modeling (both lines $>$3RMS).}
\label{fig:maps_deu_ratio}
\end{figure}

We present the [DCO$^{+}$]/[HCO$^{+}$] ratio in Figure~\ref{fig:maps_deu_ratio}. We find ratios varying between $\sim$0.01 and 0.07$\pm$0.016 around the protostellar envelopes and the surrounding gas covering the outflow towards the north of IRAS~4A. The error estimate reflects the average of the observational uncertainties, which takes its higher values as one moves to the edges of our maps where the signal from the lines becomes more comparable to the RMS. In particular, we find a $\sim$3 times higher [DCO$^{+}$]/[HCO$^{+}$] abundance ratio towards the N--NE part of the H$_{2}$D$^{+}$ slab compared to IRAS~4A and IRAS~4B while the 
N--NW part of the slab does not show an enrichment in deuteration. The shock position on the north though (R1) is characterized 
by values equal and up to two times higher in deuteration compared to IRAS~4A. The observed differences are significant given our error estimates. The observed [DCO$^{+}$]/[HCO$^{+}$] abundance ratio is 3 orders of magnitude higher than the cosmic ratio [D]/[H] of 1.5$\times$10$^{-5}$ \citep{Linsky95} and it reflects the strong deuteration occurring at those early embedded stages of star formation.

An enhancement of DCO$^{+}$ abundance requires an enhancement of H$_{2}$D$^{+}$ abundance and the presence of CO in the gas phase, since it is produced through the reaction H$_{2}$D$^{+}$ $+$ CO $\Longrightarrow$ DCO$^{+}$ $+$ H$_{2}$. As we described in Section~\ref{cold}, H$_{2}$D$^{+}$ is enhanced in very cold conditions($<$ 20--30~K). This is in agreement with the estimated kinetic temperature towards the shock position (20--30~K). Under the observed conditions CO freezes out though. To explain the observed DCO$^{+}$ abundance enhancement towards the shock position we conclude that this emission is dominated by deuterated species originally formed in the gas phase after the removal of CO from grains at low gas kinetic temperature ($<$ 20~K). Such release of CO into the gas phase at such low temperatures can occur during the passage of a shock wave.
 
\section{Chemistry of the IRAS~4A envelope} 

\subsection{RATRAN -- model setup}

\label{ratran}

In order to estimate molecular abundance profiles through the envelope of NGC~1333~IRAS~4A, we ran the Monte Carlo radiative transfer code RATRAN \citep{Hogerheijde2000} and produced synthetic line emission. RATRAN takes into account the physical structure of the source including temperature and density gradients, kinematics and 
continuum dust emission and absorption. 

The line spectra of NGC~1333 IRAS~4A are generally broad ($>$5~km~s$^{-1}$) and the wings are very prominent in many species. The physical models are determined using continuum observations that characterize the protostellar envelope. Therefore we focus on the narrow component of the lines. For this purpose we perform a multi-Gaussian fit and use only the narrow component for modeling. Some of the lines show heavy self-absorption making their Gaussian fitting very inaccurate and thus, we chose not to include them in our models. 

The H$_{2}$O 2$_{11}$--2$_{02}$ line shows only a broad component (Figure~\ref{fig:prof1}) and we observe no isotopologs, making the identification of the envelope component unreliable. The situation for SiO is similar, thus we do not model these species. We preferably model the isotopologs when present, and otherwise model the narrow component of the main species when multiple Gaussian fitting is possible (e.g., H$^{13}$CO$^{+}$, C$^{17}$O and C$^{18}$O, H$^{13}$CN). We use fixed local ISM isotopic ratios of 
$^{12}$C/$^{13}$C $=$ 60, $^{16}$O/$^{17}$O $=$ 2000, and $^{16}$O/$^{18}$O $=$ 560 \citep{Wannier80,Wilson94,Wilson1999}. In addition we assume an ortho to para ratio of 3 for the two collisional partners o--H$_{2}$ and p--H$_{2}$. Dust continuum radiation is taken into account using the dust opacity OH5 taken from \citet{Ossenkopf94}, which corresponds to dust grains with thin ice mantles. 

\begin{figure}[h]
\includegraphics[scale=0.5]{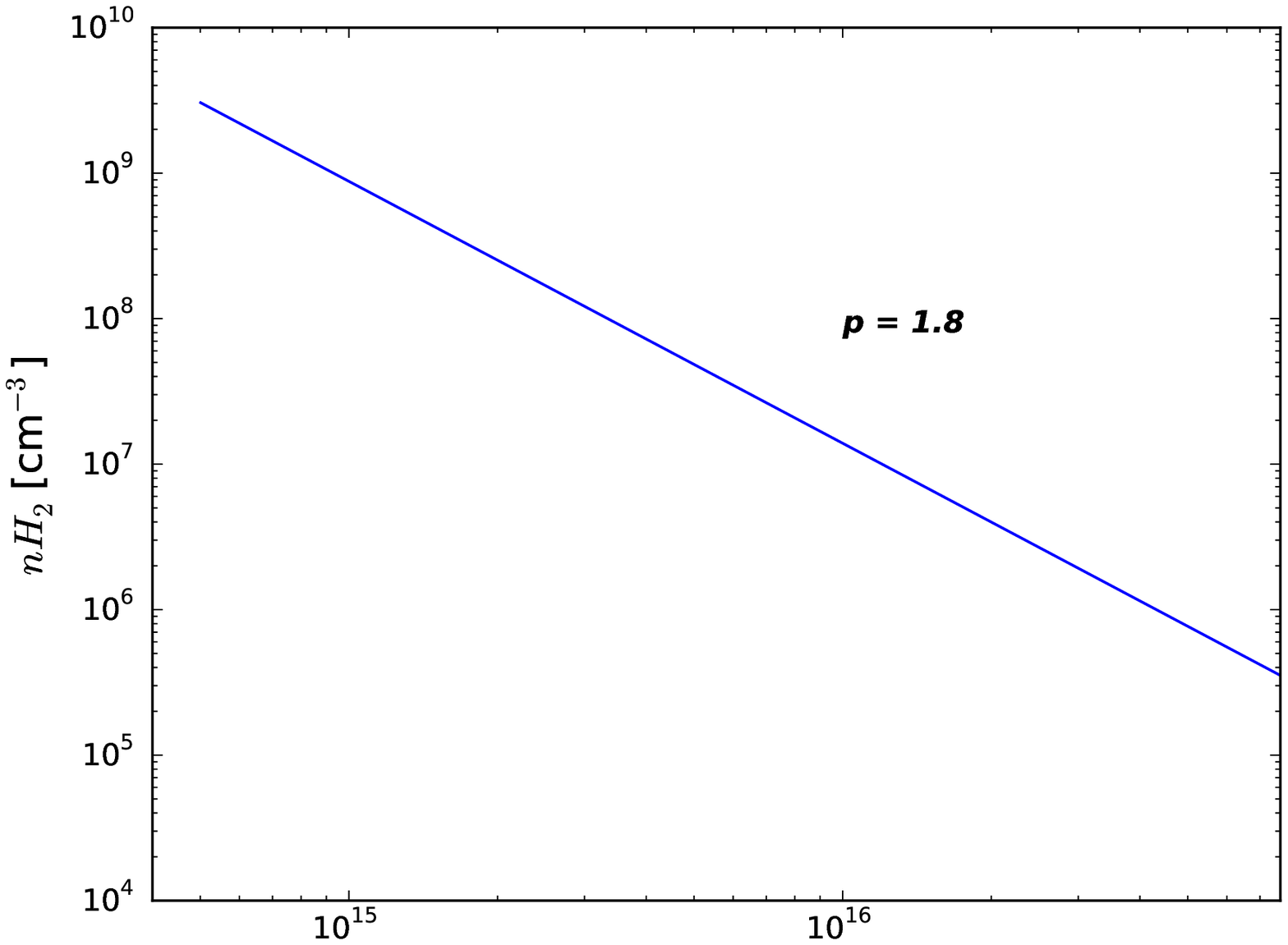} 

\includegraphics[scale=0.5]{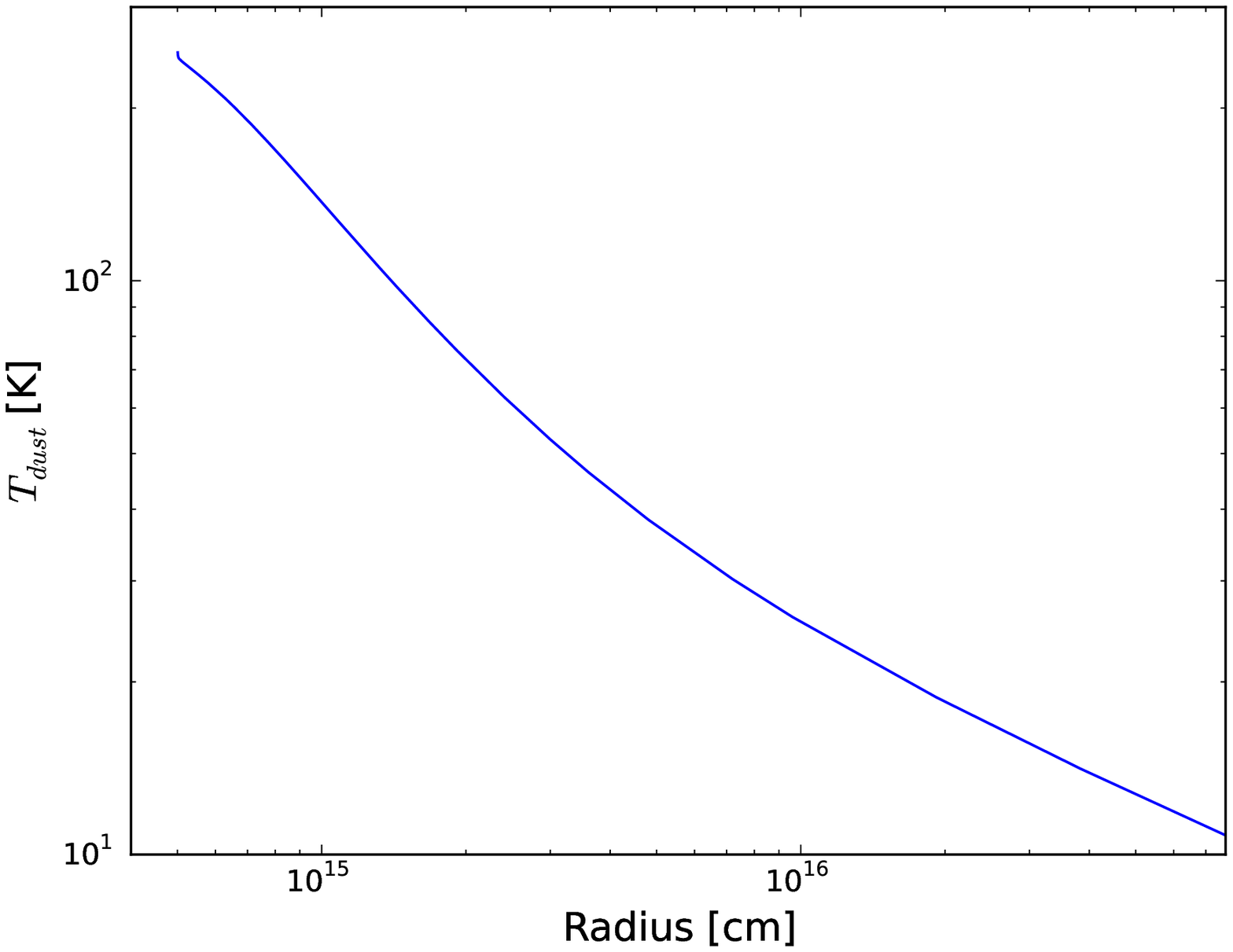}
\caption{The density (top) and temperature (bottom) profiles of NGC~1333~IRAS~4A as derived by \citet{Krist2012}.}
\label{fig:profiles}
\end{figure}

We ran RATRAN applying the density and temperature radial profiles of NGC~1333~IRAS~4A (Figure~\ref{fig:profiles}) as determined by \citet{Krist2012}. The density profile is assumed to be a power-law, n(r) $=$ n$_{0}$ $\times$(r/r$_{0}$)$^{-p}$, with a best-fit index of p $=$ 1.8. The density and temperature 
profiles were derived from the best-fit dust model assuming that the gas is entirely molecular and using a mean molecular mass of 2.4 amu and a gas-to-dust ratio of 100. For our calculations we defined a grid of 19 spherical shells from r$_{in}$ $=$ 5$\times$10$^{14}$~cm (33 au) up to r$_{out}$ $=$ 7.7$\times$10$^{16}$~cm (5147 au) where the dust temperature is 250~K and 10~K and the H$_{2}$ density is 3.05$\times$10$^{9}$cm$^{-3}$ and 3.5$\times$10$^{5}$cm$^{-3}$, respectively. 
The original model extends further out as seen in Figure~\ref{fig:profiles} but our adopted profiles better represent the size and outer temperature of the 
envelope of a protostar ($\sim$5000~au, 10~K). We also assumed thermal equilibrium between dust and gas at those high densities. 

The papers from which the collisional rate coefficients of the main isotopologs with H$_{2}$ were adopted are presented in Table~\ref{authors}. The collisional data of H$_{2}$S are scaled from the files of ortho- and para-H$_{2}$O as calculated by \citet{Dubernet2009,Daniel2010,Daniel2011}. For isotopologs and the deuterated species, the same collision data were used as for the main isotopolog.

\begin{table}[ht]
\caption{References for the collisional rate coefficients of the species modeled in this paper.}
\centering
\small\addtolength{\tabcolsep}{-4.7pt}
\begin{tabular}{c c}
\hline\hline
Species & Authors \\
\hline\hline

CO & \citet{Yang2010} \\
HCO$^{+}$ & \citet{Flower1999} \\
N$_{2}$H$^{+}$ & \citet{Flower1999} \\
HCN & \citet{Dumouchel2010} \\
CS & \citet{Lique2006, Lique2011} \\
HNC & \citet{Dumouchel2010} \\
H$_{2}$CO & \citet{Wiesenfeld2013} \\
CH$_{3}$OH & \citet{Rabli2010} \\
C$_{2}$H & \citet{Muller2005} \\
CN & \citet{Lique2006, Lique2011} \\
OCS & \citet{Green1978} \\
SO$_{2}$ & \citet{Green1995} \\
H$_{2}$CS & \citet{Wiesenfeld2013} \\
\hline
\end{tabular}
\label{authors}
\end{table}

We assumed a static envelope without infall or expansion and we used a range of abundances typically varying between 10$^{-7}$ and 10$^{-12}$ in 
order to constrain the abundance profile that best fits the observations. The turbulent line width was fixed to 1.9~km~s$^{-1}$, which is the average value we found for the narrow component of most species after fitting multiple Gaussians. Modeling the observed lines with a constant abundance is the only way when it comes to species that show only a few transitions, but might not always be a realistic approach. Several molecules have been suggested to be present in volatile ice mantles on dust grain surfaces at temperatures $<$ 20--110~K. The exact temperature depends on species and the surface composition \citep{Bisschop2006,Herbst2009}. We chose to apply jump-like abundances at 100~K for H$_{2}$CO and CH$_{3}$OH for which more transitions are available and for which a constant abundance does not result in a good fit.


\subsection{RATRAN -- model results}

The resulting abundances from the process described above are presented in 
Table~\ref{abundances} which compares this 
low-mass case with a high-mass case from the literature \citep[AFGL~2591;][]{Kazmierczak2015}. We find observed abundance profiles for the low-mass protostellar envelope (NGC~1333~IRAS~4A) that are 
systematically 1-2 orders of magnitude lower than the high-mass protostellar envelope (AFGL~2591). Although, one can extract this information from previous studies on high and low-mass protostellar envelopes \citep[e.g., H$_2$CO, CH$_3$OH][]{vanderTak2000,Maret2004,Maret2005}, this study is the first direct comparison among high-mass and low-mass protostellar envelopes with the use of datasets of the same instruments and similar methodology (radiative transfer and chemical models). We attribute the observed differences to the absence of a freeze-out zone (i.e., temperature differences) towards the high-mass protostellar envelope. Although AFGL~2591 is a much more distant object (3.3 kpc) than NGC~1333 IRAS~4A (235 pc), which could potentially affect our results, our models take the different distances into account.

Table~\ref{abundances} also presents the resulting abundances with respect to CO. In particular, we find a similarity in the observed abundances with respect to CO within a factor of a few among the low- and the high-mass protostellar envelope. \citet{Herbst2009} present complex organic molecule abundances relative to CH$_{3}$OH for low- and high-mass YSOs and find similar results. Our results suggest that gas phase CO/H$_{2}$ measurements are essential for comparison among low- and high-mass protostellar envelopes in the future. 

The observed data for CO, H$_{2}$CO, and CH$_{3}$OH do not fit well for constant abundances and are discussed below.



\begin{table}
\caption{Constant empirical abundances estimated with RATRAN for the envelope 
of NGC~1333~IRAS~4A based on HIFI and JCMT observations. The table 
contains also the abundances of AFGL~2591 (outer envelope) for direct comparison.}
\centering
\small\addtolength{\tabcolsep}{-4.7pt}
\begin{tabular}{c c c c c}
\hline\hline
Molecule & \multicolumn{2}{c}{Abundance wrt H$_{2}$} & \multicolumn{2}{c}{Abundance wrt CO} \\
 & IRAS~4A & AFGL~2591 & IRAS~4A & AFGL~2591 \\
 &  & (outer) &  &\\
\hline\hline

CO & 3 $\times$ $10^{-5}$ & 2 $\times$ $10^{-4}$ & 1 & 1\\
HCO$^{+}$ & 1 $\times$ $10^{-9}$ & 3 $\times$ $10^{-8}$ & 3.3 $\times$ $10^{-5}$ & 1.5 $\times$ $10^{-4}$\\
N$_{2}$H$^{+}$ & 8 $\times$ $10^{-11}$ & 8 $\times$ $10^{-10}$ & 2.7 $\times$ $10^{-6}$ & 4 $\times$ $10^{-6}$\\
HCN & 3 $\times$ $10^{-10}$ & 5 $\times$ $10^{-7}$ & 1 $\times$ $10^{-4}$ & 2.5 $\times$ $10^{-3}$ \\
CS & 3 $\times$ $10^{-9}$ & 4 $\times$ $10^{-8}$ & 1 $\times$ $10^{-4}$ & 2 $\times$ $10^{-4}$ \\
HNC & 8 $\times$ $10^{-11}$ & 1 $\times$ $10^{-8}$ & 2.7 $\times$ $10^{-6}$ & 5 $\times$ $10^{-5}$ \\
H$_{2}$CO & 4 $\times$ $10^{-10}$ & 1 $\times$ $10^{-8}$ & 1.3 $\times$ $10^{-5}$ & 5 $\times$ $10^{-5}$\\
CH$_{3}$OH & 1 $\times$ $10^{-8}$ & 8 $\times$ $10^{-8}$ & 3.3 $\times$ $10^{-4}$ & 4 $\times$ $10^{-4}$\\
DCO$^{+}$ & 1 $\times$ $10^{-11}$ & \ldots & 3.3 $\times$ $10^{-7}$ & \ldots\\
C$_{2}$H &  6 $\times$ $10^{-10}$ & 8 $\times$ $10^{-8}$ & 3 $\times$ $10^{-5}$ & 4 $\times$ $10^{-4}$ \\
CN & 1 $\times$ $10^{-10}$ & 1 $\times$ $10^{-9}$ & 3.3 $\times$ $10^{-6}$ & 5 $\times$ $10^{-6}$ \\
OCS & 6 $\times$ $10^{-9}$ & 4 $\times$ $10^{-8}$ & 2 $\times$ $10^{-4}$ & 2 $\times$ $10^{-4}$ \\
SO$_{2}$ &  4 $\times$ $10^{-10}$ & 5 $\times$ $10^{-9}$ & 1.3 $\times$ $10^{-5}$ & 2.5 $\times$ $10^{-5}$\\
HDCO & 5 $\times$ $10^{-11}$ & \ldots & 1.7 $\times$ $10^{-6}$ & \ldots \\
H$_{2}$CS & 4 $\times$ $10^{-10}$ & 4 $\times$ $10^{-9}$ & 1.3 $\times$ $10^{-5}$ & 2 $\times$ $10^{-5}$ \\
D$_{2}$CO & 5 $\times$ $10^{-11}$ & \ldots & 1.7 $\times$ $10^{-6}$ & \ldots \\
H$_{2}$S & 4 $\times$ $10^{-10}$ & 4 $\times$ $10^{-9}$ & 1.3 $\times$ $10^{-5}$ & 2 $\times$ $10^{-5}$ \\
\hline
\end{tabular}
\label{abundances}
\end{table}

\begin{table}
\caption{As Table~\ref{abundances}, but for jump and drop abundance profiles.}
\centering
\small\addtolength{\tabcolsep}{-4.7pt}
\begin{tabular}{c c c c c c }
\hline\hline
 & \multicolumn{2}{c}{IRAS~4A} & \multicolumn{2}{c}{AFGL~2591} & Jump/ \\Molecule & X$_{IN}$ & X$_{OUT}$ & X$_{IN}$ & X$_{OUT}$ & Drop at \\
\hline\hline

CO & 1 $\times$ $10^{-4}$ & 3 $\times$ $10^{-6}$ & \multicolumn{2}{c}{2 $\times$ $10^{-4}$ (const.)} & 25~K \\
H$_{2}$CO & 4 $\times$ $10^{-8}$ & 4 $\times$ $10^{-10}$ & 1 $\times$ $10^{-11}$ & 1 $\times$ $10^{-8}$ & 100~K \\
CH$_{3}$OH & 1 $\times$ $10^{-8}$ & 7 $\times$ $10^{-10}$ & 8 $\times$ $10^{-7}$ & 8 $\times$ $10^{-8}$ & 100~K \\
\hline
\end{tabular}
\label{abundances2}
\end{table}

\begin{figure}[h]
\includegraphics[scale=0.35, angle=90]{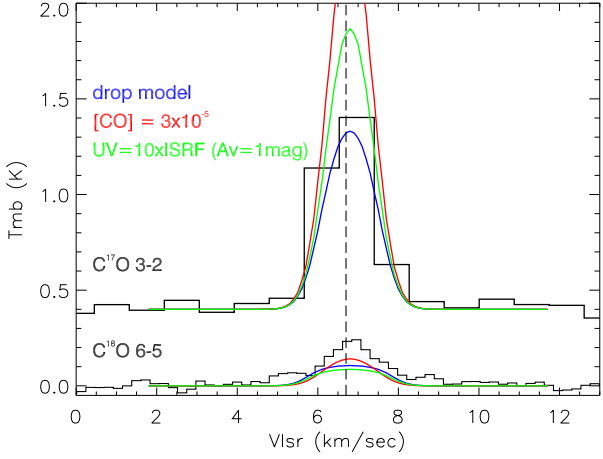} 
\caption{Observed C$^{17}$O 3-2 and C$^{18}$O 6-5 line profiles (black) overplotted with the modeled line profiles resulting in RATRAN after applying a) 
$\sim$1.5 orders of magnitude drop in CO abundance at the snowline (blue) or b) CO constant abundance of 3$\times$10$^{-5}$ (red) and the best-fit abundance profile as derived from our chemical models for UV$=$10$\times$ISRF at A$_{V}$$=$1mag (black line in Figure~\ref{fig:chem31}).}
\label{fig:co_models}
\end{figure}

\subsection{Drop models}

We chose to work with CO isotopologs (C$^{17}$O and C$^{18}$O) because they are optically thin and have narrow line-widths and thus trace the quiescent envelope. The fact that we do not use the optically thick broad $^{13}$CO lines avoids systematic errors in our solution. Figure~\ref{fig:co_models} shows the observed line profiles of C$^{17}$O 3-2 and C$^{18}$O 6-5 with the best fit abundance profile model
overplotted after adopting a constant CO abundance of 3$\times$10$^{-5}$ (scaled accordingly for the isotopologs) and a drop model, where the CO abundance profile shows a drop in the freeze-out zone and increases again, for comparison. Using a constant abundance, it was impossible to reproduce the intensities of both C$^{17}$O and C$^{18}$O lines. In particular, the constant CO abundance reproduces the C$^{18}$O 6--5 transition very well, but it overproduces the C$^{17}$O 3-2 by
a factor of $\sim$2. Having additional transitions would help to better constrain the abundance profile. The fact that the higher--J C$^{18}$O 6--5 can be reproduced by a constant abundance profile indicates that this emission does not originate from the snowline of CO, which is characterized by a drop in the abundance.

Taking into account the freeze--out zone one would expect a drop of the CO abundance but this alone is also not able to reproduce the line intensities. Thus, we use a similar drop profile as suggested by \citet{Yildiz2010,Yildiz2012}, in which the abundance of CO drops in the freeze-out zone but rises again in the outer envelope. For the evaporation temperature of CO we chose to use the lower limit from the laboratory, which is $\sim$25~K \citep{Collings2003,Munoz2010,Luna2014}. We find the best fit model is one with an inner abundance for T$>$25~K of 1$\times$10$^{-4}$ which drops to 3$\times$10$^{-6}$ at the coldest part of the envelope and rises again to the canonical value of 1$\times$10$^{-4}$ towards the outer envelope where external UV radiation becomes important. This is in good agreement with the additional C$^{18}$O transitions from \citet{Yildiz2012} who found a CO abundance of 6$\times$10$^{-5}$ for the warmest part of the envelope (50$<$T$<$250~K) dropping to 3$\times$10$^{-6}$ for the coldest part ($<$30~K) where depletion is prominent and a jump up to 
3$\times$10$^{-4}$ again for the outer envelope. 

\subsection{Jump models}

The assumption of a constant abundance does not work well for all transitions of H$_{2}$CO and CH$_{3}$OH, 
indicating that a jump model is required. In jump models an abundance enhancement by up to a few orders of magnitude is applied in the inner warmest regions, where the temperature is sufficient to evaporate the grain mantles ($>$100~K).

\begin{figure}[h]
\includegraphics[scale=0.35, angle=90]{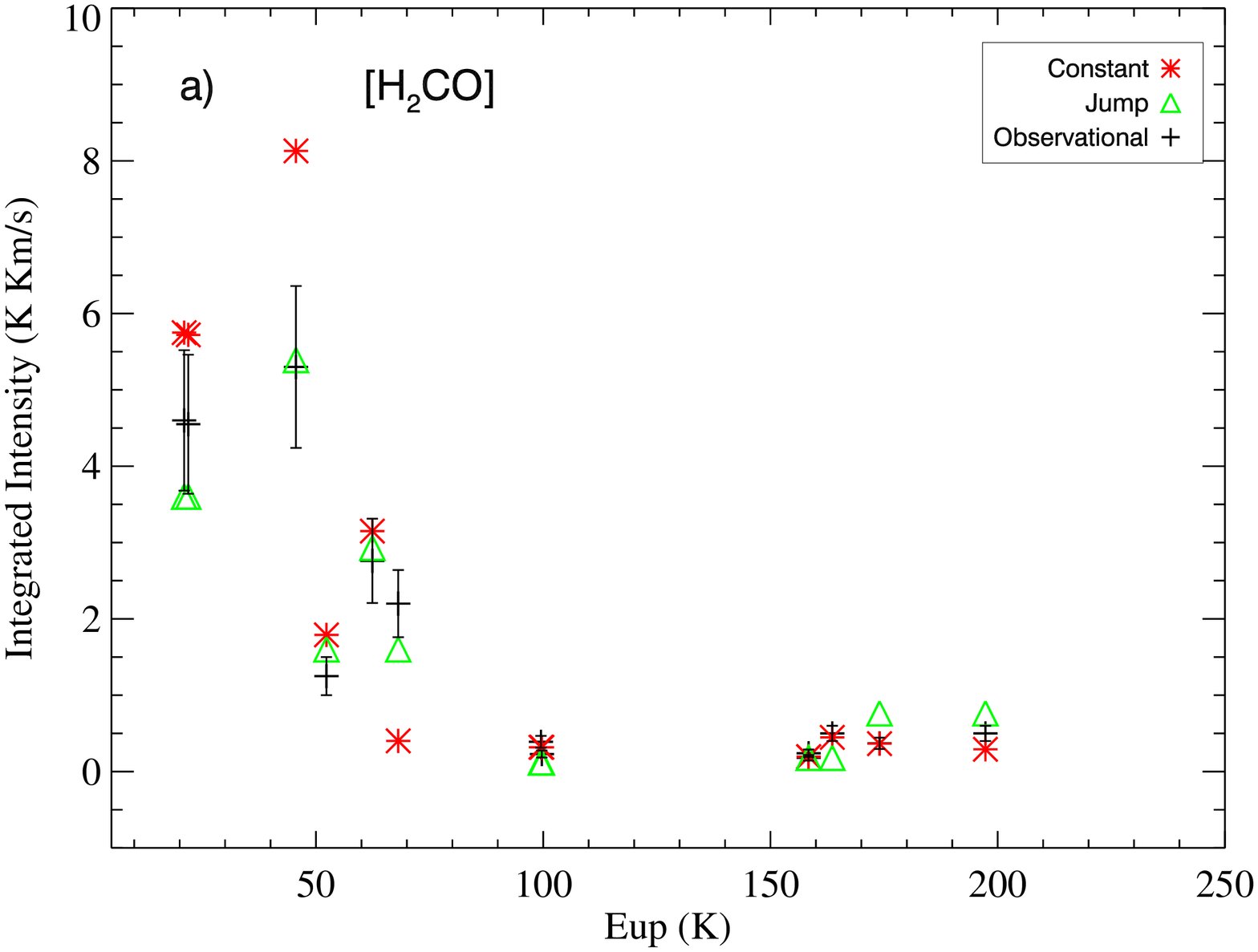}

\includegraphics[scale=0.35, angle=90]{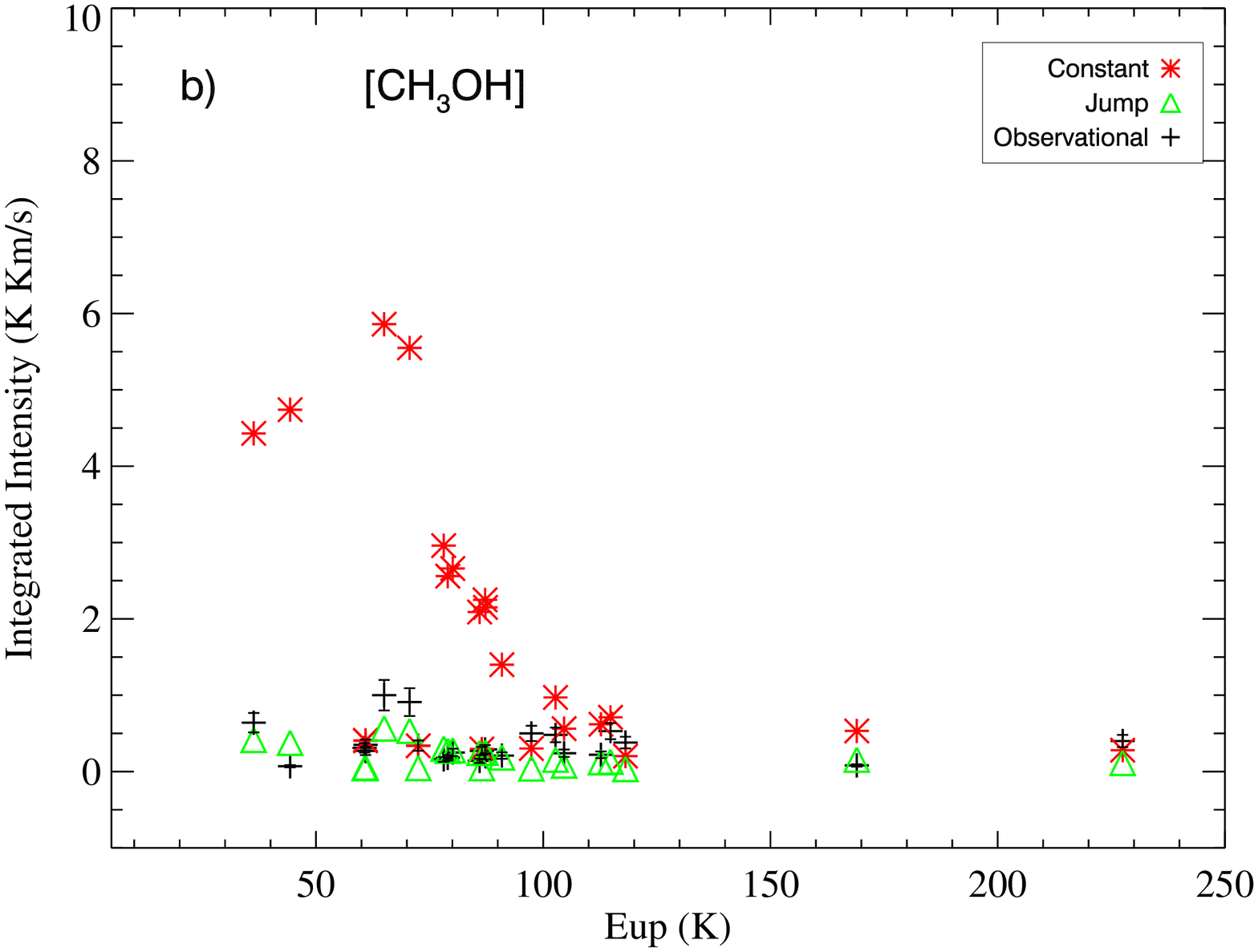}
\caption{Integrated intensities of the observed (black) and modeled a) H$_{2}$CO and b) CH$_{3}$OH transitions in the observed E$_{up}$ range using a constant 
abundance of [H$_{2}$CO] $=$ 4$\times$10$^{-10}$ and [CH$_{3}$OH] $=$ 1$\times$10$^{-8}$ (red) and a jump model (green). The errors of the observed values are $\sim$20\% of the measured value.}
\label{fig:theor_obs}
\end{figure}

 Figure~\ref{fig:theor_obs}a shows the integrated intensities of the observed H$_{2}$CO transitions using HIFI and JCMT towards IRAS~4A 
overplotted with the convolved ($\sim$15--33$\arcsec$) synthetic emission as calculated with RATRAN for a constant abundance of [H$_{2}$CO] $=$ 4$\times$10$^{-10}$ and jump models at $\sim$100~K.  
 Our jump model at $\sim$100~K results in X$_{IN}$ $=$ 4$\times$10$^{-8}$ and X$_{OUT}$ $=$ 4$\times$10$^{-10}$ and reproduces the lower-energy lines better but the higher-energy lines slightly worse compared to the constant abundance model. \citet{Maret2004} found H$_{2}$CO abundance about a factor of two lower for the inner envelope (2$\times$10$^{-8}$) and a factor of two lower for the outer envelope 
compared to our jump model. Differences up to 2 orders of magnitude between H$_{2}$CO abundance of the inner and outer low-mass protostellar envelopes have been observed before \citep[IRAS~16293--2422;][]{Ceccarelli2000}.

Figure~\ref{fig:theor_obs}b shows the integrated intensities of the observed CH$_{3}$OH transitions using HIFI and JCMT towards IRAS~4A 
overplotted with the convolved ($\sim$15--33$\arcsec$) synthetic emission as calculated with RATRAN for [CH$_{3}$OH] $=$ 1$\times$10$^{-8}$. 
\citet{Maret2005} found an upper limit of 1$\times$10$^{-8}$ for the CH$_{3}$OH abundance in the inner envelope and the same abundance as we do in the outer envelope. 

The models we used are characterized by the same power law for the density profiles and the temperature profiles also follow a similar pattern. The main differences in the two methods are the additional HIFI transitions we use, the new collisional data, and the new physical model derived by \citet{Krist2012} including PACS observations, which were the smallest-scale used data. \citet{Maret2004,Maret2005} used the physical model derived by \citet{Jorgensen2002}.

\subsection{Deuteration}

Using the abundances reported in Table~\ref{abundances} we find an HDCO/H$_{2}$CO ratio of 
$\sim$10\% which is comparable with the one found by \citet{Loinard2000,Loinard2001} towards IRAS~16293--2422 ($\sim$10--15\%) 
and the value reported for Orion KL ($\sim$15\%) \citep{Liu2011}. We also find a very high D$_{2}$CO over H$_{2}$CO ratio ($\sim$10\%) although this is based on only one 
transition. Very high values between $\sim$5--10\% of the relative D$_{2}$CO 
abundance have been found before towards the low-mass protostar 
IRAS~16293--2422 \citep{Ceccarelli1998,Loinard2000}, which is approximately 2 orders of 
magnitude higher than in Orion KL. The average gas kinetic temperature 
of IRAS~4A is $\sim$45~K \citep{Koumpia2016} and thus such high 
deuteration cannot be explained if D$_{2}$CO is only formed in the gas-phase. 
On the other hand, D$_{2}$CO may have been enriched in the dust grains 
in low-mass protostellar environments during the cold, 
dense pre-collapse period  
followed by its evaporation thus it is probably formed via the 
grain surface reactions. Lastly, we find DCO$^{+}$/HCO$^{+}$ $=$ 1\% in agreement with what is reported by \citet{Loinard2000,Loinard2001} for IRAS~16293--2422.

\section{Chemical modeling} \label{chem}

\subsection{Pseudo time-dependent model}
\label{sec:chem_model}

In this section we compare the empirical abundance profiles of the studied species obtained with RATRAN with those resulting 
from chemical models for the same adopted physical structure of NGC~1333~IRAS~4A. Our goal is to a) better understand the chemical processes that take place in NGC~1333~IRAS~4A and b) constrain its age. 

To compare the observationally derived abundance profiles, we used the same 1D physical model as in Section~\ref{ratran} and the same gas-grain
time-dependent chemical model as in \citet{Kazmierczak2015}. This allows us to compare chemistry
modeling results between a low-mass and a high-mass protostellar envelope. The chemical model is based on the chemical kinetics
ALCHEMIC code of \citet[]{Semenov2010} and a gas-surface network with deuterium fractionation
\citep{Albertsson2013,Albertsson2014}.
The original, non-deuterated chemical network stems from the osu.2007 ratefile developed by
the group of Eric~Herbst~\footnote{http://web.archive.org/web/20081204232936/} \citep{Garrod2006}.
The network is supplied with a set of approximately $ 1\,000$ high-temperature neutral-neutral reactions
from \citet{Harada2010,Harada2012} and updated as of June 2013, using the KIDA database\footnote{http://kida.obs.u-bordeaux1.fr}.

All H-bearing species reactions in this network were cloned by adding D, with the exception of molecules
with the $-$OH functional group.
Primal isotope exchange reactions for H$_3^+$ as well as CH$_3^+$ and C$_{2}$H$_{2}^+$ from \citet{Roberts2000,Roberts2004,Roueff2005} were included.
In cases where the position of the deuterium atom in a reactant or in a product was ambiguous, a statistical branching approach
was used \citep[for further details please consult][]{Albertsson2013}. In \citet{Albertsson2014} this deuterium network was
further extended by adding ortho- and para- forms of H$_2$, H$_2^+$ and H$_3^+$ isotopologs
and the related nuclear spin-state exchange processes.

We adopted the cosmic ray ionization rate of \mbox{$5 \times
10^{-17}$~s$^{-1}$,} as derived by \citet{vanderTak2000c} and \citet{Indriolo2015}. In addition, we ran chemical simulations with a
higher value of \mbox{$10^{-16}$~s$^{-1}$}.
The photodissociation and photoionization rates in the model are adopted for
a 1D slab model from \citet{vanDishoeck2006}. To mimic the presence of an outflow cavity in this source,
we also considered several other models with enhanced UV irradiation. For that, we use a scaled interstellar UV radiation field in \citet{Draine1978} units and moderate dust extinctions of 10, 3, 2, and 1 mag. The pre-computed photoionization and dissociation rates for a 1D plane-parallel slab model were used \citep[see Eq. 5;][]{Semenov2010}.
The self- and mutual-shielding of CO and H$_2$ from external dissociating radiation was calculated as in \citet{Semenov2011}. Water self-shielding is neglected. 

The grains are assumed to be uniform and spherical and made of amorphous olivine with a density of $3$~g\,cm$^{-3}$ and a radius of $0.1\,\mu$m. If grains were on average bigger than $0.1\,\mu$m it would slow down freeze-out which would increase the CO gas/ice ratio, especially for short timescales (1000 years; Figure~\ref{fig:chem2}). The same effect would be seen for all other species that can freeze out. This would enhance the overall gas-phase molecular abundances with respect to our standard model. 

Each grain provides $\approx 1.88\times10^6$ surface sites
\citep[][]{Biham2001} for freeze-out of gaseous molecules. We use the desorption energies E$_{des}$ from \citep{Garrod2006}. We compute the diffusion energies of surface reactants by multiplying their binding (desorption) energies by 0.4. In the current literature, factors of 0.3--0.8 are commonly used, motivated by existing data on stable species \citep[e.g.,][]{Cuppen2017}, although a firm physical basis for such a scaling is lacking.

The gas-grain interactions include sticking of neutral species and electrons to dust grains with 100\%
probability and desorption of ices by thermal, cosmic ray, and UV-driven processes.
In our model we use a single E$_{des}$ for H$_2$ $=$ 450~K, which properly describes H$_2$ diffusion over the dust surface and the H$_2$ binding to water or silicate surfaces. However, the H$_2$--H$_2$ binding energy is much lower (23~K). This means that as soon as there is a H$_2$ monolayer on a surface, further freeze-out of H$_2$ would be compensated by immediate desorption of H$_2$ back to the gas phase. Our models cannot capture this process well and thus we simulate it by not allowing H$_2$, HD, and D$_2$ to stick to grains. There is still enough H$_2$, HD, and D$_2$ forming on grains though, and thus their surface abundances are not zero \citep[See also;][]{Hincelin2015,Wakelam2016}.

The UV photodesorption yield of $10^{-3}$ was used
\citep[e.g.,][]{Oberg2009,Fayolle2011,Fayolle2013}. This is consistent for CO and some other light species, but it drops to $\sim$10$^{-5}$ or lower for anything bigger \citep[e.g., CH$_{3}$OH;][]{Mart2016,Berti2016}. This is one of the limitations of our method.
Photodissociation processes of solid species are taken from \citet{Garrod2006}.

Surface recombination is assumed to proceed through the classical Langmuir-Hinshelwood
mechanism \citep[e.g.,][]{HHL92}. We do not allow tunneling of surface species via the potential wells of the adjacent surface sites.
To account for hydrogen tunneling through barriers of surface reactions, we have employed Eq.~(6) from \citet{HHL92}, which describes a tunneling probability through a rectangular barrier with a thickness of 1~\AA.

For each surface recombination, 
we assume there is a 1\% probability for the products to leave the grain due to the partial
conversion of the reaction exothermicity into breaking the surface-adsorbate bond
\citep{Vasyunin2013}.
Following experimental studies on the formation of molecular hydrogen on amorphous dust grains by \citet{Katz1999}, the
standard rate equation approach to the surface chemistry is utilized.
In addition, dissociative recombination and
radiative neutralization of molecular ions on charged grains and grain re-charging are taken into account.

With this network and $10^{-5}$ relative and $10^{-20}$ absolute tolerances, the 1D IRAS~4A model
takes about 1~minute using Core-i7 2.5~GHz CPU (OS X 10.11, gfortran~6-x64) to compute over
\mbox{$10^6$}~years. This time span encompasses the likely age of this object.
%
%
\subsection{Initial abundances}
\label{sec:in_abunds}
%
%

\begin{table}
\setlength{\extrarowheight}{2pt}
\caption{Initial abundances for chemical modeling.}             
\label{tab:init_abunds_lm}
\centering                          
\begin{tabular}{ll}        
\hline\hline
Species & Abundance \\
\hline                        
\multicolumn{2}{c}{``Low metals''  (LM), ${\rm C/O} = 0.44$}\\
\hline
o-H$_2$ &   $3.75\, \times$ 10$^{-1}$  \\
p-H$_2$ &   $1.25\, \times$ 10$^{-1}$  \\
HD      &   $1.55\, \times$ 10$^{-5}$  \\
He      &   $9.75\, \times$ 10$^{-2}$  \\
C       &   $7.86\, \times$ 10$^{-5}$  \\
N       &   $2.47\, \times$ 10$^{-5}$  \\
O       &   $1.80\, \times$ 10$^{-4}$  \\
S       &   $9.14\, \times$ 10$^{-8}$   \\
Si      &   $9.74\, \times$ 10$^{-9}$   \\
Na      &   $2.25\, \times$ 10$^{-9}$   \\
Mg      &   $1.09\, \times$ 10$^{-8}$   \\
Fe      &   $2.74\, \times$ 10$^{-9}$   \\
P       &   $2.16\, \times$ 10$^{-10}$   \\
Cl      &   $1.00\, \times$ 10$^{-9}$    \\
\hline                        
\end{tabular}
\end{table}

\begin{table}
\setlength{\extrarowheight}{2pt}
\caption{The top 25 of the initially abundant molecules for the NGC~1333~IRAS4A chemical modeling.}
\label{tab:init_abunds_afgl}
\centering                          
\begin{tabular}{ll}        
\hline\hline
Species & Abundances   \\
\hline                        
\multicolumn{2}{c}{PSC-LM model with ${\rm C/O} = 0.44$} \\
\hline                        
p-H$_2$     &   $3.77\, \times$ 10$^{-1}$  \\ 
o-H$_2$     &   $1.23\, \times$ 10$^{-1}$  \\ 
He          &   $9.75\, \times$ 10$^{-2}$   \\
H           &   $5.25\, \times$ 10$^{-4}$   \\
H$_2$O$^*$  &   $5.53\, \times$ 10$^{-5}$  \\ 
CO$^*$      &   $4.05\, \times$ 10$^{-5}$   \\
CO          &   $3.26\, \times$ 10$^{-5}$   \\
O$_2$       &   $1.79\, \times$ 10$^{-5}$   \\
HD          &   $1.52\, \times$ 10$^{-5}$   \\
N$_2$       &   $7.39\, \times$ 10$^{-6}$    \\
NH$_3^*$    &   $5.64\, \times$ 10$^{-6}$   \\
O           &   $5.59\, \times$ 10$^{-6}$   \\
O$_2^*$     &   $4.12\, \times$ 10$^{-6}$   \\
CH$_4^*$    &   $3.64\, \times$ 10$^{-6}$  \\ 
N$_2^*$     &   $1.76\, \times$ 10$^{-6}$  \\ 
H$^*$       &   $6.03\, \times$ 10$^{-7}$   \\
C$_3$H$_2^*$&   $4.48\, \times$ 10$^{-7}$   \\
OH          &   $3.43\, \times$ 10$^{-7}$  \\ 
H$_2$O      &   $2.79\, \times$ 10$^{-7}$   \\
HNO$^*$     &   $2.40\, \times$ 10$^{-7}$   \\
NO          &   $2.22\, \times$ 10$^{-7}$   \\
N           &   $1.36\, \times$ 10$^{-7}$   \\
HDO$^*$     &   $1.35\, \times$ 10$^{-7}$    \\
CO$_{2}$      &   $1.32\, \times$ 10$^{-7}$   \\
CO$_{2}^*$    &   $1.19\, \times$ 10$^{-7}$   \\
\hline                        
\hline                        
\end{tabular}
  \begin{list}{}{}
        \item[* -- ] denotes frozen species.
   \end{list}
\end{table}

To set the initial abundances, we calculated the chemical evolution of a 0D molecular cloud
with $n_{\rm H} = 2\times10^4$~cm$^{-3}$, T $ = 10$~K, and $A_{\rm V}=10$~mag over 1~Myr (model ``PSC-LM'').
For that, the neutral ``low metals'' (LM) elemental
abundances of \citet{Graedel1982,Agundez2013} were used, with the solar ${\rm C/O} = 0.44$,
initial ortho/para H$_2$ of 3:1, hydrogen being fully in molecular form, and deuterium locked up in HD (see
Table~\ref{tab:init_abunds_afgl}).

%
%
\subsection{Error estimations} \label{chem-errors}
%
%
The problem of uncertainties of the calculated abundances is well known in chemical studies of various astrophysical environments, ranging from dark clouds to hot cores \citep[see, e.g., ][]{Dobrijevic2003,Vasyunin2004,Wakelam2005,Vasyunin2008}.
The error budget of the theoretical abundances is determined by both the uncertainties in physical conditions in the object and, to a larger degree, by uncertainties in the adopted reaction rate coefficients and their barriers. Poorly known initial conditions for chemistry may also play a role here.

In order to estimate the chemical uncertainties rigorously, one needs to perform a Monte Carlo modeling by varying reaction rates within their error bars and re-calculating the chemical evolution of a given astrophysical environment. We do not attempt to perform such a detailed study and use the estimates from previous works.

In previous studies of chemical uncertainties it was found that the uncertainties are in general larger for bigger molecules as their evolution involves more reactions compared to simpler molecules. For simpler, key species such as CO and H$_2$ involved in a limited cycle of reactions it is easier to derive the reaction rates with a high accuracy of $\sim 25\%$. In addition these species are formed in the gas, which is better known than the surface part.
Consequently, their abundances are usually accurate within $10-30\%$ in modern astrochemical models.  On the other hand, for other
diatomic and triatomic species such as CN, HCO$^+$, HCN, CCH, and so on, uncertainties are usually about a factor of 3-4 \citep[see][]{Vasyunin2004,Vasyunin2008,Wakelam2010}.
The chemical uncertainties are higher for more complex molecules like methanol due to the fact that gas phase reactions are less known and the surface chemistry less well understood. These uncertainties can reach orders of a magnitude, with the factor of 10 being a likely lower limit.

Moreover, for S-bearing species, for which many reaction rates have not been properly measured or calculated or included in the networks, these intrinsic uncertainties and hence the uncertainties in their resulting abundances are higher, $\ga $ factor of 10 even for simple species such as SO, OCS, and SO$_2$ \citep{Loison2012}.
Also, the incompleteness of astrochemical networks with regard to the chemistry of Cl- and F-bearing molecules makes their calculated abundances rather unreliable.

In our study we assume that the uncertainties in the abundances of ortho- and para-H$_2$ and CO are within a factor of 30\%. For
HCO$^+$, H$_2$CO, CN, N$_2$H$^+$, C$_2$H, NO, OH, C, C$^+$, O, CH, NH$_3$, H$_2$O, HCN, and HNC the uncertainties are within a
factor of three, and  for S-bearing species, CH$_3$OH, and HCl they are within a factor of ten.

The result of this process is the abundance profiles from the species of interest. This means that we get the abundance of species 
over the radius adopted from the physical models for a range of timescales. In our study we do not use the abundance profiles 
of SO$_{2}$, SO, and OCS from the chemical model because the chemical network for S-bearing species is too inaccurate with respect to these species and thus lacks predictive power.

\subsection{Results}

\subsubsection{Standard approach}

\begin{figure*}[h]
\begin{center}$ 
\begin{array}{cc}
\includegraphics[scale=0.35, angle=90]{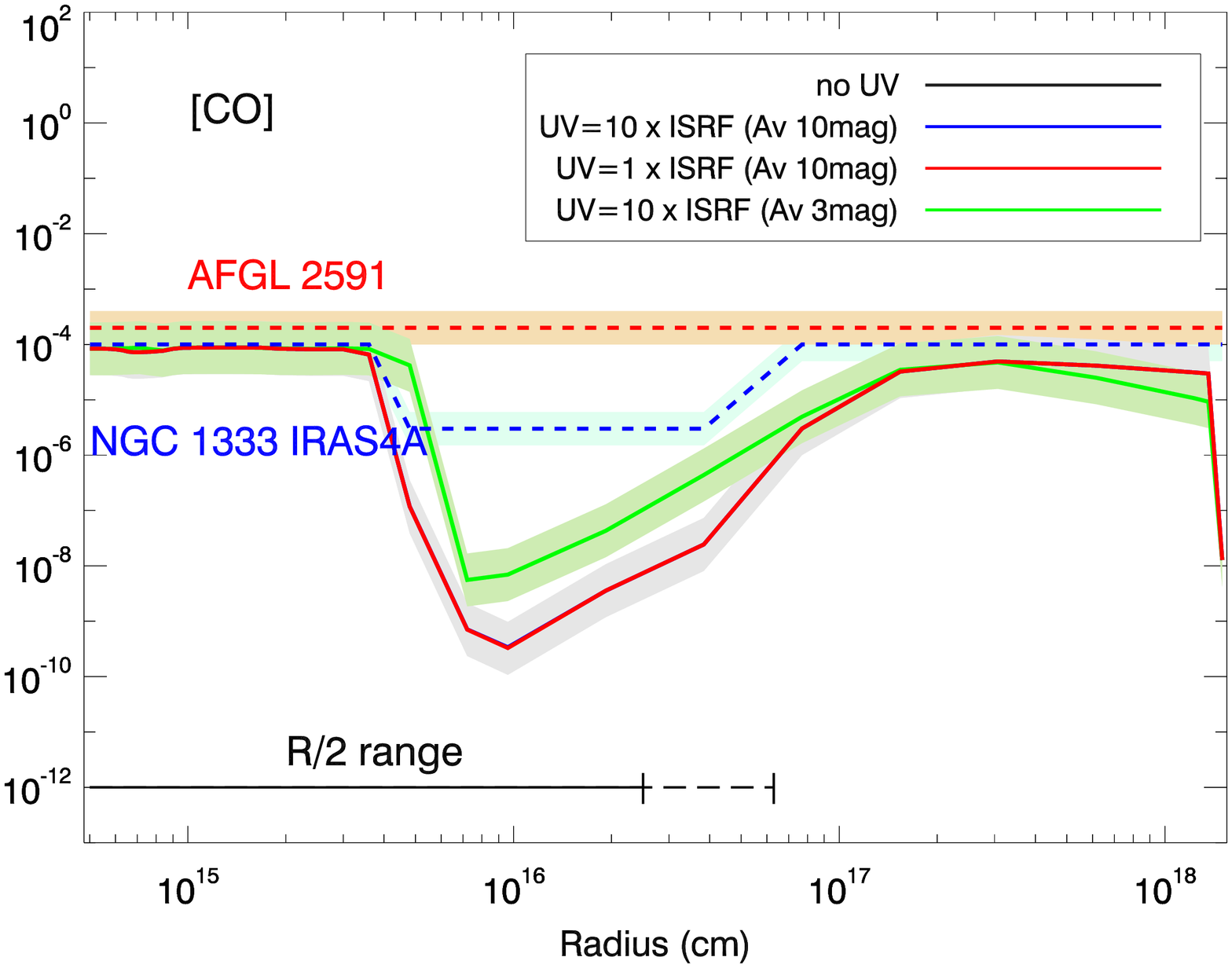} & 
\includegraphics[scale=0.35, angle=90]{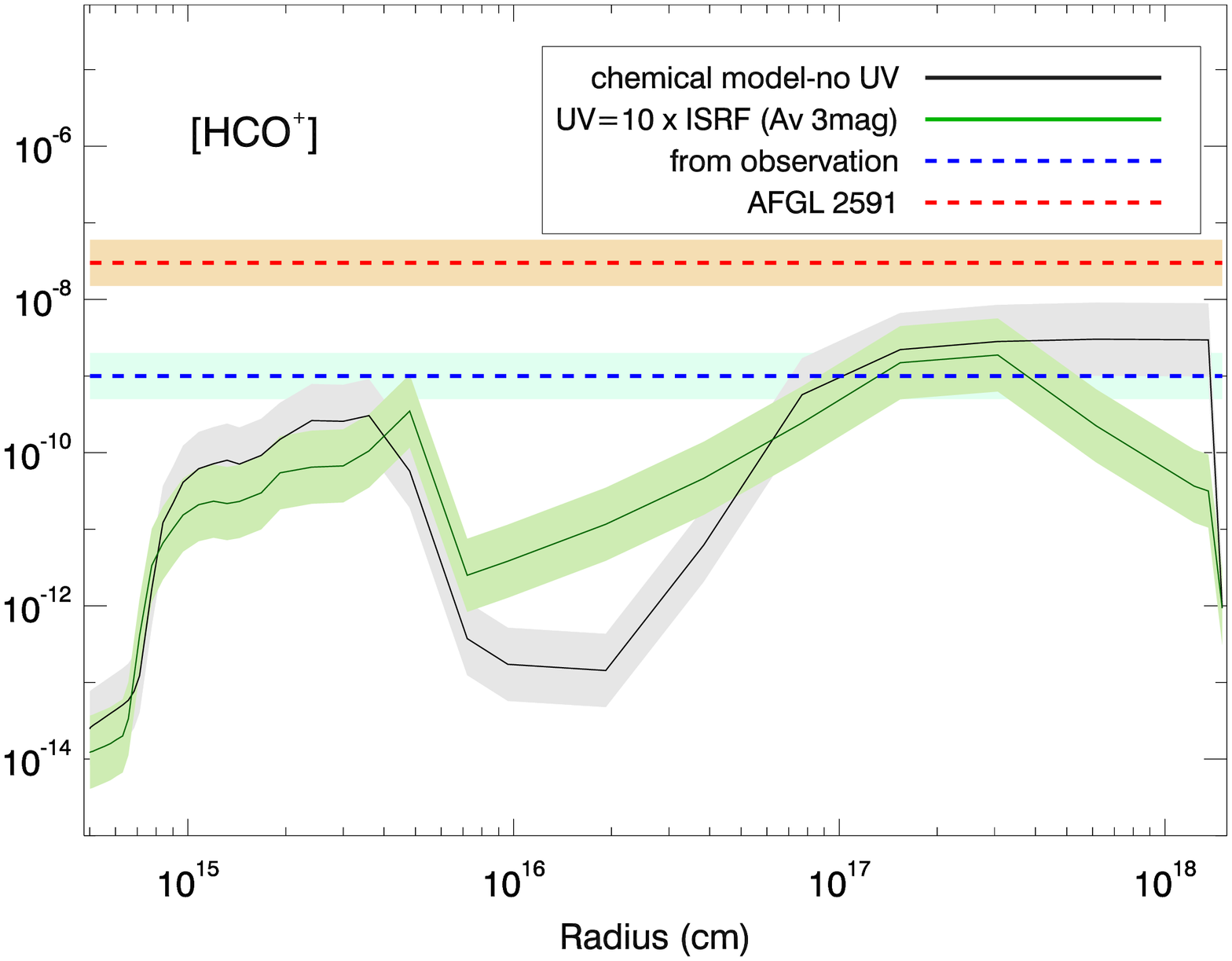} \\
\includegraphics[scale=0.35, angle=90]{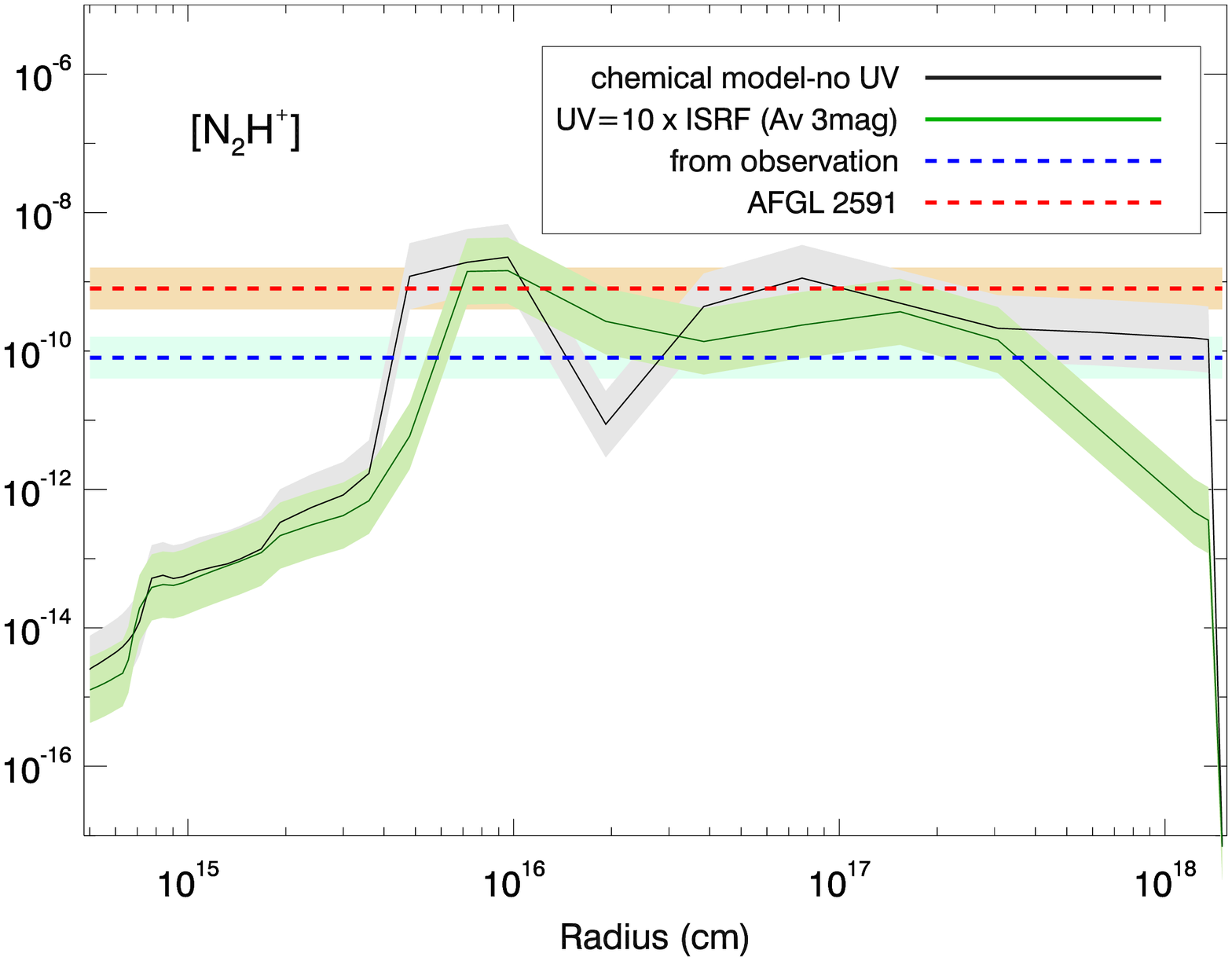} & \includegraphics[scale=0.35, angle=90]{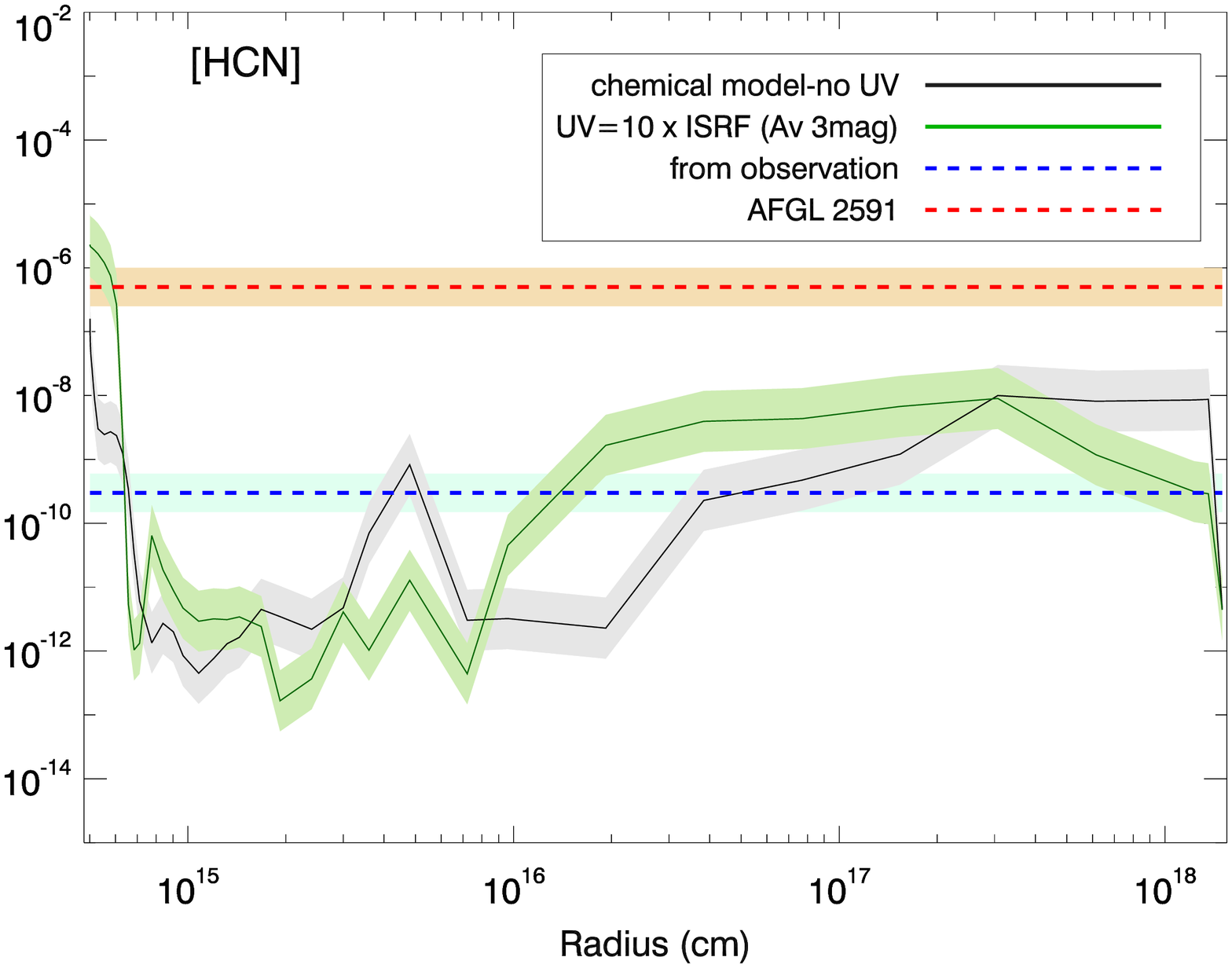} \\\includegraphics[scale=0.35, angle=90]{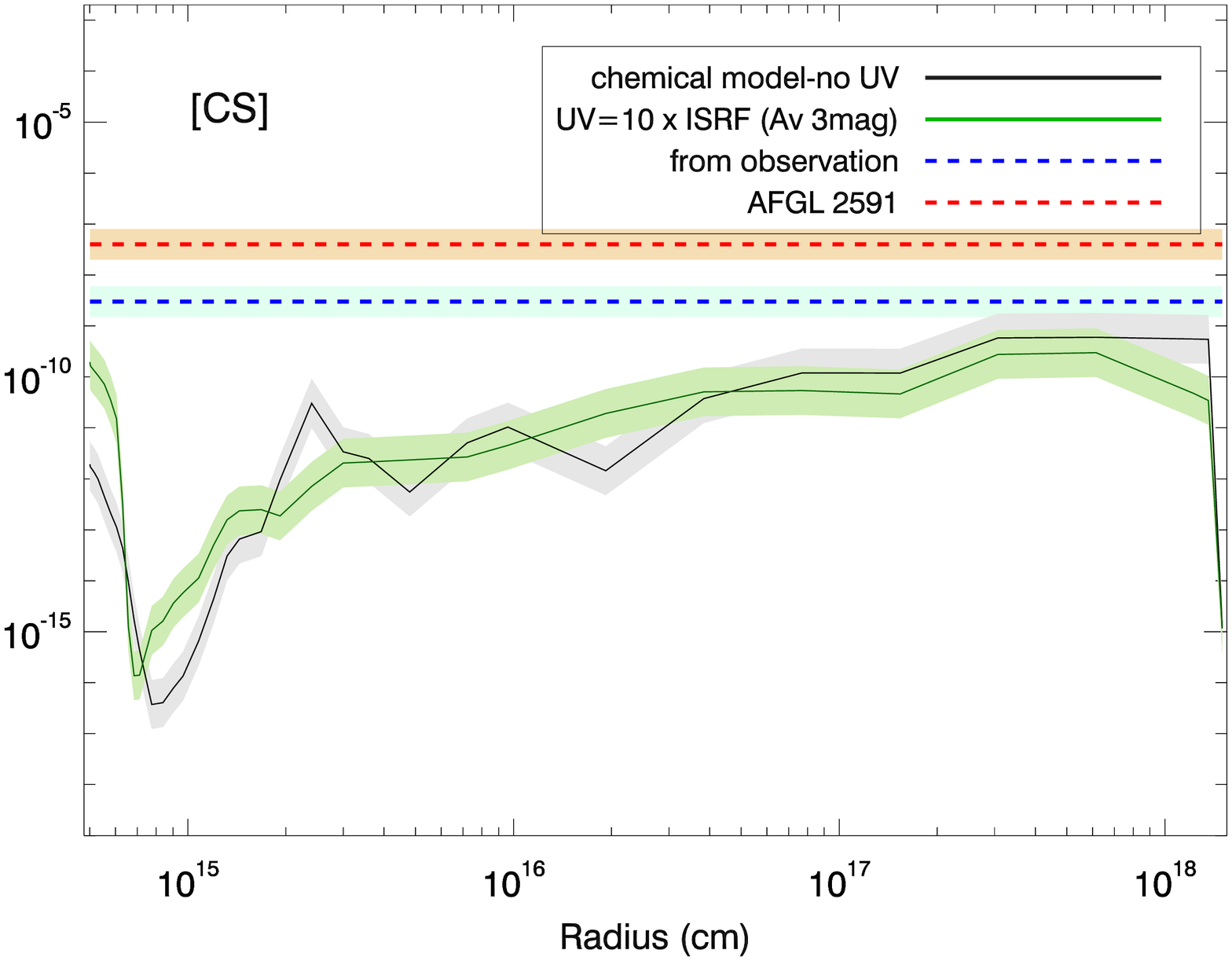} & \includegraphics[scale=0.35, angle=90]{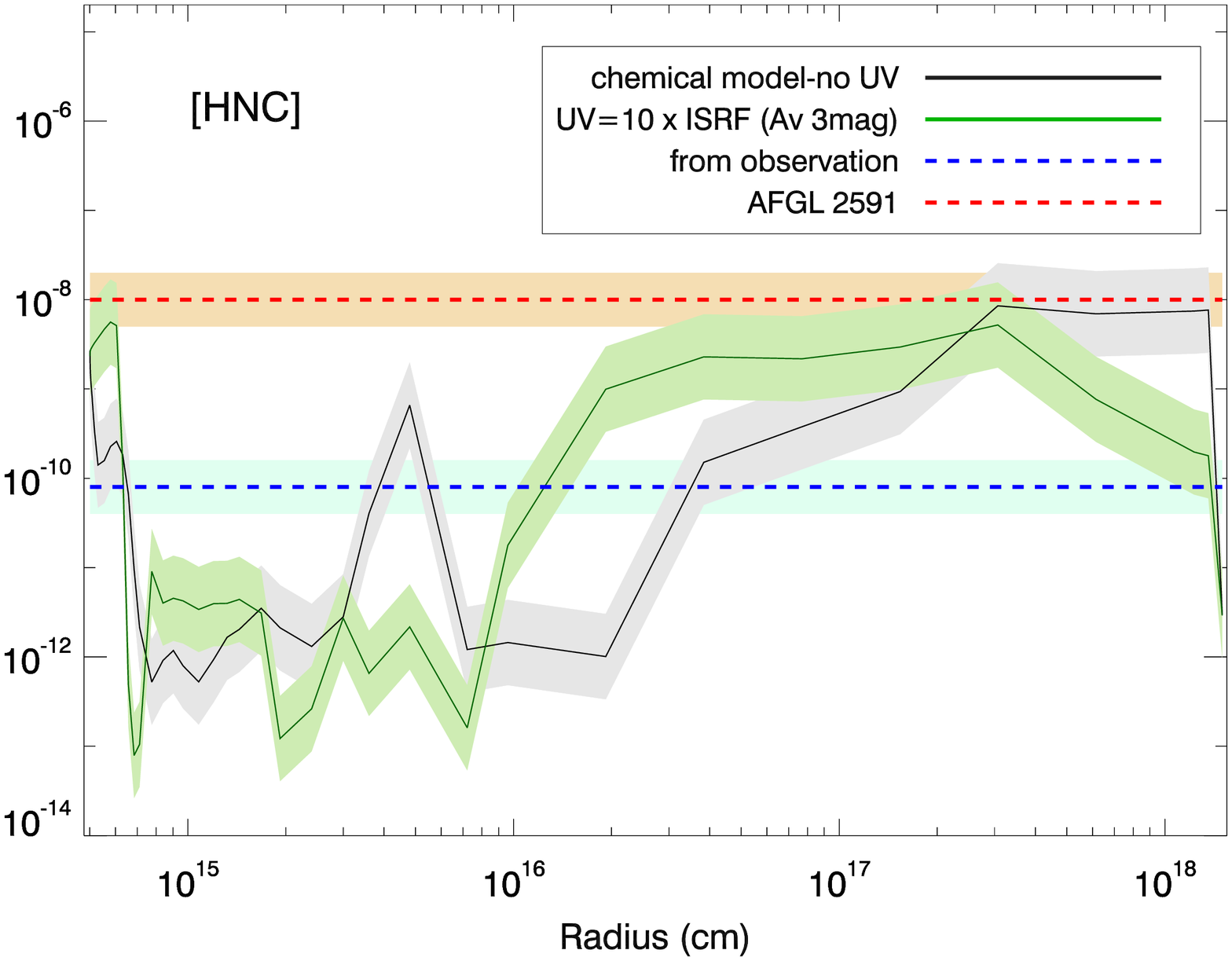} \\
\end{array}$
\end{center}
\caption{Observed and modeled abundance profiles of CO, HCO$^{+}$, N$_{2}$H$^{+}$, CS, HCN and HNC at the minimum representative timescale of 4$\times$10$^{4}$ yrs as predicted from the time-dependent CH$_{3}$OH models. The red dashed lines show the abundance profile of the outer envelope of the high-mass case, AFGL~2591 \citep{Kazmierczak2015} for comparison with NGC~1333~IRAS~4A (blue). The black solid lines represent the abundance profiles from the 1D chemical model. 
The green solid lines represent the abundance profiles from the 1D chemical model that aims to take into account outflow cavities 
by applying an extra UV radiation of 10$\times$ISRF at A$_{V}$=3~mag. The angular resolution of the observations varies between $\sim$15$\arcsec$ and $\sim$35$\arcsec$, which corresponds to 2.5 -- 6.3 $\times$ 10$^{16}$ cm (1670--4210 au) in the models.}
\label{fig:chem3}
\end{figure*}

Figures~\ref{fig:chem3}--\ref{fig:chem5} present the results of the standard chemical modeling compared to the observed abundances of NGC~1333~IRAS~4A 
from Sec.~\ref{ratran} and 
the high-mass protostellar envelope AFGL~2591 from \citet{Kazmierczak2015}. The observed abundances for most species appear to be in agreement with 
the modeled abundances in the outer envelope while they are systematically 1 to 2 orders of magnitude lower than the high-mass protostellar envelope. 
In contrast, the predicted CN, HCN, and HNC abundances are 1 to 2 orders of magnitude higher than the observed values for the outer envelope. Our chemical models do not take into account shielding of CN by H$_{2}$, as well as 
FUV scattering, which can be important. In addition our models use the CN photodissociation rates taken by van Dishoeck et al. (2006). A more recent study by \citet{el_Qadi2013} presents CN cross-sections with values 
several times smaller than those from van Dishoeck et al. (2006). From all the parameters, it seems that 
the strongest effect in those modeled abundances is due to the assumed
FUV intensity and the C/O ratio. The C/O $= 1.1$ gives X(CN) of 3$\times$10$^{-7}$ while X(HCN) $\sim$ X(HNC) $\sim$ 8$\times$10$^{-10}$.
If one assumes a strong FUV field of 10$^{4}$ of the ISRF UV with a modest extinction
of Av$=$1 mag, the fit is much better: X(CN) $\sim$1--4$\times$10$^{-10}$, X(HCN) $\sim$$<$$10^{-11}$, X(HNC) $\sim$$>$ 2$\times$$10^{-11}$.
Without any additional FUV one gets X(CN) $=$ 10$^{-14}$, X(HCN) $\sim$ 2$\times$10$^{-9}$,
X(HNC) $\sim$ 7$\times$10$^{-11}$. The standard model alone cannot explain the observed abundances for more than one of the modeled species, making 
the development of a more advanced model necessary.

\subsubsection{The necessity of an outflow cavity}

We observe a drop of only $\sim$2 orders of magnitude towards the snowline of CO, compared to 
$\sim$6 orders of magnitude predicted by our chemical models. A plausible explanation for such a discrepancy between our chemical model and the observations is the presence of the outflow, which is not accounted for in the 1D model \citep{Bruderer2009}. The way 
to approximately simulate the outflow, the UV-irradiated outflow walls, and the envelope in the 1D approximation is to add more UV radiation to the chemical model. 
For that, additional FUV components with intensities of 1 and 10 Draine's units and moderate dust extinctions of 10 and 3 mag were
considered. We find that only lower extinction influences the resulting CO abundance, increasing it by approximately 1 order of magnitude. The standard model without extra UV radiation and extinction of 10 mag produces the same abundance profiles.    
   
Figures~\ref{fig:chem31}--\ref{fig:chem51} present the results of the models with additional 10 Draine UV fields and for dust extinctions of 1, 2 and 3 mag. 
The extreme case of a 100 Draine UV field and A$_V=1$ was also considered resulting in an almost constant CO abundance profile. Our 
observed CO abundance profile appears to be reproduced reasonably well by UV$=$ 10$\times$ISRF and A$_V=$1~mag (green line; Figure~\ref{fig:co_models}). Although such a model makes the overall fit better 
for CO, HCO$^{+}$ and DCO$^{+}$, it does not really improve the fit for other species (e.g., CS, CN, C$_{2}$H) and it actually makes the overall fit worse 
for H$_{2}$CO, HDCO, D$_{2}$CO and CH$_{3}$OH, by decreasing the abundance of the outer envelope by up to 4 orders of magnitude. Thus, we do not consider it 
to be our best--fit (or standard) model.  

The modeled HCO$^+$ abundances generally follow those of its parent molecule, CO, and show
a strong decline for radii between $\sim$ 334~au and 5350~au, where the CO freeze-out zone is located. HCO$^+$ abundances also drop strongly in the inner, dense and dark
envelope region at $r\la$134~au, where the ionization degree drops due to fast
recombination processes. Not surprisingly, the modeled N$_2$H$^+$ abundances also drop in the very inner
envelope, like those of HCO$^+$. In contrast to HCO$^+$, N$_2$H$^+$ thrives in the
CO freeze-out zone, where a key destruction reaction of N$_2$H$^+$ ions by CO molecules is
no longer effective. In Figures~\ref{fig:chem4} and~\ref{fig:chem41} the DCO$^+$ abundances are compared, and the modeled
DCO$^+$ profile follows the case of HCO$^+$ and agrees with observations in two areas: 1) The inner envelope at $\sim$67--267~au and 2) the outer envelope
with $r\ga$6685~au. In contrast to HCO$^+$, DCO$^+$ is slightly overproduced in the
no-UV chemical model in the inner part of the envelope, but is well fitted by the model with
additional UV due to the outflow cavity.

The poor fit and dependence on UV irradiation of CN, HCN, and HNC have already been discussed above.
Their formation mainly proceeds via neutral-neutral gas-phase reactions involving light
hydrocarbons like C$_2$H and other N-bearing species (e.g., NO). Thus the no-UV model
that fails to fit the CN, HCN, and HNC data is also not able to fit the C$_2$H observed
abundances.

The modeled CS abundance profile shows a poor fit to the data as well. As we also mentioned
above, this is due to a general lack of predictive power of current astrochemical models
for S-bearing species. Still, H$_2$S modeled values are in good agreement with the observed data.

The observationally-driven H$_2$CO abundances are only well reproduced
in the outer envelope at $r\ga$2005~au, and are lower than the observed values
by up to 3-5 orders of magnitude in the inner part. This is also likely caused by the same
approximation of the outflow and UV-irradiated outflow cavity walls in our 1D chemical
model as for CO. Alternatively, our observations may lack the necessary resolving power and sensitivity (that interferometers can provide) to uncover and unbiasedly constrain the underlying physical structure of the inner NGC 1333 IRAS~4A envelope, which comes out in the lack of agreement between the data and
the chemical predictions. A similar behavior is shown by H$_2$CO isotopologs and
the chemically-related methanol molecule.

In Figures~\ref{fig:chem31}--\ref{fig:chem51} we show the effect of including an additional UV component in the chemical model,
as our attempt to represent the UV-irradiated outflow cavity material. As discussed above,
additional UV radiation lowers the degree of CO depletion and brings us much closer to 
agreement between the observed and modeled CO abundances. The same effect is seen for our N-bearing
species (CN, HCN, HNC). Unfortunately, for all other observed species
(HCO$^+$, N$_2$H$^+$, CS, H$_2$CO isotopologs, CH$_3$OH, C$_2$H, S-bearing species)
the modeled abundance profiles have poorer agreement with observations than in the standard model.
The enhanced UV irradiation leads to overly rapid destruction of less tightly bound molecules than CO, and limits the efficiency of surface chemistry by desorbing ices too efficiently.
The potential solution to such a chemical discrepancy is to perform chemical modeling
using a more realistic 2D or 3D physical structure of the NGC~1333 IRAS~4A envelope, including the
outflow and outflow cavity wall and performing UV radiative transfer.

\subsubsection{Time dependence and different input parameters}

\begin{figure*}[h]
\begin{center}$ 
\begin{array}{cc}
\includegraphics[scale=0.35, angle=90]{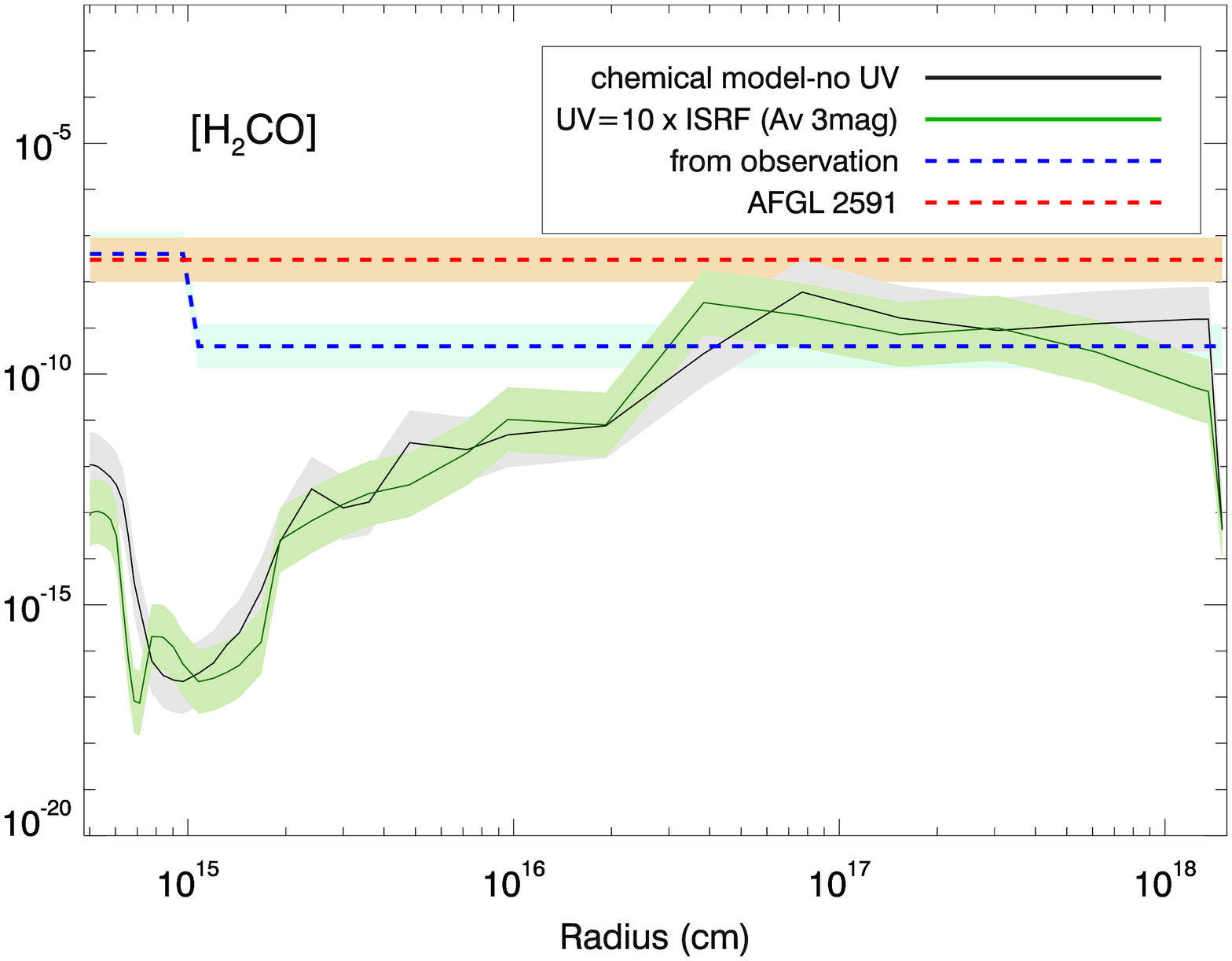} & 
\includegraphics[scale=0.35, angle=90]{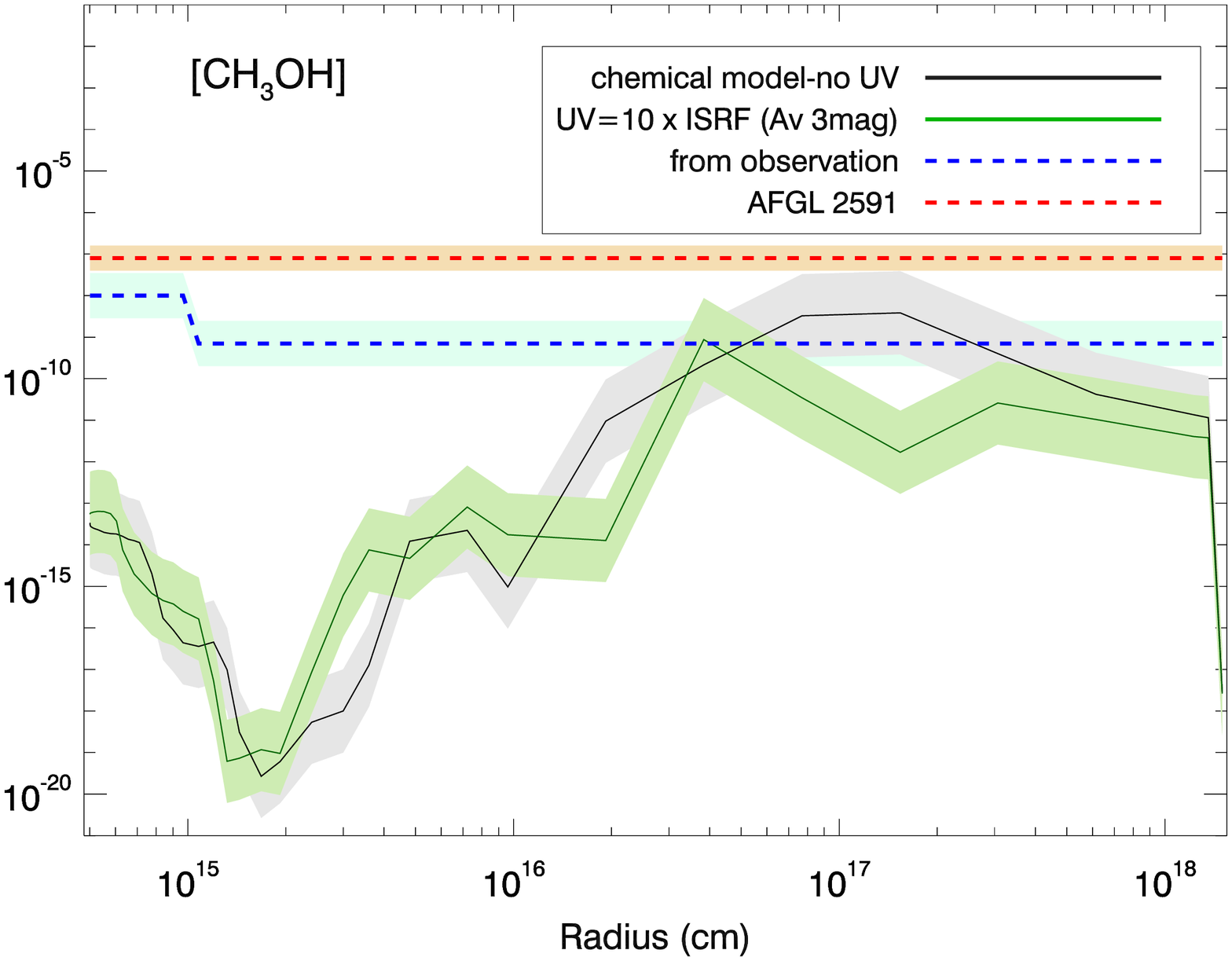} \\\includegraphics[scale=0.35, angle=90]{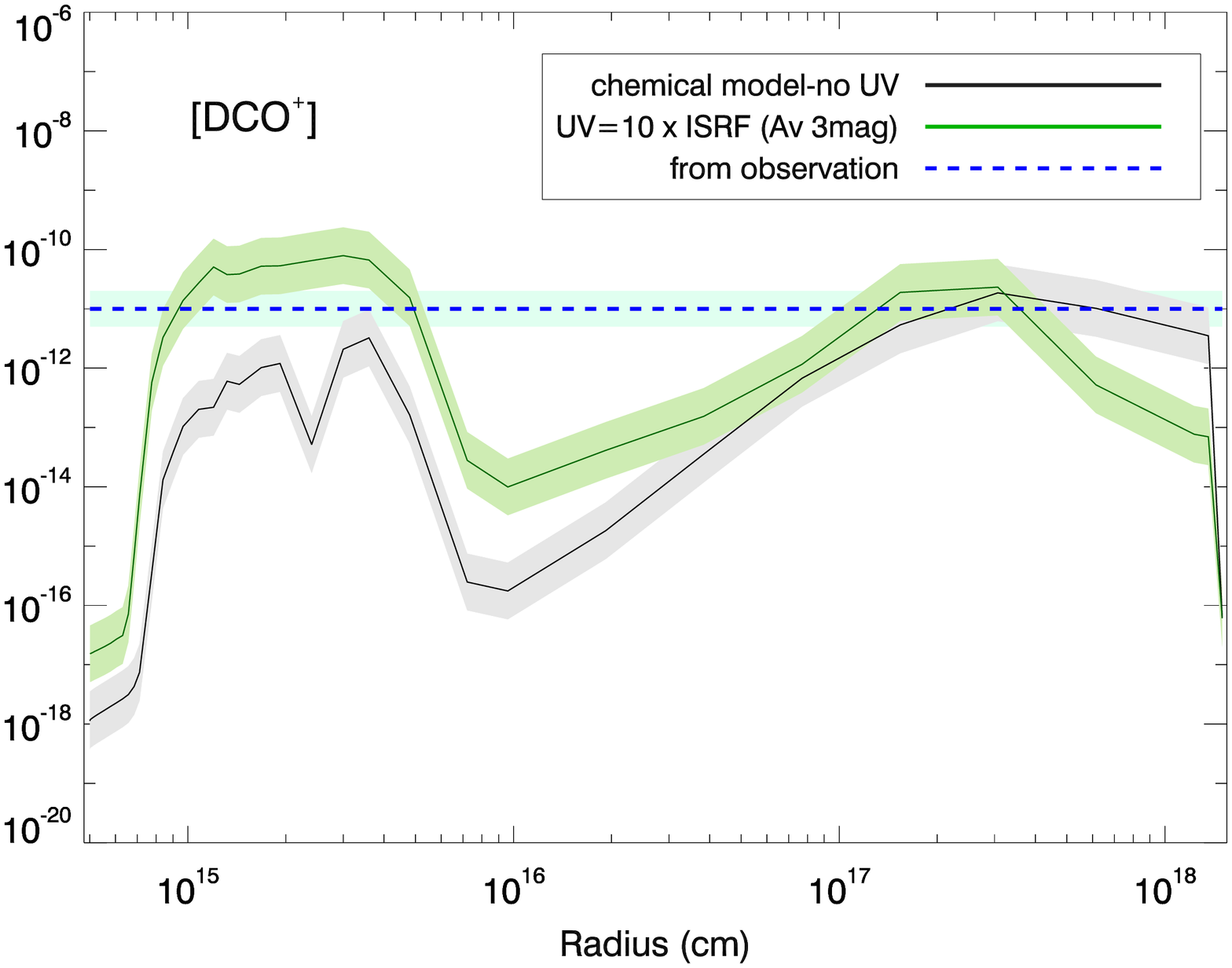} & \includegraphics[scale=0.35, angle=90]{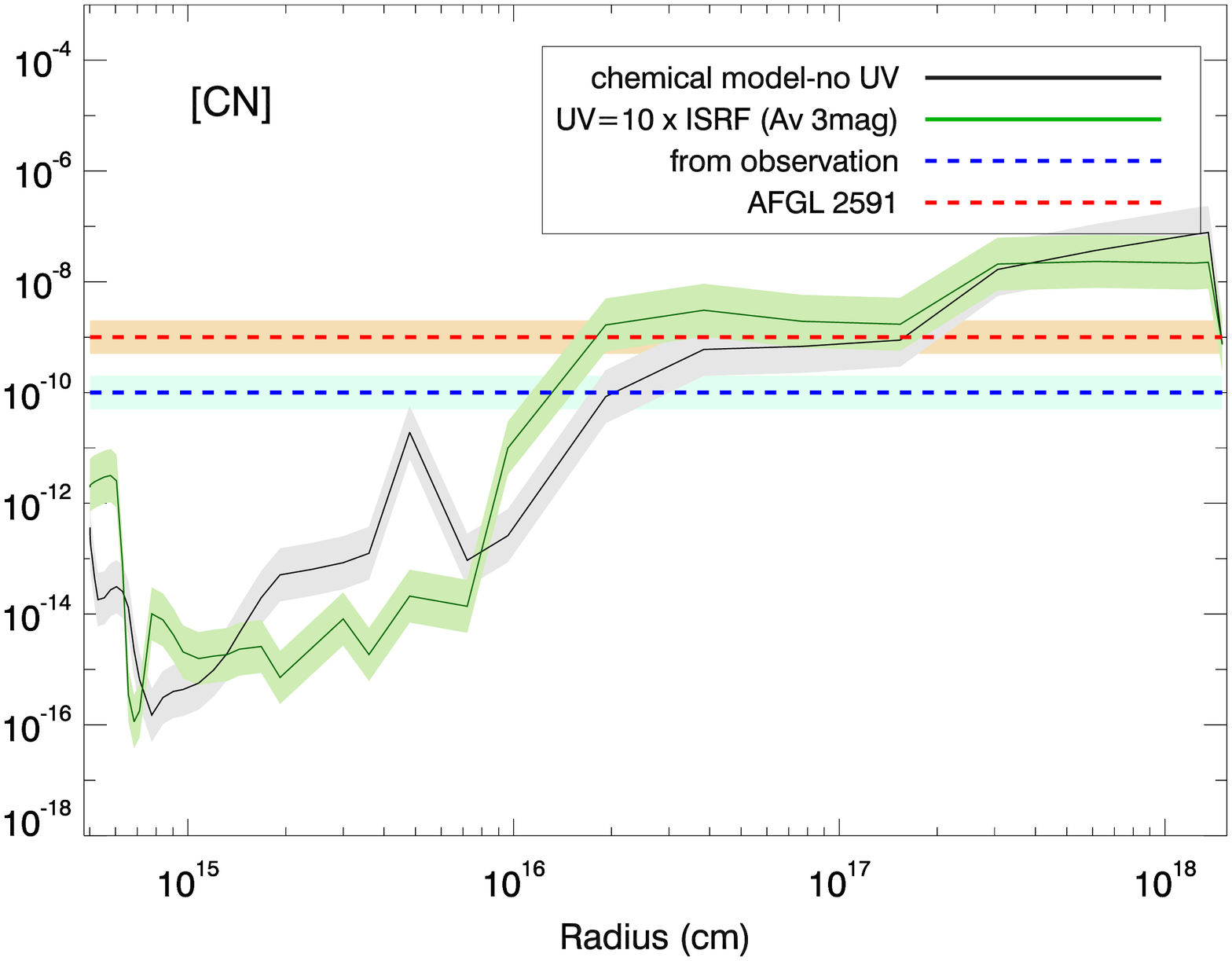} \\
\includegraphics[scale=0.35, angle=90]{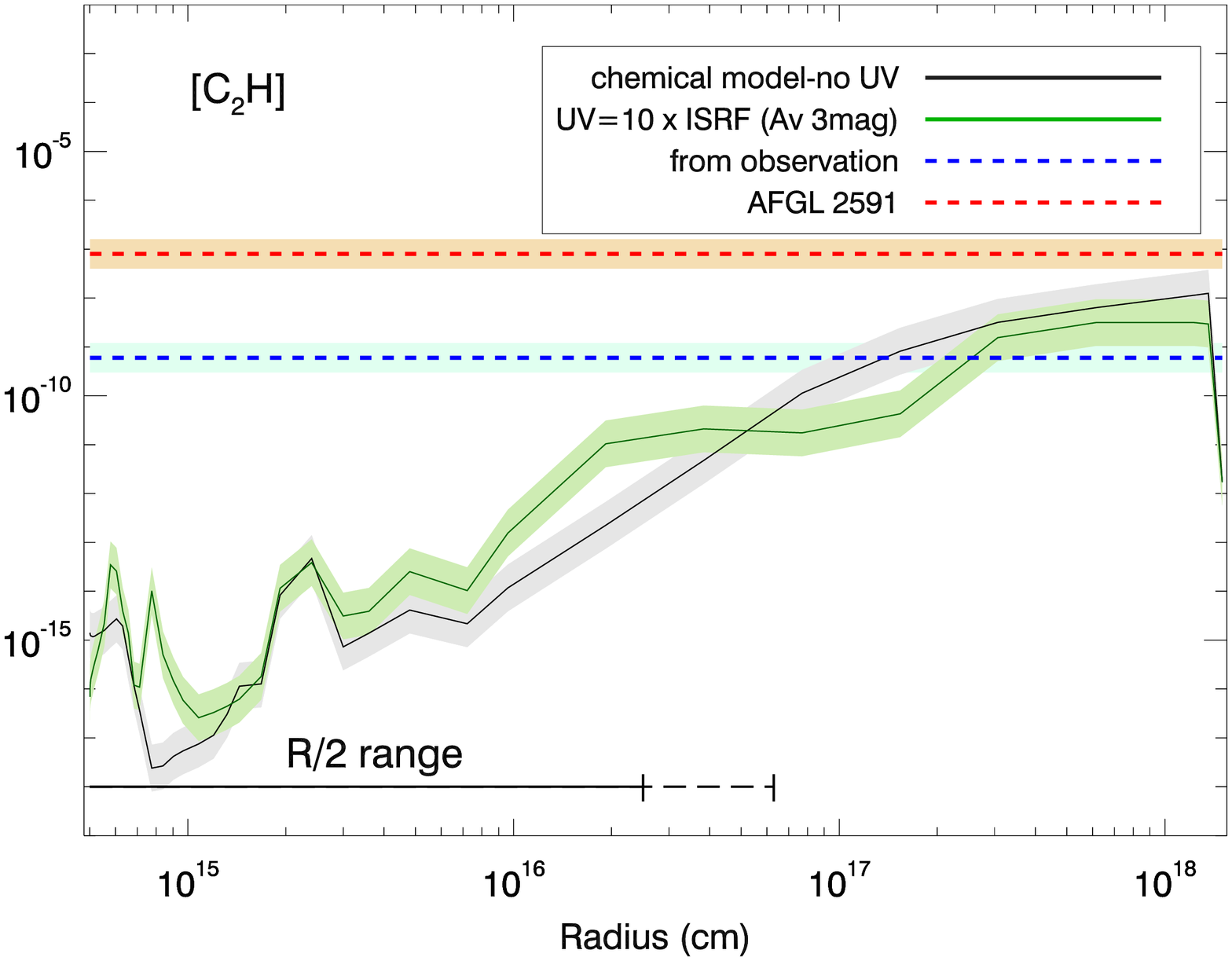} & \includegraphics[scale=0.35, angle=90]{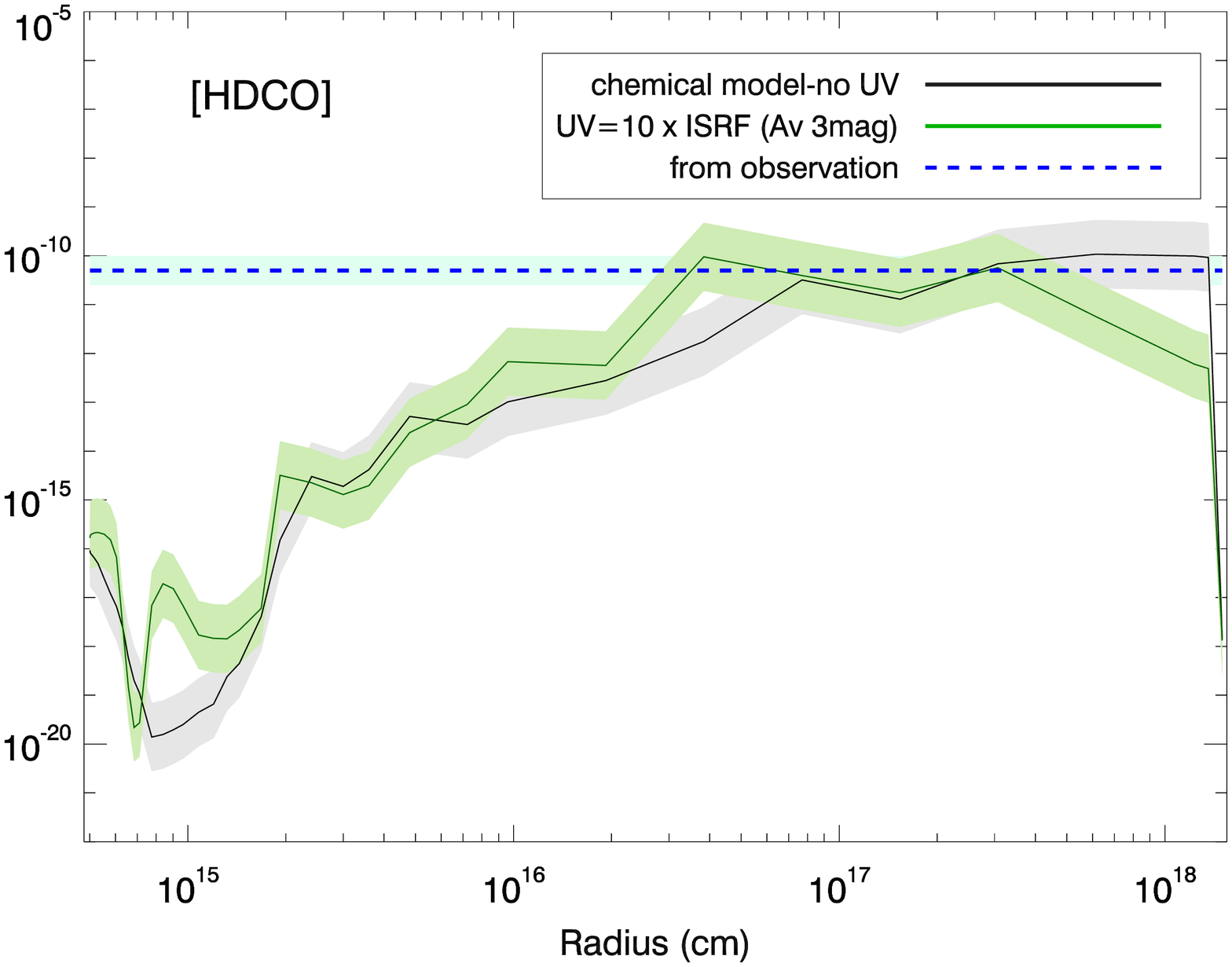} \\
\end{array}$
\end{center}
\caption{As Fig.~\ref{fig:chem3}, but for H$_{2}$CO, CH$_{3}$OH, C$_{2}$H, CN, HDCO, CO and DCO$^{+}$. The deuterated species, HDCO and DCO$^{+}$ were not observed towards AFGL~2591.}
\label{fig:chem4}
\end{figure*}

\begin{figure*}[h]
\begin{center}$ 
\begin{array}{cc}
\includegraphics[scale=0.35, angle=90]{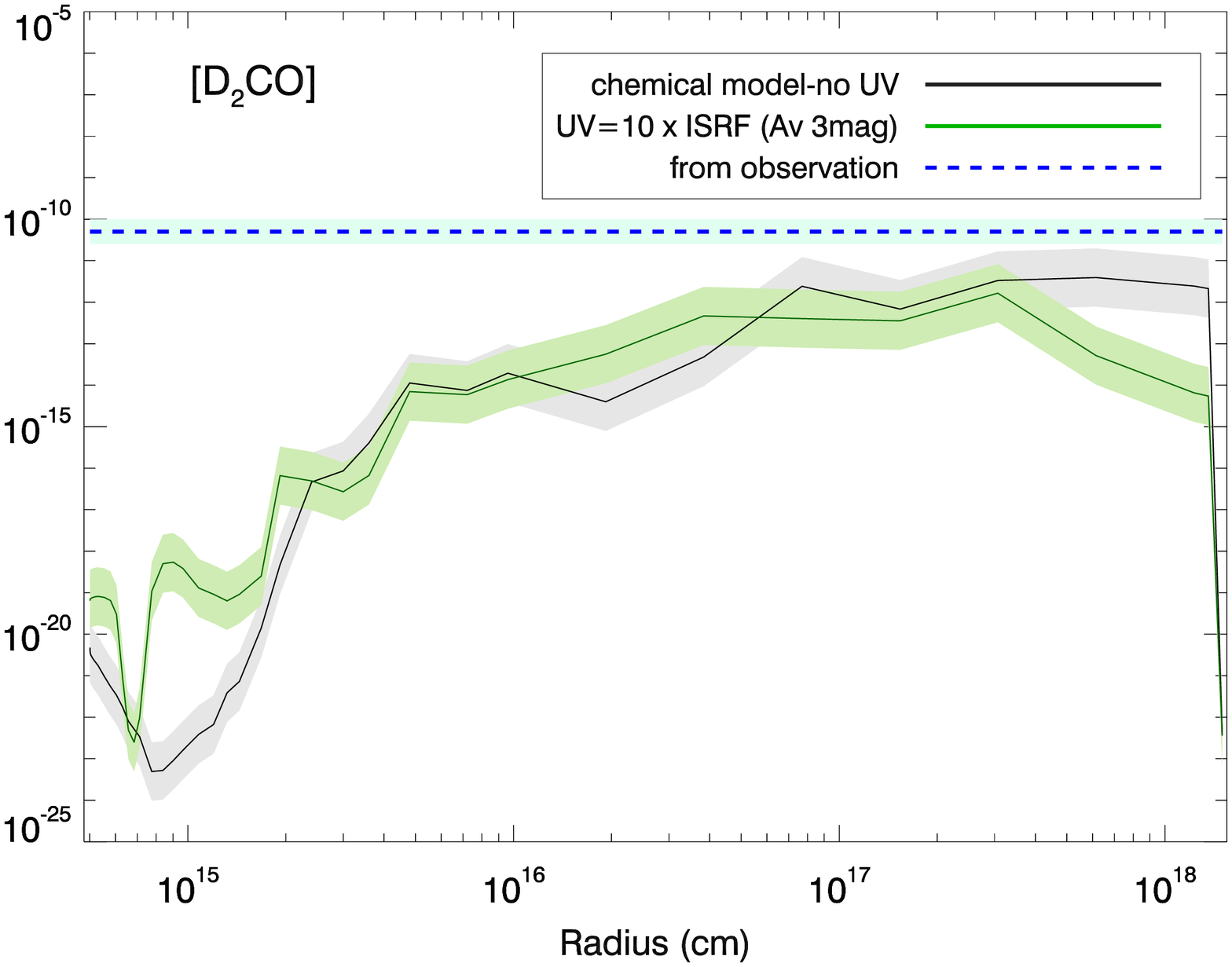} & \includegraphics[scale=0.35, angle=90]{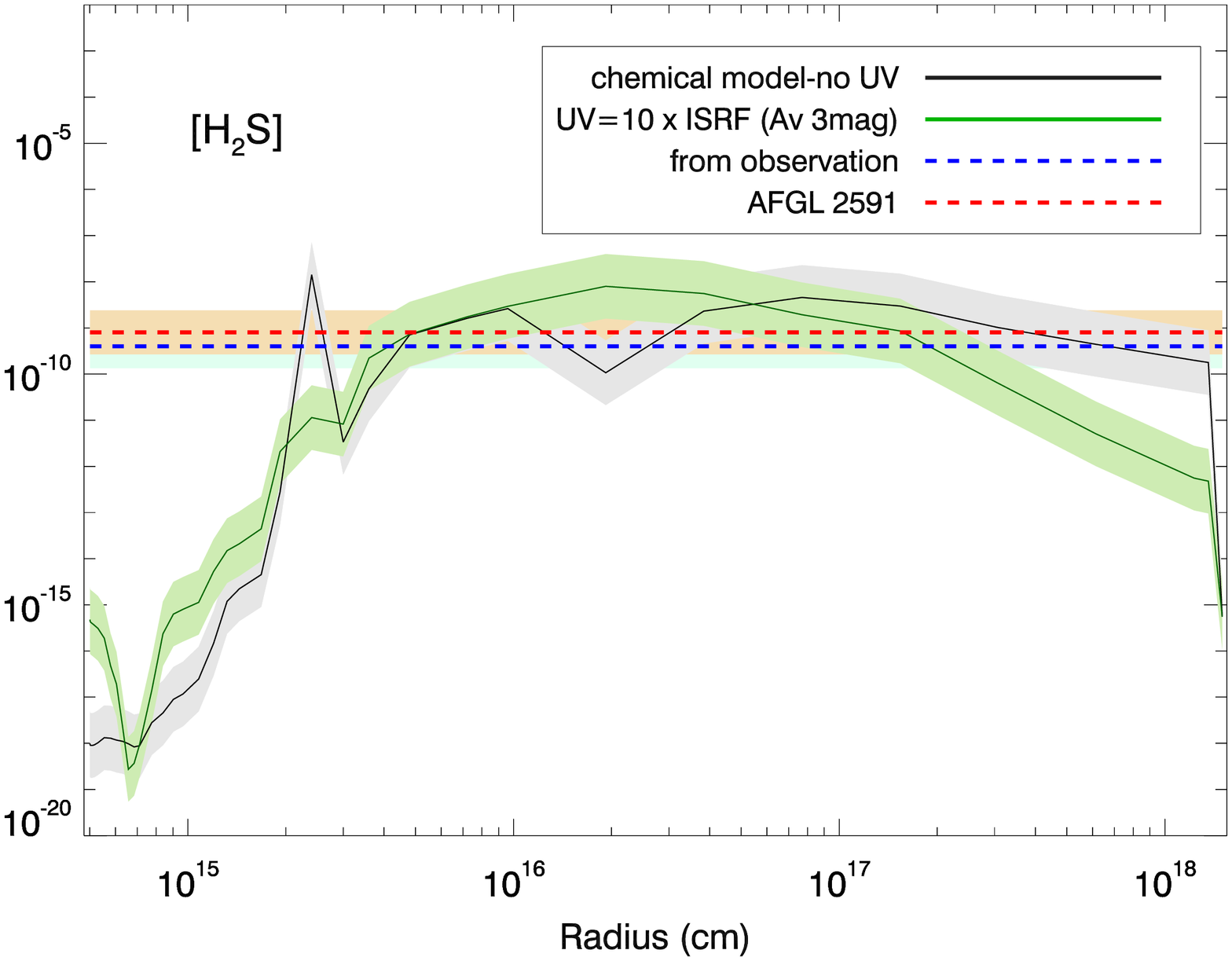} \\
\end{array}$
\end{center}
\caption{As Fig.~\ref{fig:chem5}, but for H$_{2}$S and D$_{2}$CO. D$_{2}$CO was not observed towards AFGL~2591.}
\label{fig:chem5}
\end{figure*}

\begin{figure*}[h]
\begin{center}$ 
\begin{array}{cc}
\includegraphics[scale=0.35, angle=90]{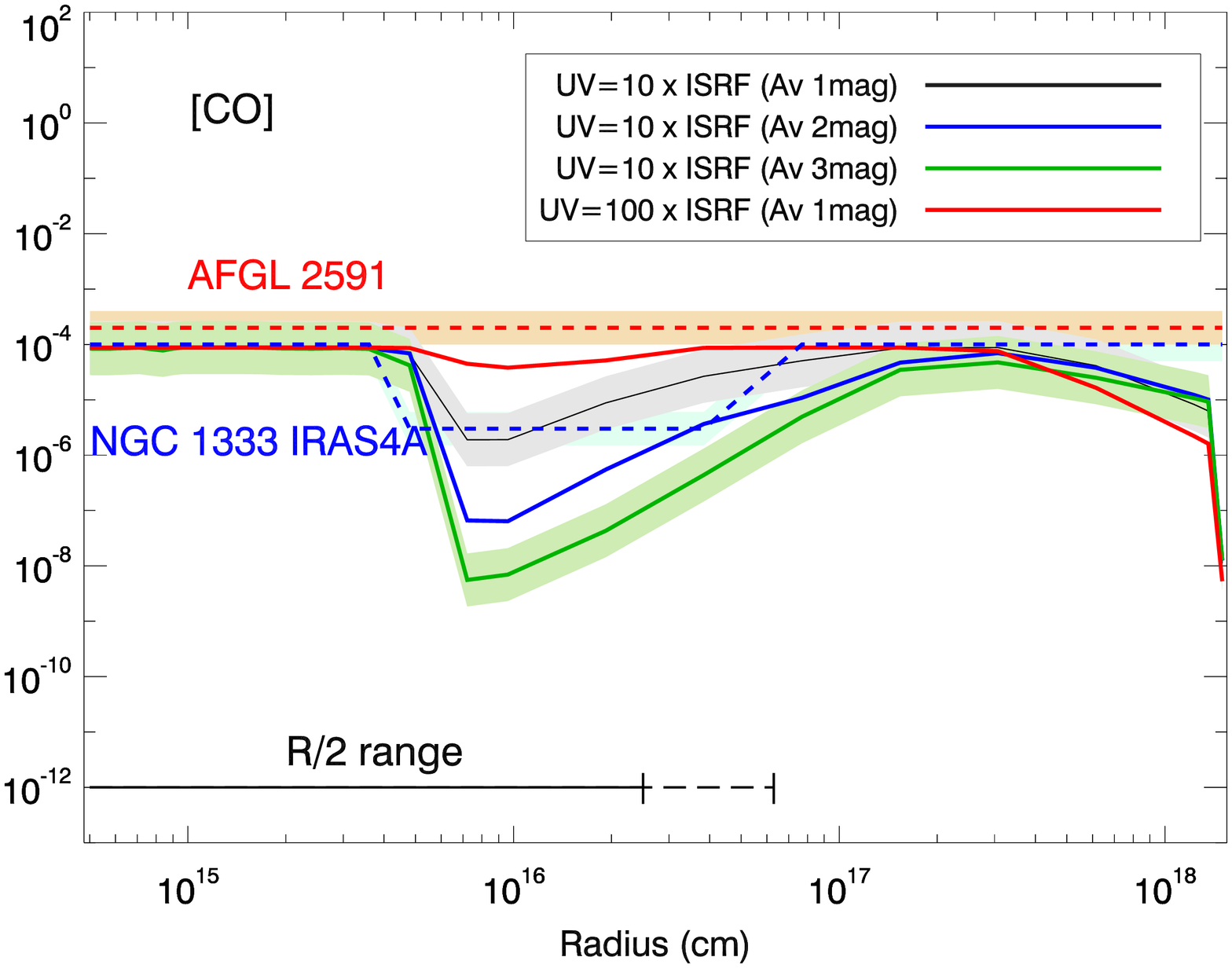} & 
\includegraphics[scale=0.35, angle=90]{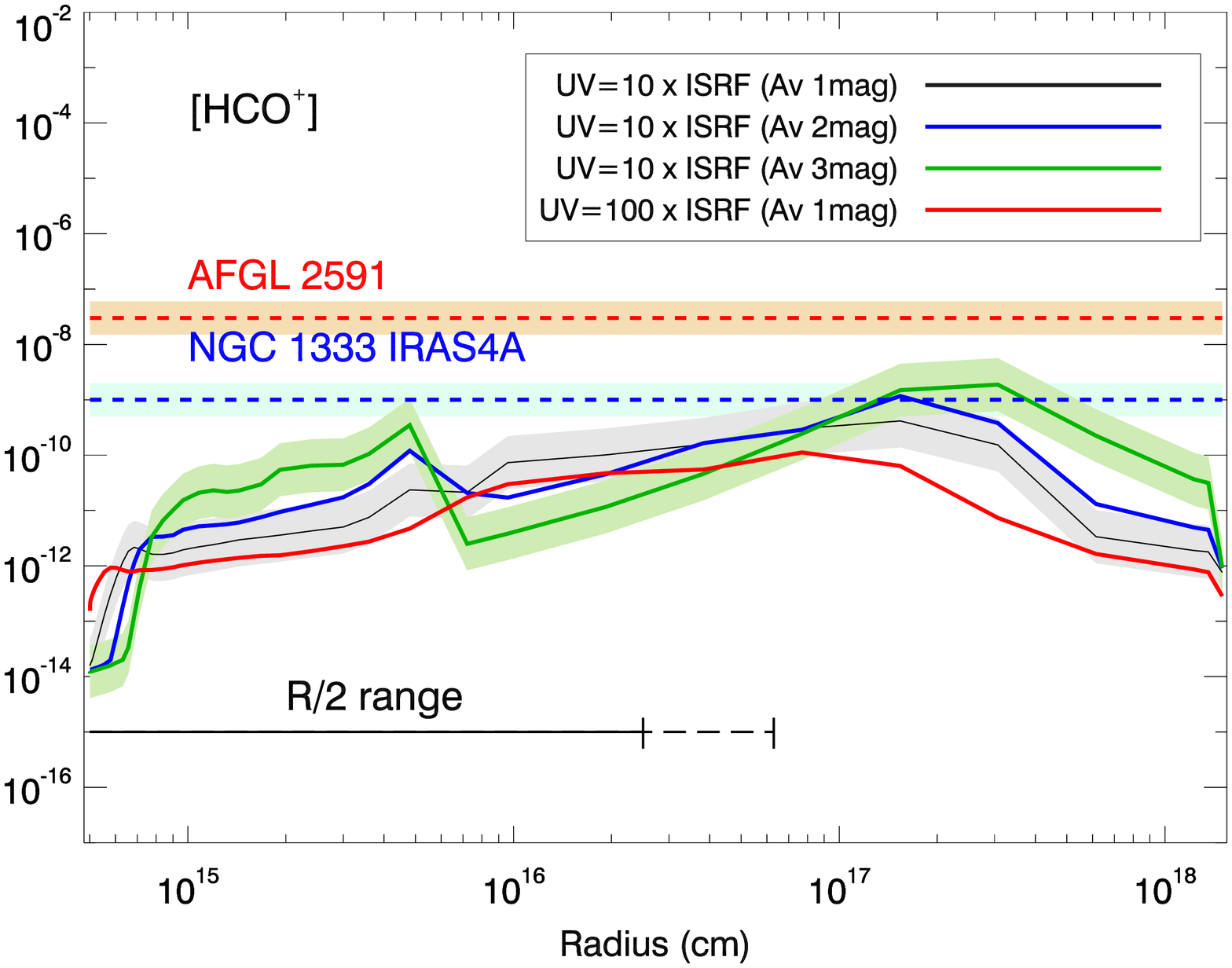} \\
\includegraphics[scale=0.35, angle=90]{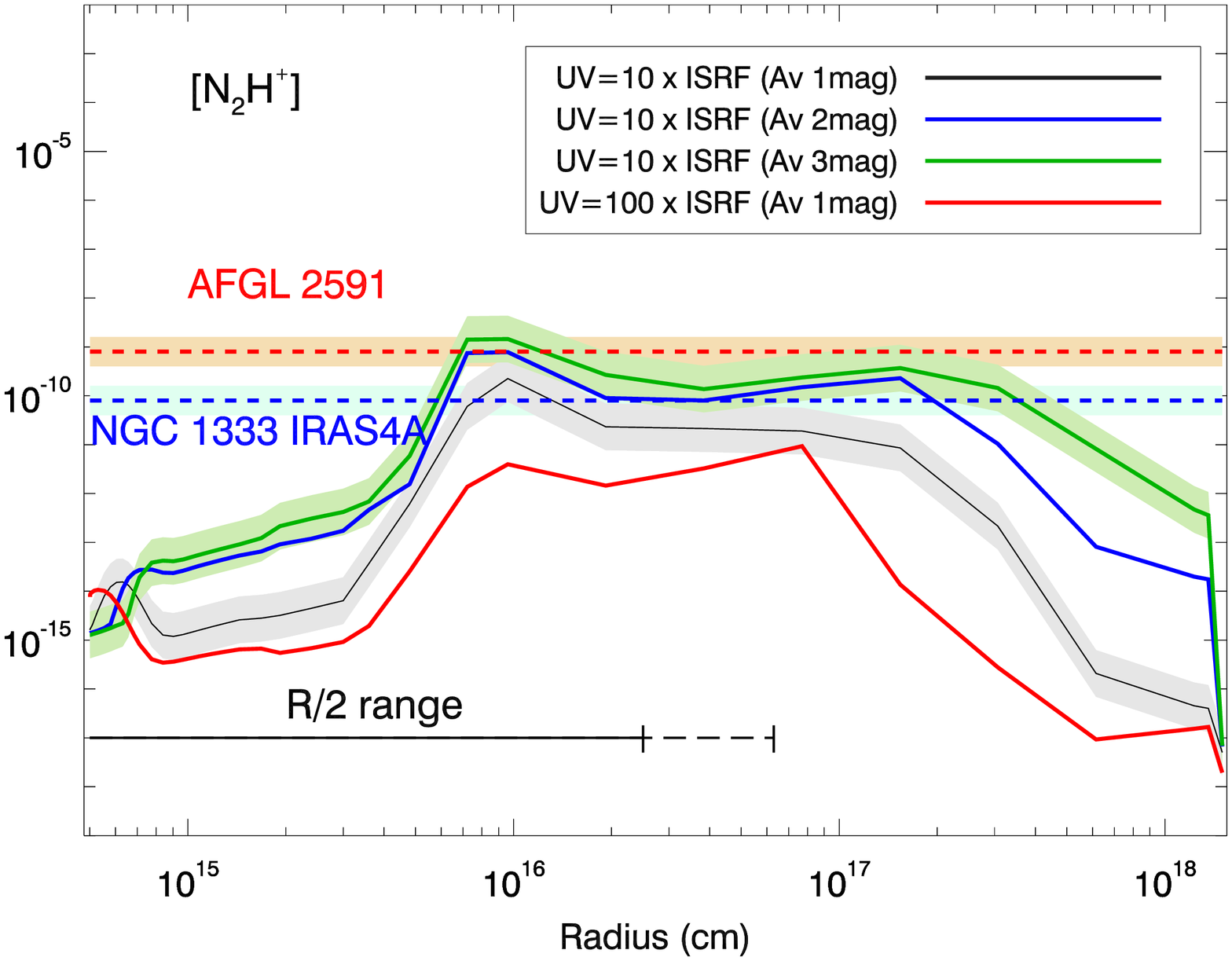} & \includegraphics[scale=0.35, angle=90]{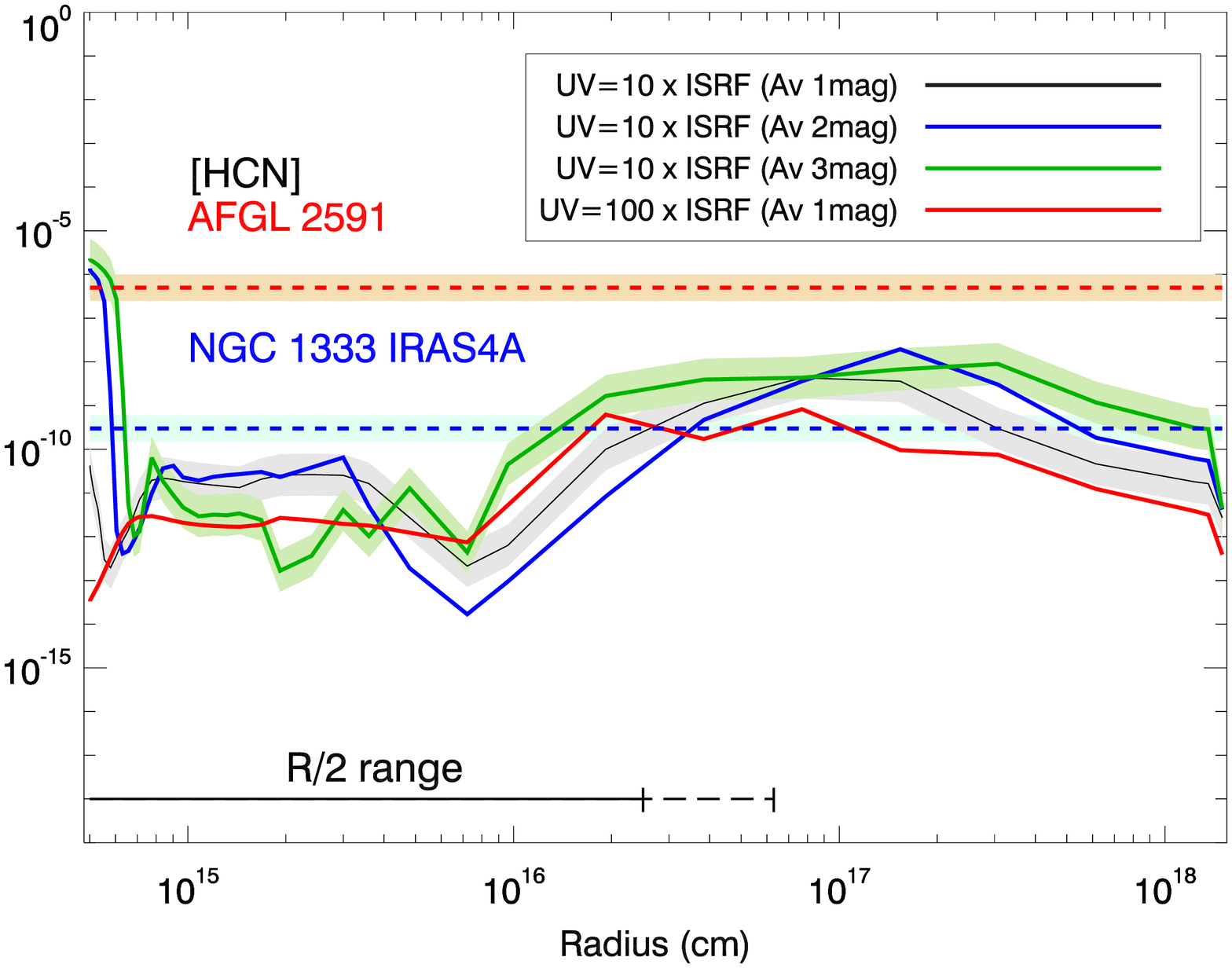} \\
\includegraphics[scale=0.35, angle=90]{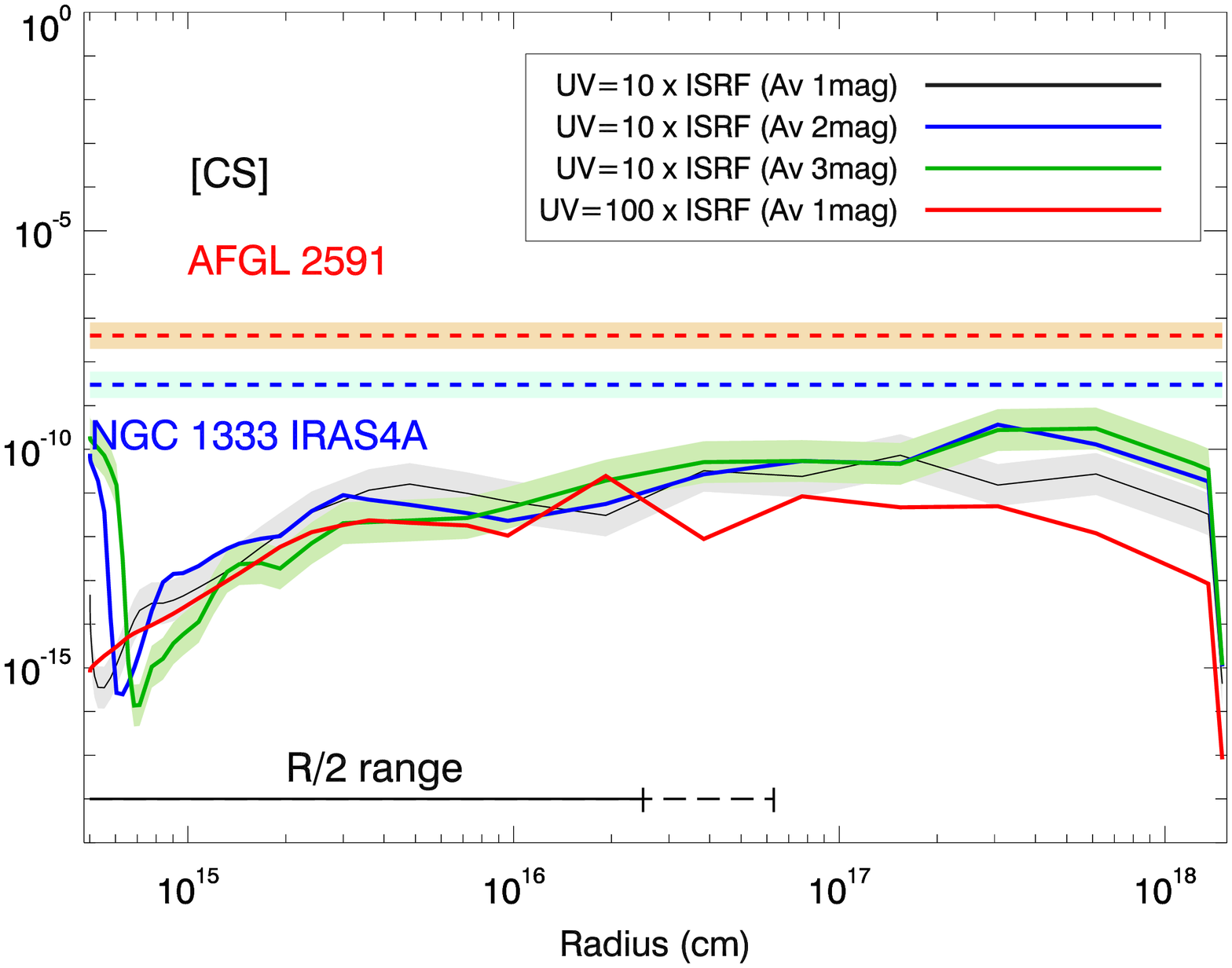} & \includegraphics[scale=0.35, angle=90]{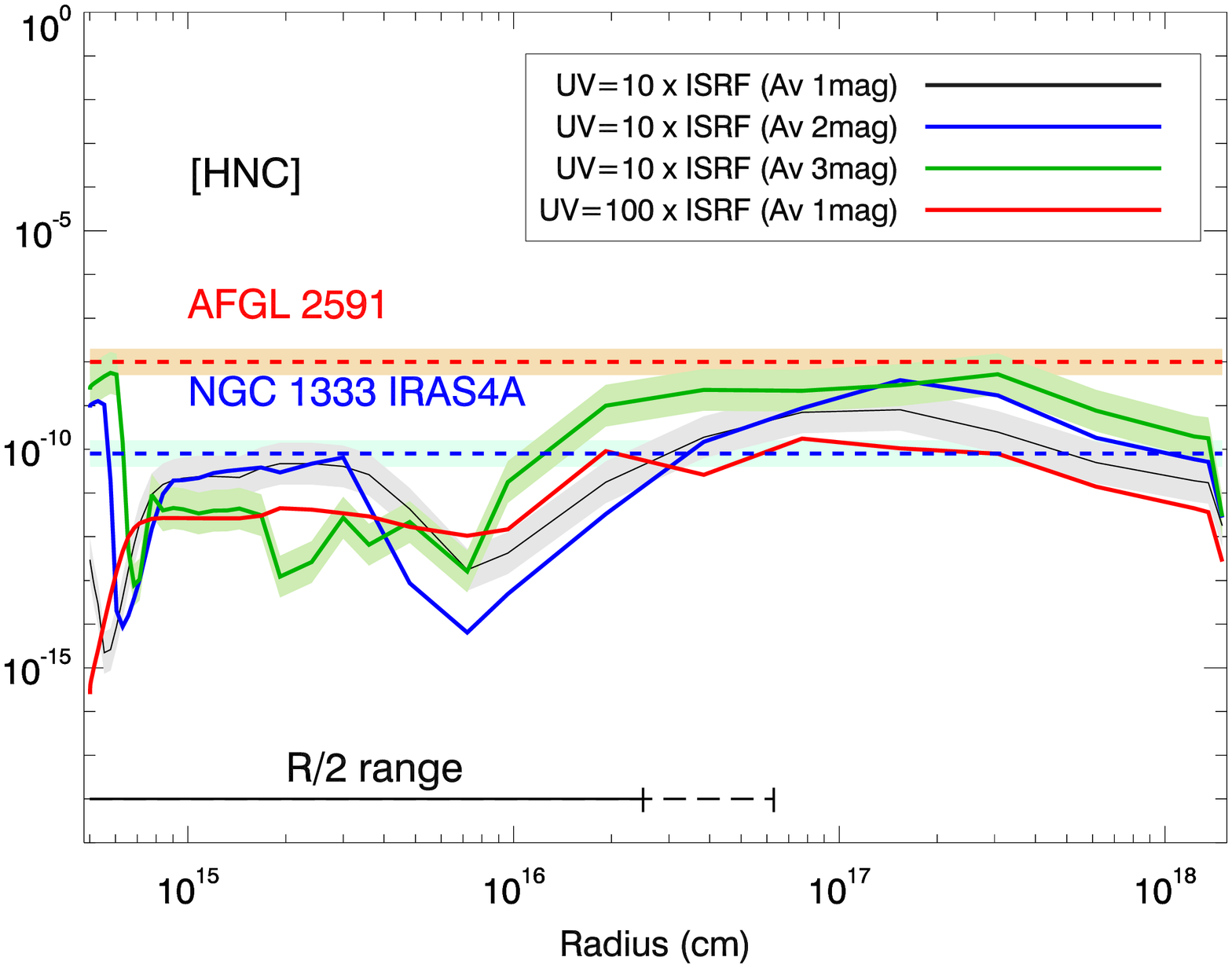} \\
\end{array}$
\end{center}
\caption{Observed and modeled abundance profiles of CO, HCO$^{+}$, N$_{2}$H$^{+}$, CS, HCN and HNC at the minimum representative timescale of 4$\times$10$^{4}$ yrs as predicted from the time-dependent CH$_{3}$OH models. The red dashed lines show the abundance profile of the outer envelope of the high-mass case, AFGL~2591 \citep{Kazmierczak2015} for comparison with NGC~1333~IRAS~4A (blue). The solid lines represent the abundance profiles from the 1D chemical model applying an extra UV radiation of 10$\times$ISRF A$_{V}$=1mag, 2mag and 3mag, and the extreme case of 
100$\times$ISRF and A$_{V}$=1mag. The angular resolution of the observations varies between $\sim$15$\arcsec$ and $\sim$35$\arcsec$, which corresponds to 2.5 -- 6.3 $\times$ 10$^{16}$ cm (1670--4210 au) in the models.}
\label{fig:chem31}
\end{figure*}

\begin{figure*}[h]
\begin{center}$ 
\begin{array}{cc}
\includegraphics[scale=0.35, angle=90]{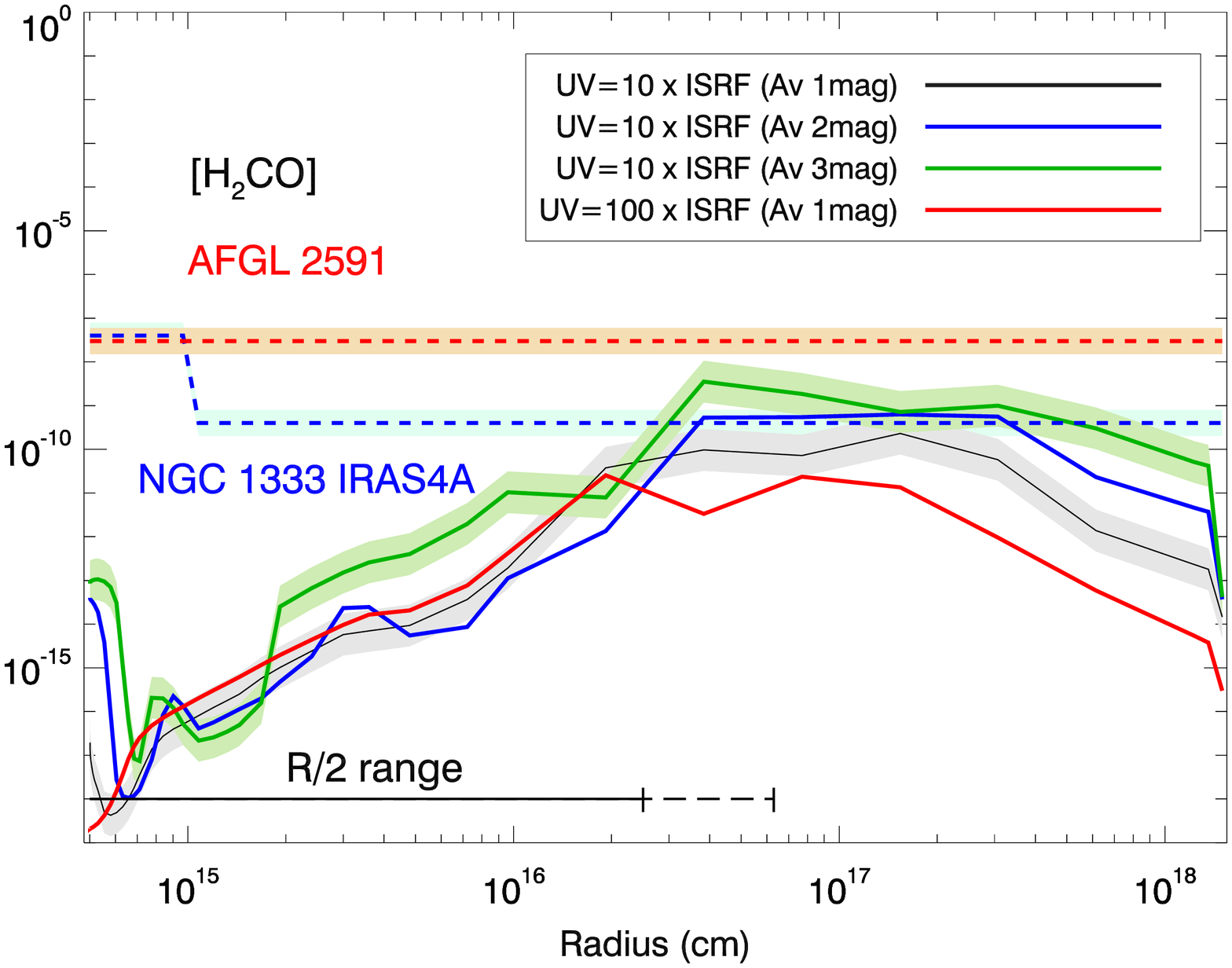} & 
\includegraphics[scale=0.35, angle=90]{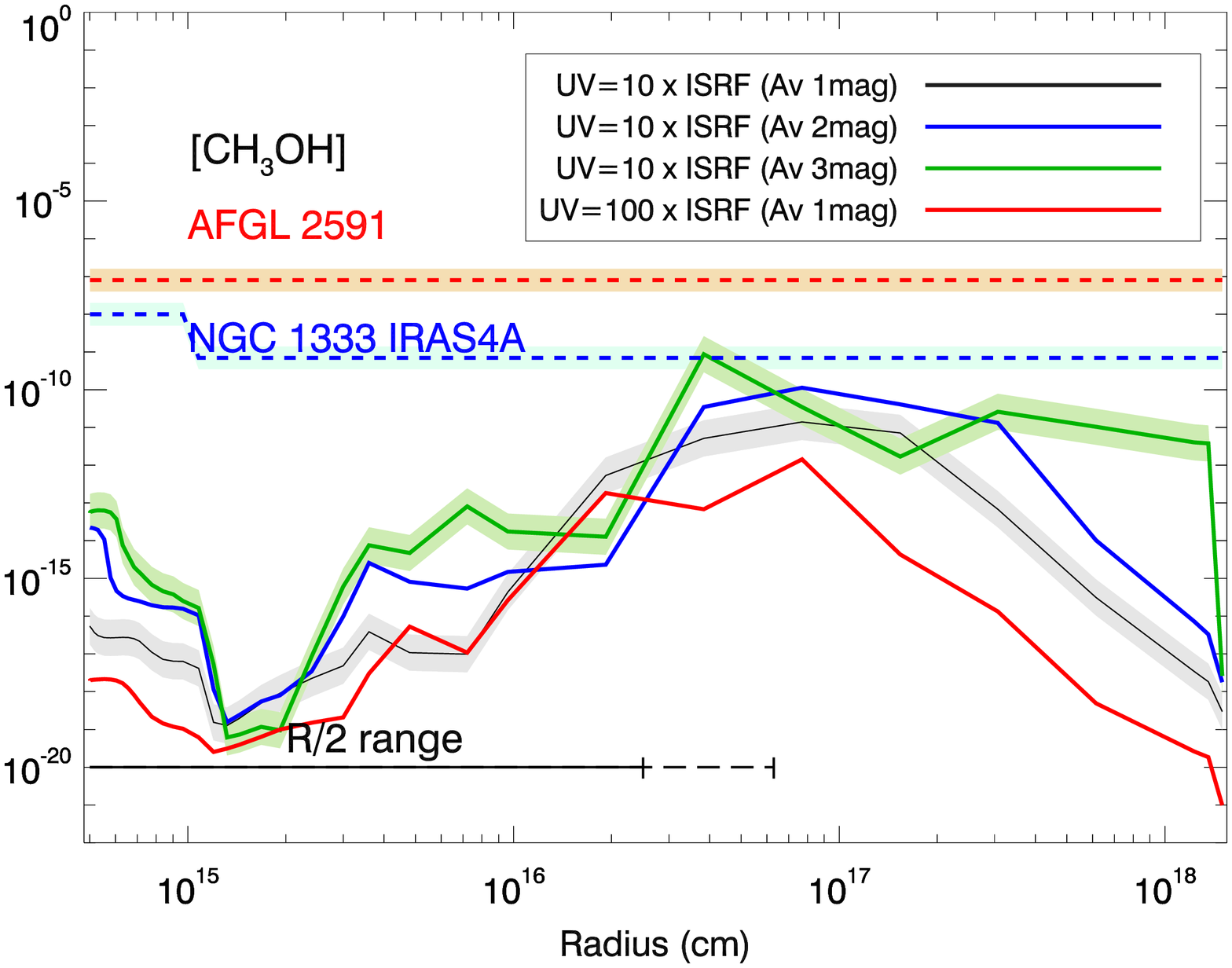} \\
\includegraphics[scale=0.35, angle=90]{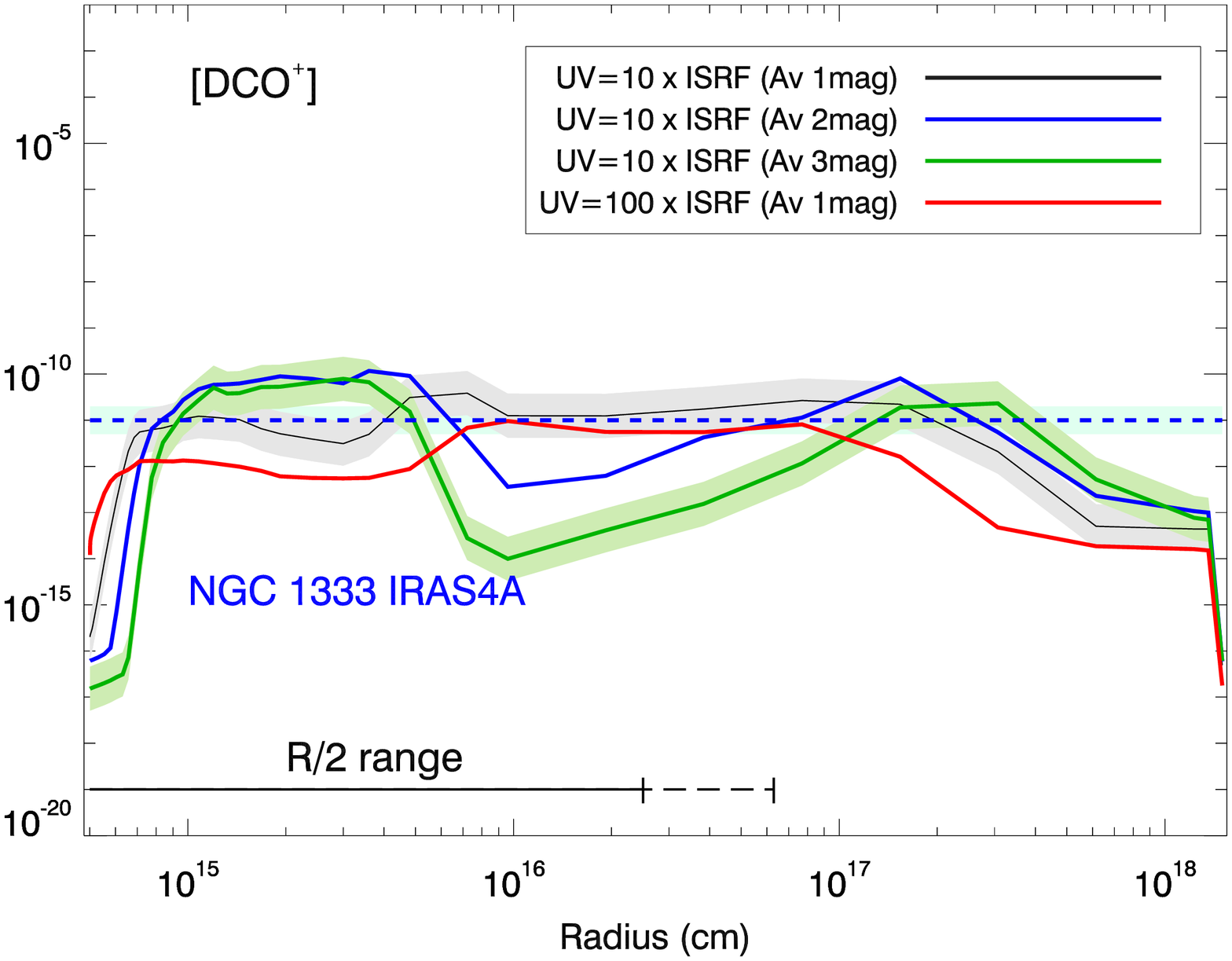} & \includegraphics[scale=0.35, angle=90]{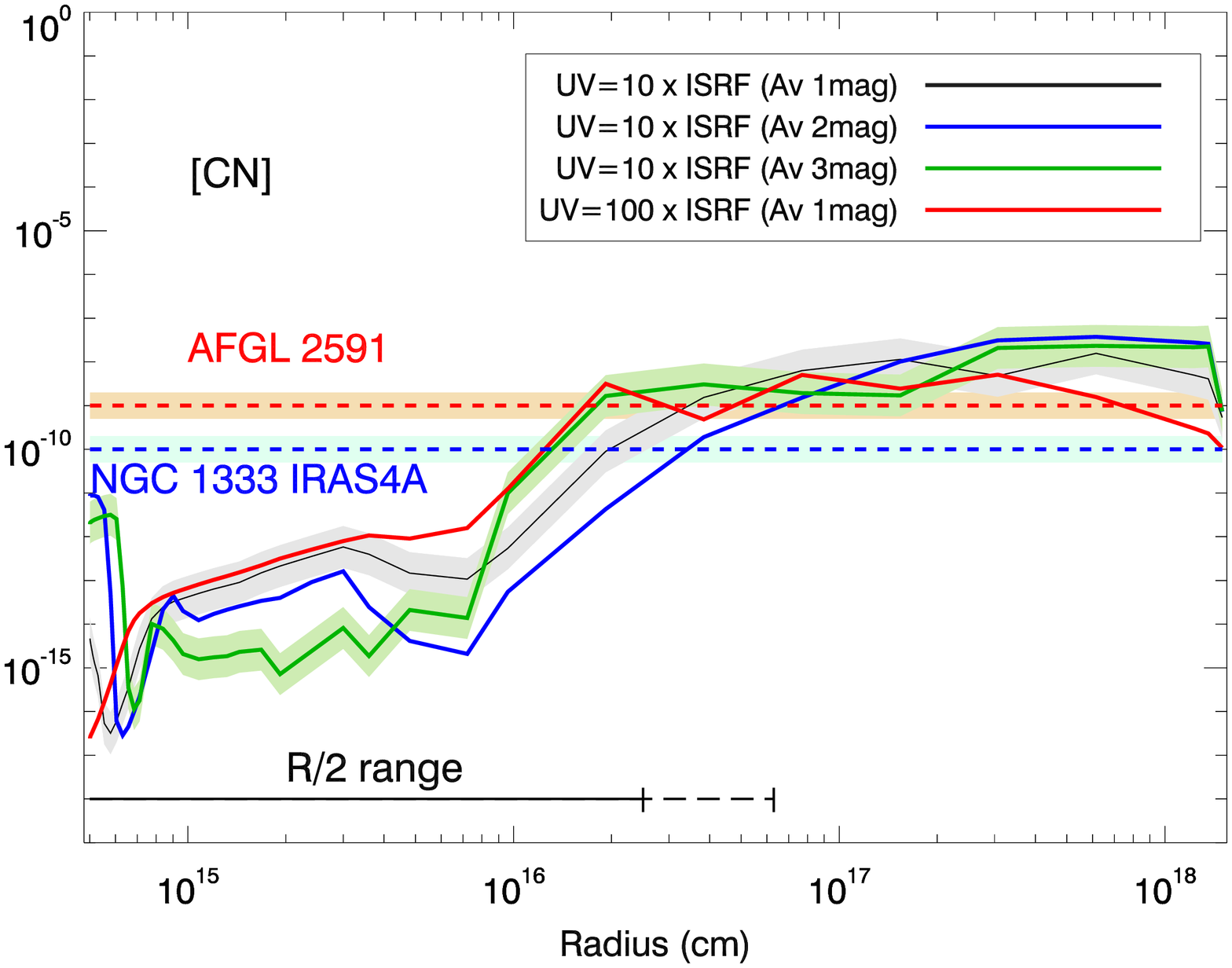} \\
\includegraphics[scale=0.35, angle=90]{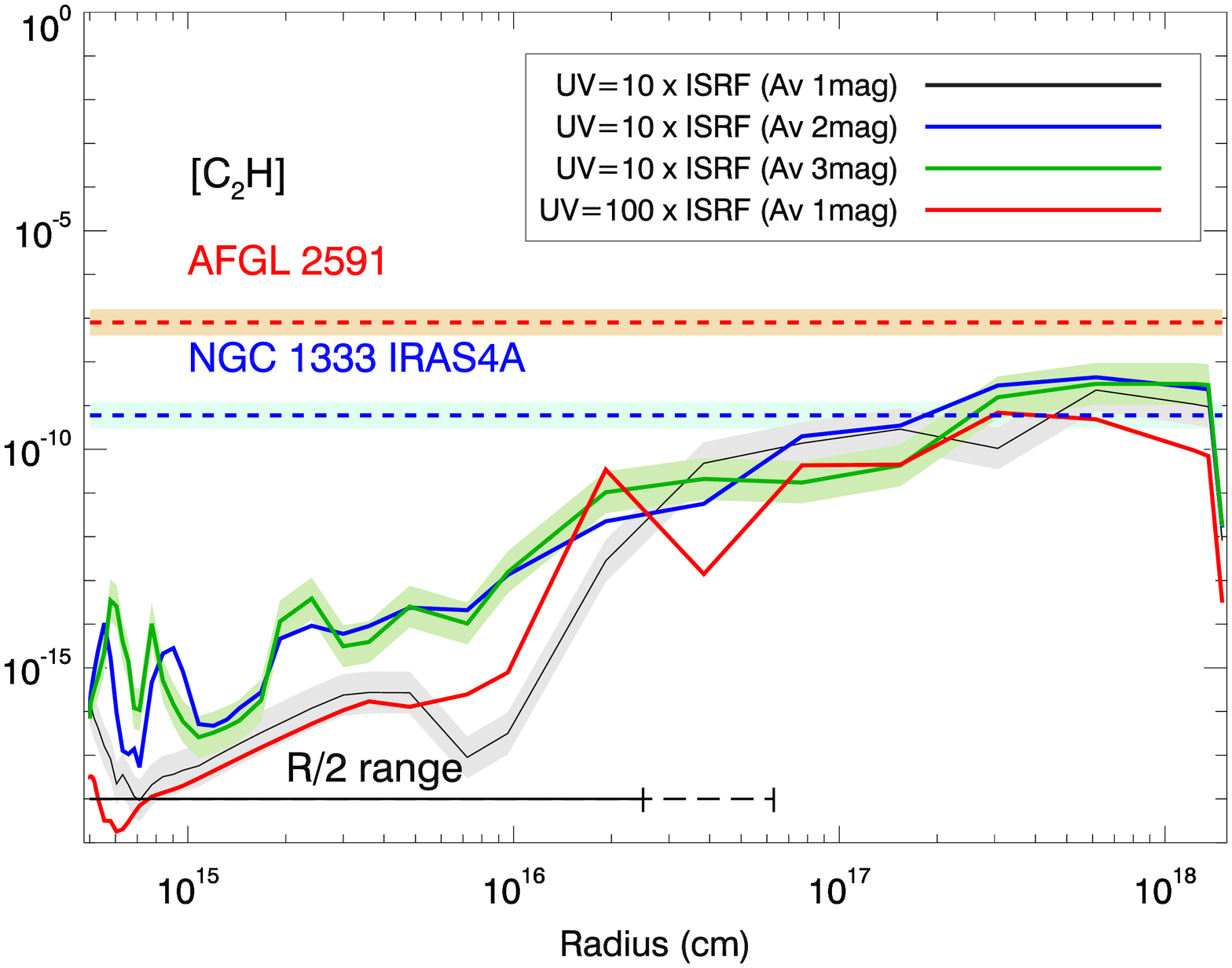} & \includegraphics[scale=0.35, angle=90]{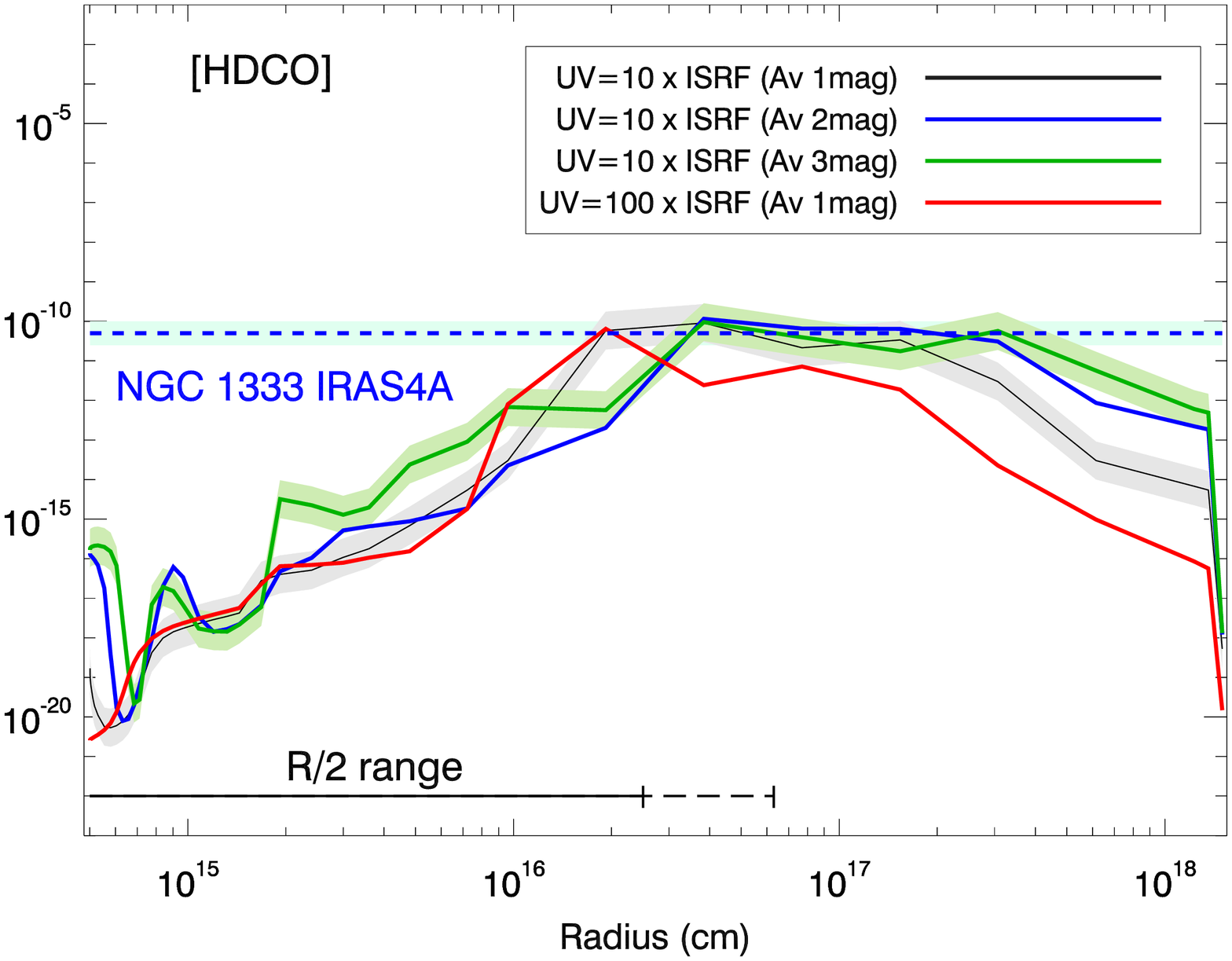} \\
\end{array}$
\end{center}
\caption{As Fig.~\ref{fig:chem31}, but for H$_{2}$CO, CH$_{3}$OH, C$_{2}$H, CN, HDCO, CO and DCO$^{+}$. The deuterated species, HDCO and DCO$^{+}$ were not observed towards AFGL~2591.}
\label{fig:chem41}
\end{figure*}

\begin{figure*}[h]
\begin{center}$ 
\begin{array}{cc}
\includegraphics[scale=0.35, angle=90]{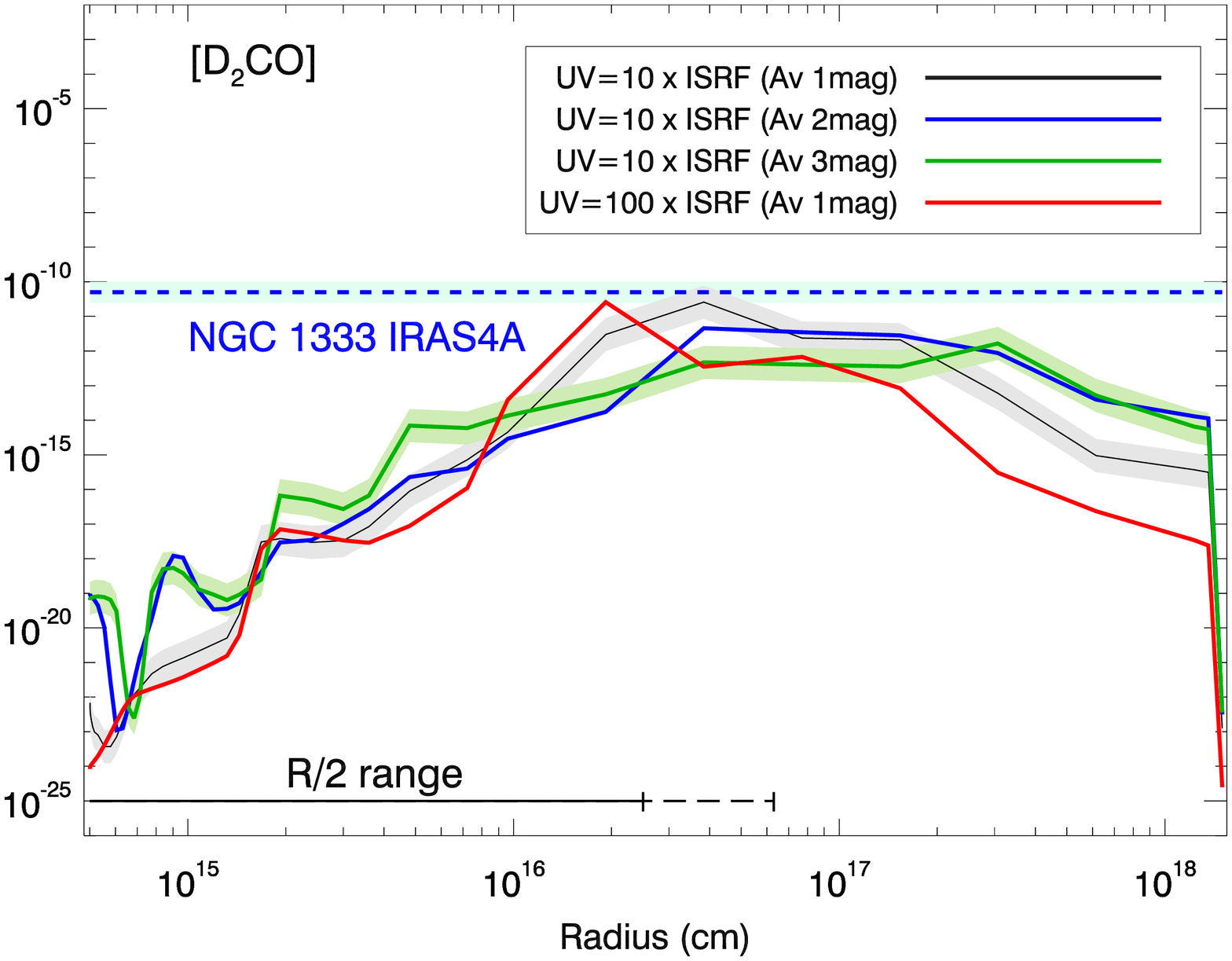} & \includegraphics[scale=0.35, angle=90]{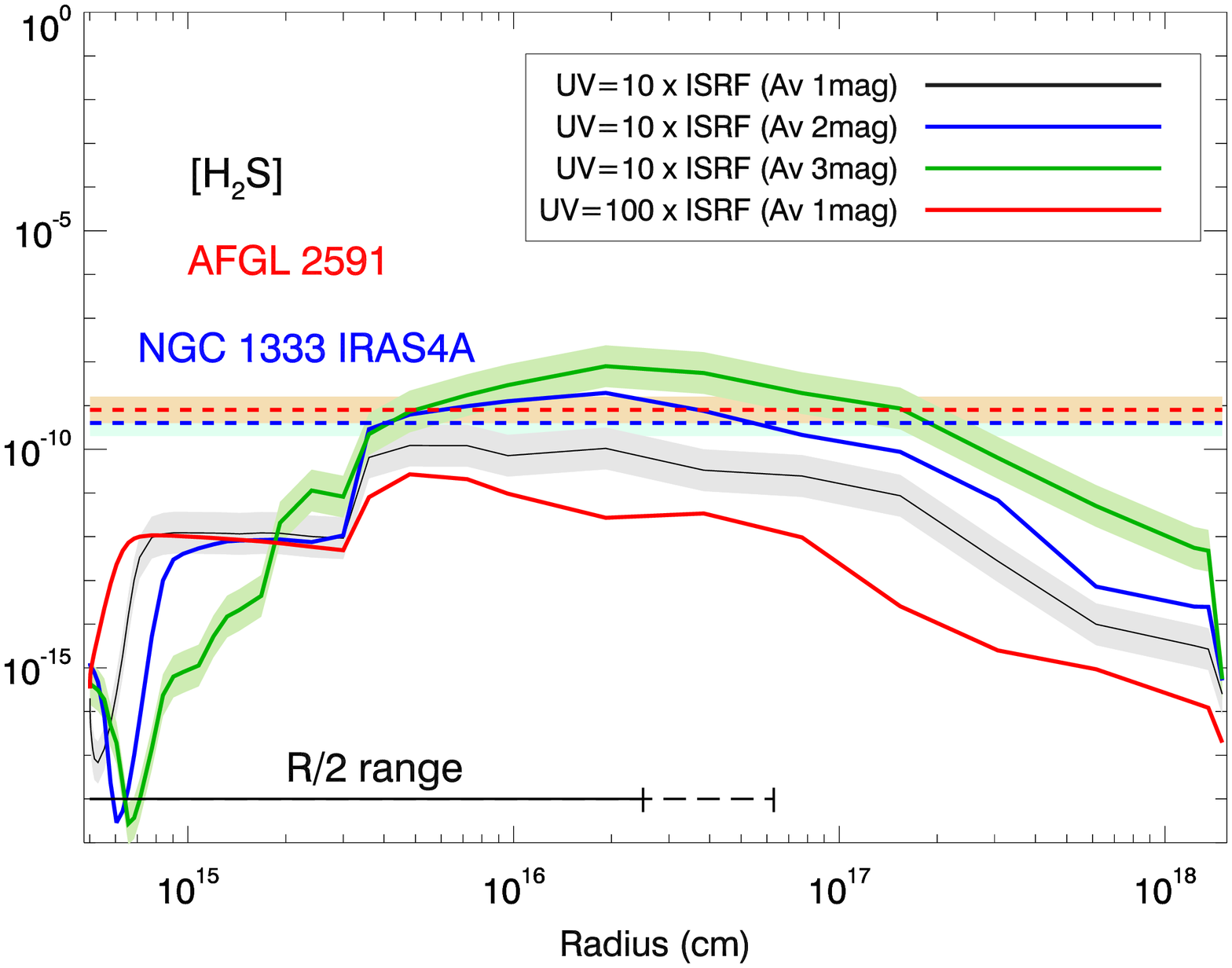} \\
\includegraphics[scale=0.35, angle=90]{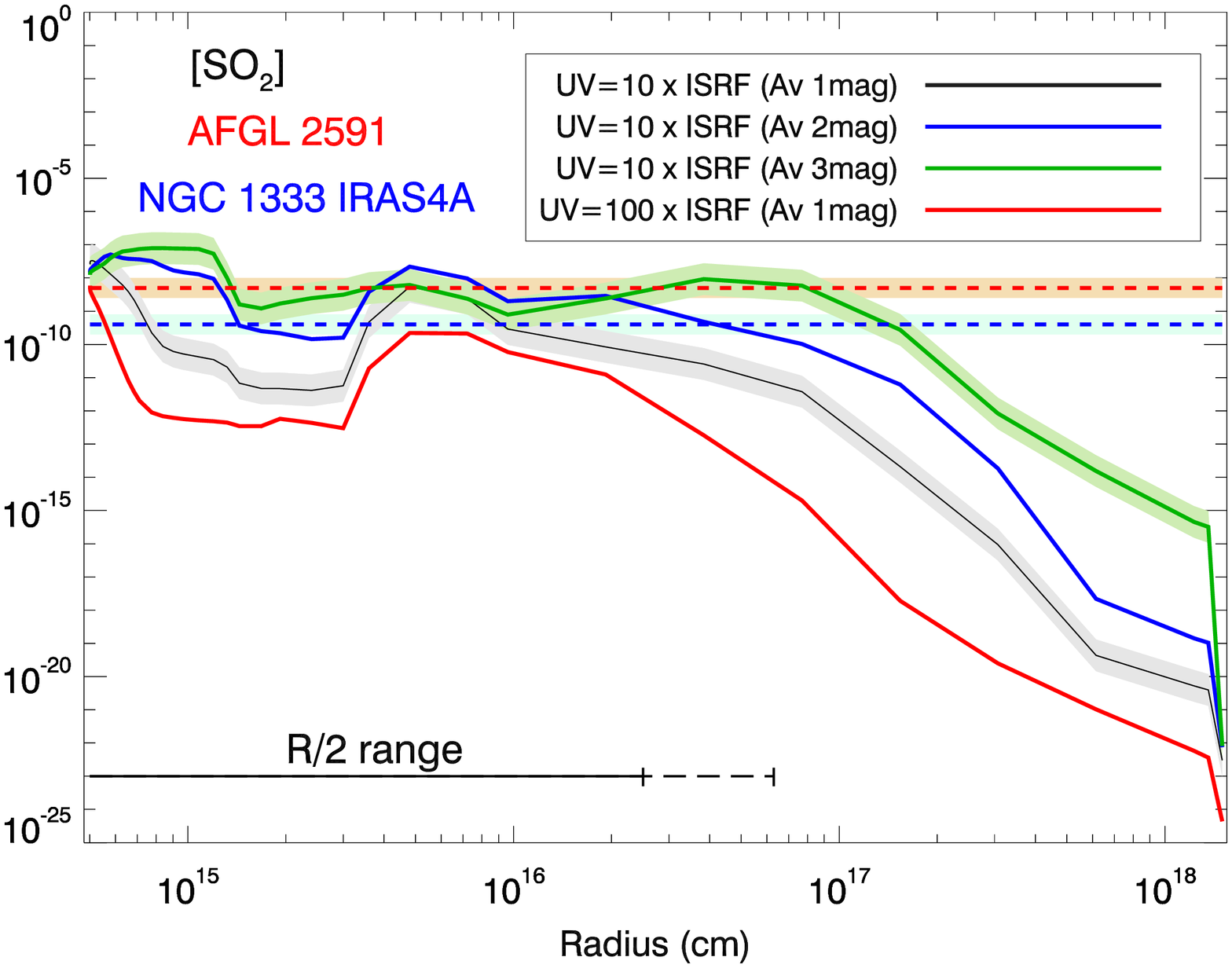} & \includegraphics[scale=0.35, angle=90]{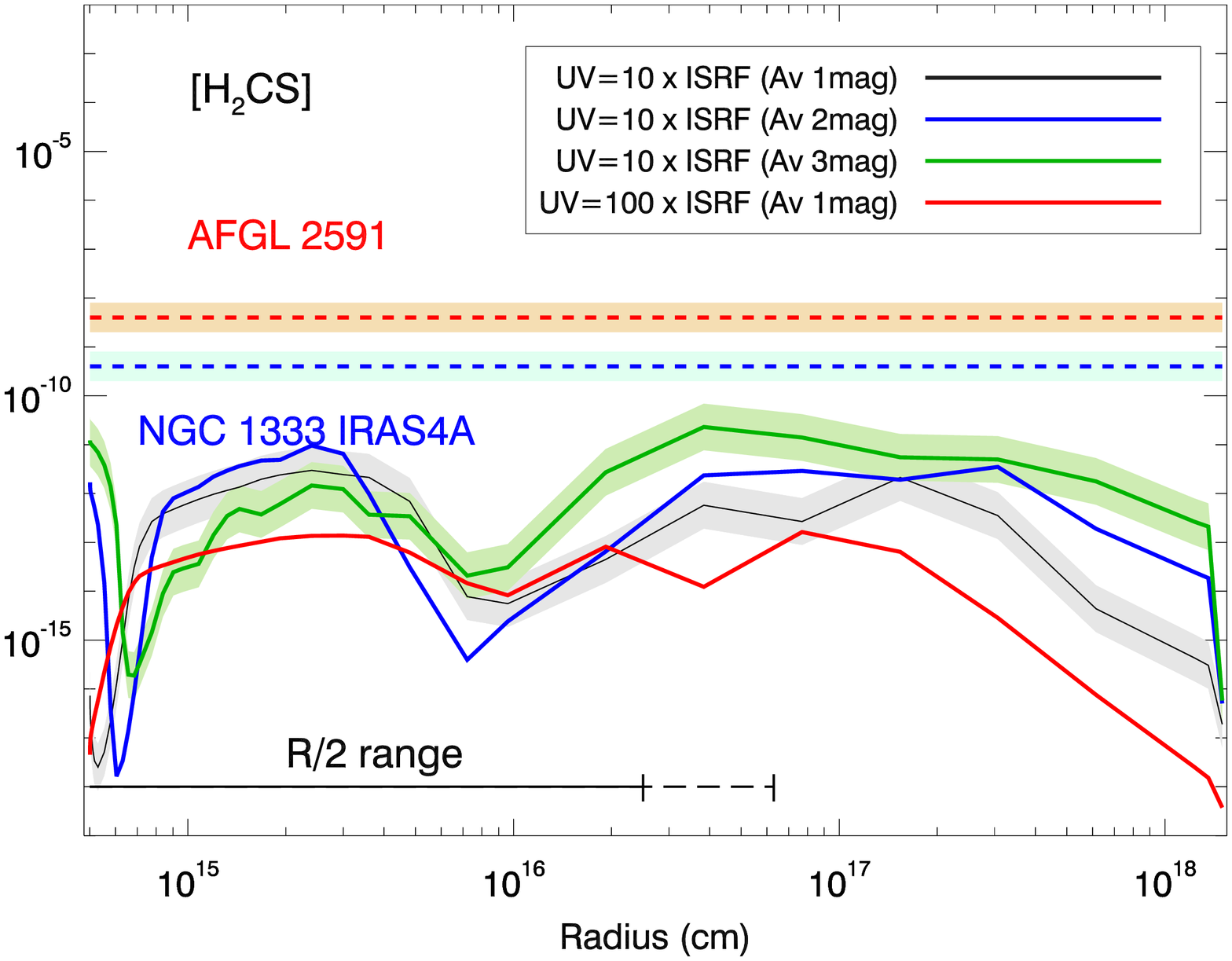} \\
\end{array}$
\includegraphics[scale=0.35, angle=90]{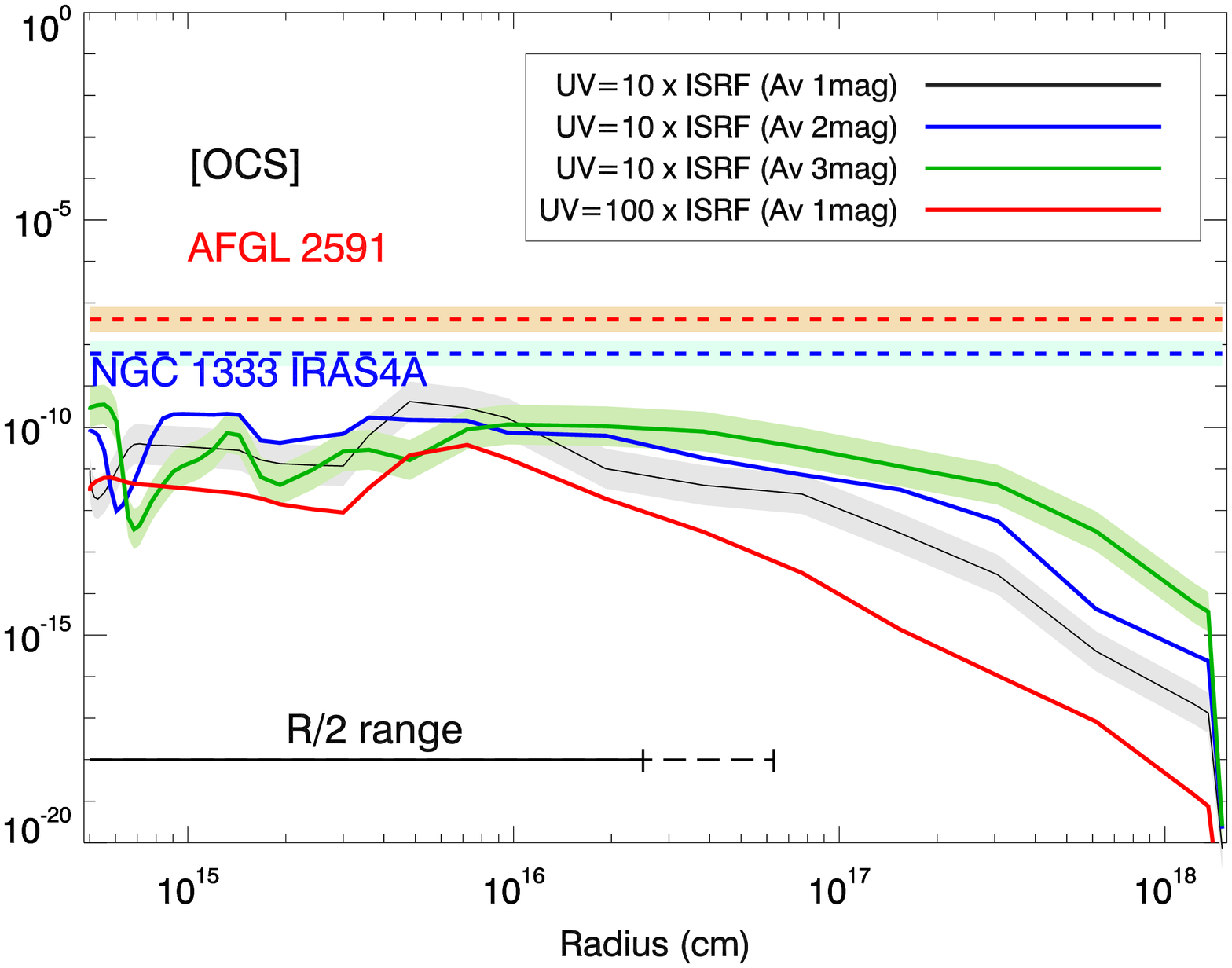}
\end{center}
\caption{As Fig.~\ref{fig:chem31}, but for H$_{2}$S, D$_{2}$CO, SO$_{2}$, H$_{2}$CS and OCS. D$_{2}$CO was not observed towards AFGL~2591.}
\label{fig:chem51}
\end{figure*}

Our models are time dependent so we also investigate the influence of different timescales on the produced chemical abundance profiles. 
Figure~\ref{fig:chem} in the appendix presents the time dependent abundance profiles of several species, demonstrating the insignificant influence 
of time in the short timescales that characterize a Class~0 object \citep[10$^{4}$-10$^{5}$ yrs; e.g.][]{Enoch2009}. Our modeled methanol abundance though is in better agreement with the observed abundance for $\ge$ 4$\times$10$^{4}$ yrs (Figures~\ref{fig:chem4},~\ref{fig:chem}), while other species do not show significant abundance variation on these timescales and thus we cannot use them as additional constraints for age. We provide a lower limit to the age of IRAS~4A which is at least four times older than the one given by \citet{Maret2002} and potentially in agreement with the value of 9$\times$10$^{4}$ yrs given by \citet{Goncalves2008} based on the morphology of the observed magnetic field. We should point out that the derived best--fit age of our object is dependent on the framework of model that we use. If in reality conditions are different (e.g., presence of a disk), or our chemical network misses some key reactions, the best-fit age value can vary significantly. A rigorous way to do such modeling would require running numerous models with varying temperature, density, CR ionization rate, and reaction rates, which would give a best-fit chemical age plus its error bars. A previous study towards young high-mass star-forming regions took these factors into account and found chemical ages that were characterized by uncertainties of a factor of 2--3 \citep{Gerner2014}.

To test the dependence of our results on the adopted physical conditions, we have run models with a twice higher cosmic ray ionization rate, a grain growth up to 0.5~$\mu$m, and different initial abundances for timescales between 10$^{3}$ and 
10$^{6}$ yrs. To set the different initial abundances for the chemical modeling, we
calculated the chemical evolution of a 0D model of an infrared dark cloud with n$_{H}$ $=$ 2$\times$ 10$^{5}$ cm$^{-3}$, T $=$ 15~K, $\zeta$$_{CRP}$ $=$ 5$\times$10$^{-17}$ s$^{-1}$, H$_{2}$ OPR $=$ 3:1 and A$_{V}$ $=$ 10 mag over 1~Myr. The neutral “low metals”  elemental abundances of \citet{Graedel1982} and \citet{Agundez2013} 
were used. The resulting abundance profiles of this process can be seen in Figure~\ref{fig:chem2}.

We find that the CO abundance profiles are not strongly affected by increasing the cosmic ray ionization rate. In contrast, the different initial abundances can cause a decrease of 1--2 orders of magnitude in the abundance profiles at the inner or outer envelope but do not significantly affect the CO abundance at the snow line. Only in combination with the short timescale of 10$^{3}$ yrs do we see 0.5--1 order of magnitude higher abundance at the snow line compared to the other timescales. This timescale is too short for a Class~0 object though. Lastly, the model with bigger grain sizes of 0.5~$\mu$m shows an increase in CO abundances by 1--2 orders of magnitude at the snow line. This is because the total dust surface area per unit gas volume is smaller in this model compared to the standard case of 0.1~$\mu$m grains, lowering the pace and hence the degree of the CO freeze-out. The effect is particularly dramatic at shorter chemical ages of $\sim 10^3$~years, where the difference in CO abundances between the 0.1~$\mu$m and 0.5~$\mu$m models is $\sim 3-4$ orders of magnitude. When grains are on average bigger than $0.1\,\mu$m it slows down freeze-out which would increase the CO gas/ice ratio especially for short timescales (1000 years; Figure~\ref{fig:chem2}). The same effect would be seen for all other species that can freeze out. This would enhance the overall gas-phase molecular abundances with respect to our standard model. Although IRAS~4A is still a very young object ($\sim$10000 yrs) its lifespan cannot be only 1000 years. Even if we were to assume such a short lifespan, it could not explain a grain growth up to $0.5\,\mu$m, but rather an ISM-like size up to 0.2--0.3~$\mu$m \citep{Bianchi2003}. Some studies of mm observations have shown that larger, mm-sized grains may exist in the envelopes of Class~0 protostars \citep{Kwon2009,Chiang2012}. However, Class 0 objects are highly embedded, making it difficult to eliminate the optically thick emission, which can cause an overestimation on the grain size. In addition our modeled abundances cannot explain the observed ones for timescales of 10000 yrs only by applying a grain growth.

In conclusion, the time-dependent models show no significant differences in abundance profiles for the timescales that are relevant to our object. Therefore, adopting an age of 10000 yrs does not introduce significant systematic errors. The same applies for the variation of the cosmic ray ionization rate.
In contrast, we find that grain growth and different initial abundances have a more significant influence in the resulting abundance profiles, especially at short timescales (1000 yrs). Such short chemical age in combination with significant grain growth is rather unlikely though for a Class~0 object such as IRAS~4A.

\section{Conclusions}

We used HIFI and JCMT data to constrain the chemical structure of a low-mass protostellar envelope and compare it with a high-mass equivalent. 

\subsection{Results}

\begin{itemize}
  \item Constant or jump/drop-like empirical abundance profiles reproduce well our single-dish submillimeter observations. The abundance in the outer envelope is supported for most species by the predictions of our 1D time-dependent gas-grain chemical model.
  \item The presence of an outflow cavity with a strong UV field can explain the observed abundances of several species (e.g., CO, HCO$^{+}$, DCO$^{+}$) in the outer low-mass protostellar envelope, while passive heating is sufficient towards a high-mass protostellar envelope.
  \item The empirical abundance (with respect to H$_{2}$) profiles for the low-mass protostellar envelope (NGC~1333~IRAS~4A) are systematically 1 to 2 orders of magnitude lower than the high-mass protostellar envelope (AFGL~2591). The overall warmer temperature profile of high-mass protostellar envelopes seems to drive this result. We find similar empirical abundances as soon as we estimate them with respect to CO.  
  \item Population diagrams for H$_{2}$CO indicate 20\% lower T$_{ex}$ and 30\% lower H$_{2}$CO column density for the envelope compared to the outflow. 
  \item High D$_{2}$CO over H$_{2}$CO ratio (10\%) towards IRAS 4A points towards formation via 
grain surface reactions during the cold phase and not gas-phase chemistry. This is in agreement with what has 
been observed before towards IRAS 16293-2422.
  \item We find an enrichment of DCO$^{+}$ over HCO$^{+}$ ratio towards the shock position compared to the protostellar envelope. We attribute this result to the CO originally formed in the grains and later released into the gas phase (at T$<$20~K) during the passage of the shock wave.
  \item H$_{2}$D$^{+}$ shows a different spatial distribution compared to the other deuterated species and a peak velocity at $\sim$8 km~s$^{-1}$. The most prominent explanation is that it is located in a different layer of gas than the clump that contains the protostars.
  \item The abundance profile of CH$_{3}$OH provides a lower limit for the age of NGC~1333~IRAS~4A of 4$\times$10$^{4}$ yrs. 
  
  \end{itemize}

\subsection{Discussion}

The modeled chemical abundance profiles of the inner envelope are a few orders of magnitudes lower than the observed ones for all species. We find a decrease of about 2 orders of magnitude in the abundance of species with more observed transitions such as CO, H$_2$CO and CH$_3$OH in the CO depletion zone (outer envelope). Similar drops have been seen before by interferometric studies of other low-mass protostars \citep[e.g., IRAS 16293--2422, L1448--C;][]{Schoier2004,Jorgensen2005}. \citet{Jorgensen2007} suggest that the emission of H$_{2}$CO and CH$_{3}$OH are related to the shocks caused by the protostellar outflows rather than being the result of compact low-mass ‘‘hot corinos’’. More transitions of the other observed species will help us in the future to better constrain their observed profile. 

Furthermore, we tried to simply simulate an outflow cavity by increasing the UV radiation that the observed species 
are exposed to. We found that this approach improved the fit among the theoretical abundance profiles and 
the observed for several species (e.g., CO, HCO$^{+}$), thus a more detailed 2D/3D chemical modeling that takes into account disk structure and outflow 
cavities is expected to be more accurate. Such a model requires more transitions and interferometric (high sensitivity and high spatial resolution) observations that trace the inner region of the protostellar envelope. 

Lastly, we attribute the observed abundance difference with respect to H$_{2}$ among the low- and the high-mass protostellar envelope to the higher temperatures that characterize the high-mass case and the absence of a freeze-out zone. For safer comparison, further studies of the same nature between high-mass and low-mass protostellar envelopes are necessary. In particular, the similarity in the observed abundances with respect to CO, suggests that gas-phase CO/H$_{2}$ measurements are essential, as all other species are off by within factors of a few as the one of CO.  

\begin{acknowledgements}
D.S. acknowledges support from the Heidelberg Institute of Theoretical Studies for the project Chemical kinetics models and visualization tools: Bridging biology and astronomy. The authors thank Lars Kristensen for useful discussions. We also thank Inga Kamp and Veronica Allen for careful reading who helped in improving the clarity of this paper.  
\end{acknowledgements}

\bibliographystyle{aa} 
\bibliography{ref}

\begin{appendix}

\section{Line detections towards NGC~1333-IRAS~4A}

\begin{table*}[tp]
\scriptsize\addtolength{\tabcolsep}{-2.0pt}
\caption{Molecular Line Detections for NGC~1333-IRAS~4A - HIFI data}
\begin{tabular}{c c c c c c c c c c c}
\toprule
 & & & &   &\bf Component 1 & &  & \bf Component 2 &\\
\hline\hline
Molecule & Transition & Rest Frequency & E$_{up}$ & V$_{lsr}$ & FWHM & T$_{mb}$ & V$_{lsr}$ & FWHM & T$_{mb}$ \\  &  & (MHz) & (K) & (km~s$^{-1}$) & (km~s$^{-1}$) & (K) & (km~s$^{-1}$) & (km~s$^{-1}$) & (K) \\
\midrule
CO & 6-5 & 691473.08 & 116.2 & $7.5\pm0.1$ & $8.04\pm0.07$ & $2.78\pm0.01$ & $6.6\pm0.6$ & $0.8\pm0.2$ & $5.2\pm1.2$ \\
$^{13}$CO & 6-5 & 661067.28 & 111.0 & $6.7\pm0.1$ & $10.7\pm0.4$ & $0.19\pm0.01$ & $6.93\pm0.01$ & $1.60\pm0.04$ & $0.74\pm0.01$ \\
$^{13}$CO & 7-6 & 771184.12 & 148.1 & $7.4\pm0.3$ & $9.0\pm1.0$ & $0.17\pm0.02$ & $6.78\pm0.03$ & $1.4\pm0.1$ & $0.48\pm0.03$ \\
C$^{18}$O & 6-5 & 658553.28 & 110.6 & $6.8\pm0.1$ & $2.4\pm0.3$ & $0.14\pm0.02$ & $6.81\pm0.05$ & $0.6\pm0.2$ & $0.11\pm0.03$ \\
HCO$^{+}$ & 8-7 & 713341.23 & 154.1 & $6.1\pm0.6$ & $7.0\pm2.0$ & $0.17\pm0.04$ & $7.07\pm0.04$ & $1.9\pm0.1$ & $0.68\pm0.05$ \\
N$_{2}$H$^{+}$  & 7-6 & 652095.57 & 125.2 & $7.7\pm0.2$ & $5.3\pm0.3$ & $0.05\pm0.08$ & $7.5\pm0.1$ & $1.5\pm0.4$ & $0.15\pm0.08$ \\
N$_{2}$H$^{+}$  & 8-7 & 745209.87 & 161.0 & $7.9\pm0.1$ & $2.3\pm0.3$ & $0.13\pm0.02$ & $7.73\pm0.05$ & $0.5\pm0.10$ & $0.11\pm0.02$ \\
H$_{2}$O & 2(1,1)-2(0,2) & 752033.14 & 136.9 & $8.6\pm0.6$ & $30.2\pm0.7$ & $0.34\pm0.01$ & $0.2\pm0.2$ & $11.4\pm0.9$ & $0.24\pm0.02$ \\
H$_{2}$S & 2(1,2)-1(0,1) & 736034.10 & 55.1 & $6.9\pm0.2$& $4.1\pm0.2$& $0.13\pm0.01$ & $6.6\pm0.1$ & $0.5\pm0.1$& $0.4\pm0.1$\\
\hline
 & &  &  & & \bf Single component &  &  &  &  \\
\hline\hline
HCN & 8-7 & 708877.21 & 153.1 & $6.4\pm0.5$ & $8.9\pm0.5$ & $0.10\pm0.01$ & & & \\
HCN & 9-8 & 797433.66 & 191.4 & $4.7\pm0.6$ & $6.4\pm0.6$ & $0.09\pm0.01$ & & & \\
CS & 13-12 & 636531.84 & 213.9 &$6.8\pm0.5$ & $6.2\pm0.5$& $0.05\pm0.02$ & & & \\
CS & 14-13 & 685434.76 & 246.8 &$4.5\pm0.5$ & $6.2\pm0.5$ & $0.05\pm0.2$ & & & \\
CS & 15-14 & 734324.00 & 282 & $5.5\pm0.6$ & $6.1\pm0.6$ & $0.04\pm0.01$ & & & \\
H$_{2}$CO & 9(1,9)-8(1,8) & 631702.81 & 163.6 & $7.0\pm0.2$& $2.7\pm0.3$& $0.18\pm0.02$ &  &  & \\
H$_{2}$CO & 9(0,9)-8(0,8) & 647081.73 & 156.2   &$6.5\pm0.2$ & $3.8\pm0.5$&$0.09\pm0.01$ & &  &\\
H$_{2}$CO & 9(5,5)-8(5,4) & 655212.10 & 451.2   &$6.2\pm0.2$ &$2.8\pm0.6$ & $0.08\pm0.02$ &  &  &\\
H$_{2}$CO & 9(1,8)-8(1,7) & 674809.78 & 174.0 & $6.3\pm0.2$ & $4.6\pm0.6$ & $0.13\pm0.01$ & & & \\
H$_{2}$CO & 10(1,10)-9(1,9) & 701370.46 & 197.3 & $6.9\pm0.1$ & $4.7\pm0.2$ & $0.19\pm0.01$ & & & \\
CH$_{3}$OH  & 5(-2,4)-4(-1,4) & 665442.45 & 60.7 & $6.5\pm0.8$ & $5.2\pm0.8$ & $0.05\pm0.01$ & & & \\
CH$_{3}$OH  & 4(2,3)-3(1,2)-- & 673745.93 & 60.9 & $6.9\pm0.4$ & $4.2\pm0.4$ & $0.07\pm0.03$ & & & \\
CH$_{3}$OH  & 8(1,8)-7(0,7)++ & 674990.42 & 97.4 & $5.9\pm0.3$ & $5.0\pm0.3$& $0.09\pm0.01$& & & \\
CH$_{3}$OH  & 4(2,2)-3(1,3)++ & 678785.45 & 60.9 & $6.6\pm0.5$ & $6.1\pm0.5$ & $0.05\pm0.01$& & & \\
CH$_{3}$OH  & 9(1,9)-8(0,8)++ & 719664.88 & 118.1  & $6.9\pm0.2$& $3.7\pm0.3$& $0.09\pm0.01$&  & &\\
CH$_{3}$OH  & 5(2,3)-4(1,4)++ & 728862.52 & 72.5 & $5.2\pm0.5$ & $5.0\pm0.5$ & $0.07\pm0.01$& & & \\
CH$_{3}$OH  & 5(3,2)-4(2,2) & 772453.80 & 82.5 & $6.4\pm0.6$ & $5.6\pm0.6$& $0.05\pm0.01$& & & \\
CH$_{3}$OH  & 6(2,4)-5(1,5)++ & 779380.51 & 86.5  & $6.9\pm0.2$&$3.8\pm0.2$ &$0.15\pm0.01$ & & &\\
\hline
 & &  &  &  & \bf Absorption component &  &  &  &  \\
\hline\hline
CO & 6-5 & 691473.08 & 116.2 & $7.5\pm0.3$ & $0.7\pm0.1$ & $-4.7\pm2.1$ &  &  &  \\
H$_{2}$S & 2(1,2)-1(0,1) & 736034.10 & 55.1 & $7.6\pm0.3$ & $0.9\pm0.1$ & $-0.3\pm0.1$ &  & &  \\
\bottomrule
\end{tabular}
\label{HIFI_lines}
\end{table*}

\section{Chemical models for various input parameters}

\begin{figure*}[h]
\begin{center}$ 
\begin{array}{cc}
\includegraphics[scale=0.32, angle=90]{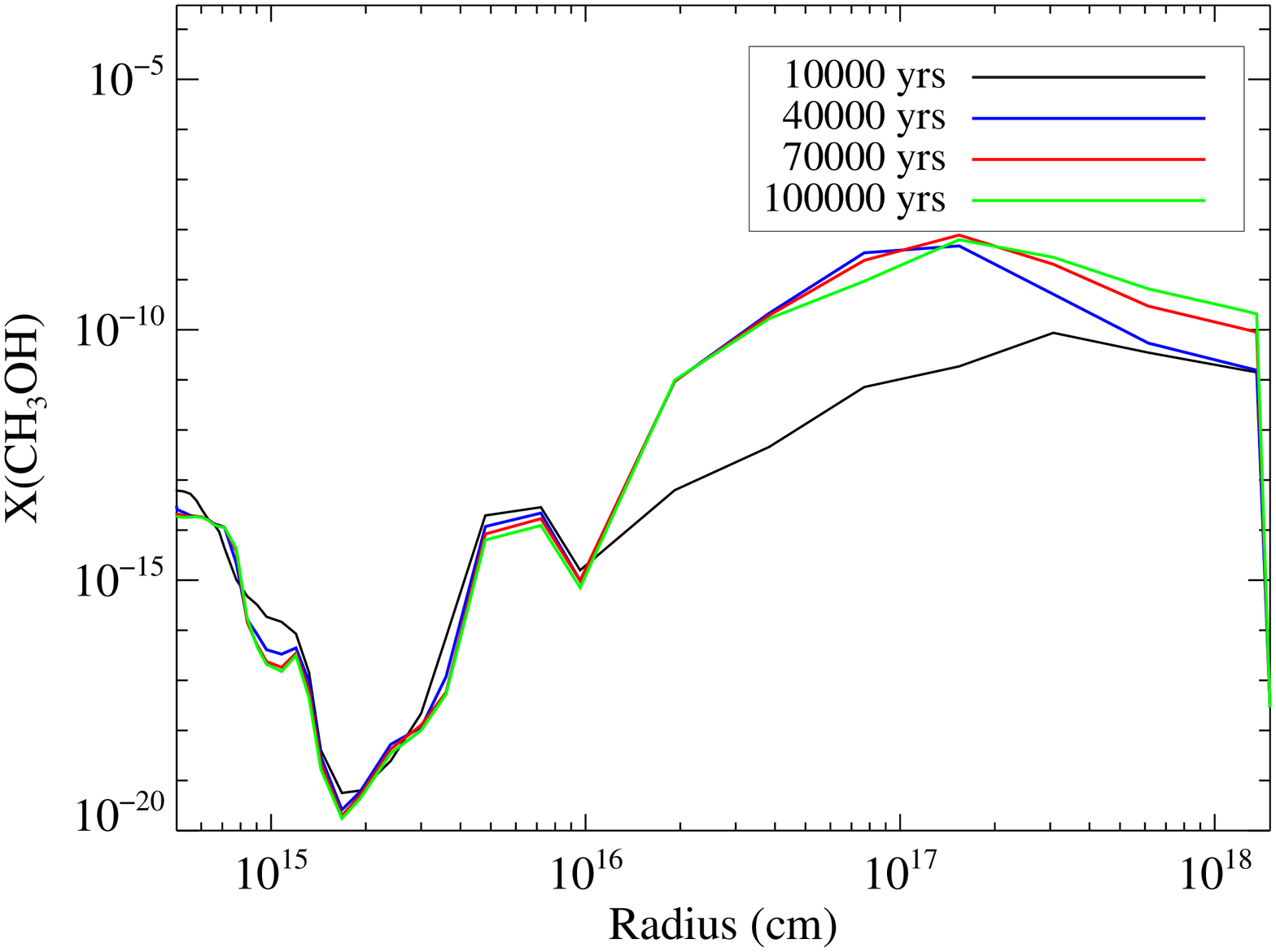} & 
\includegraphics[scale=0.32, angle=90]{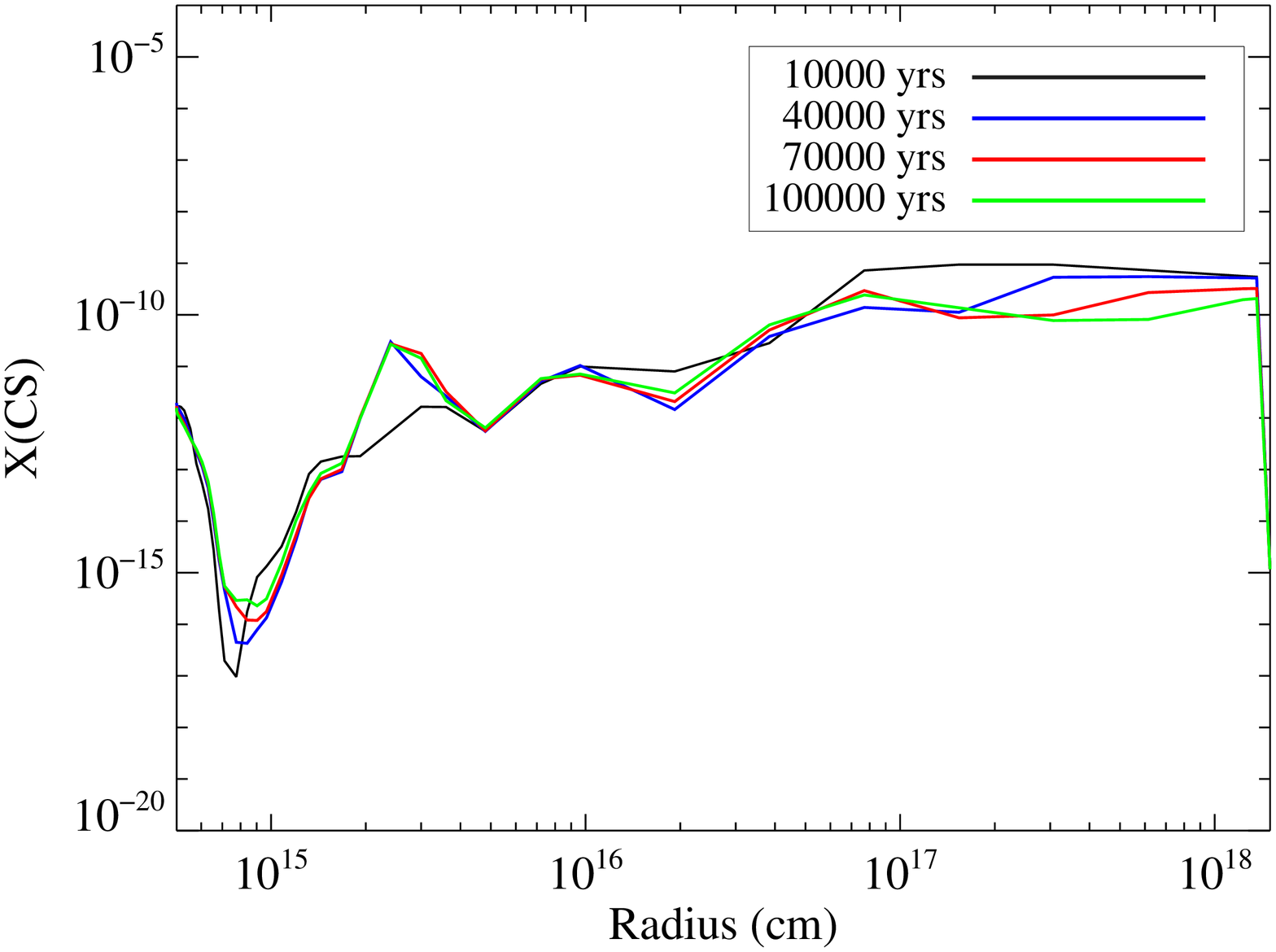} \\
\includegraphics[scale=0.32, angle=90]{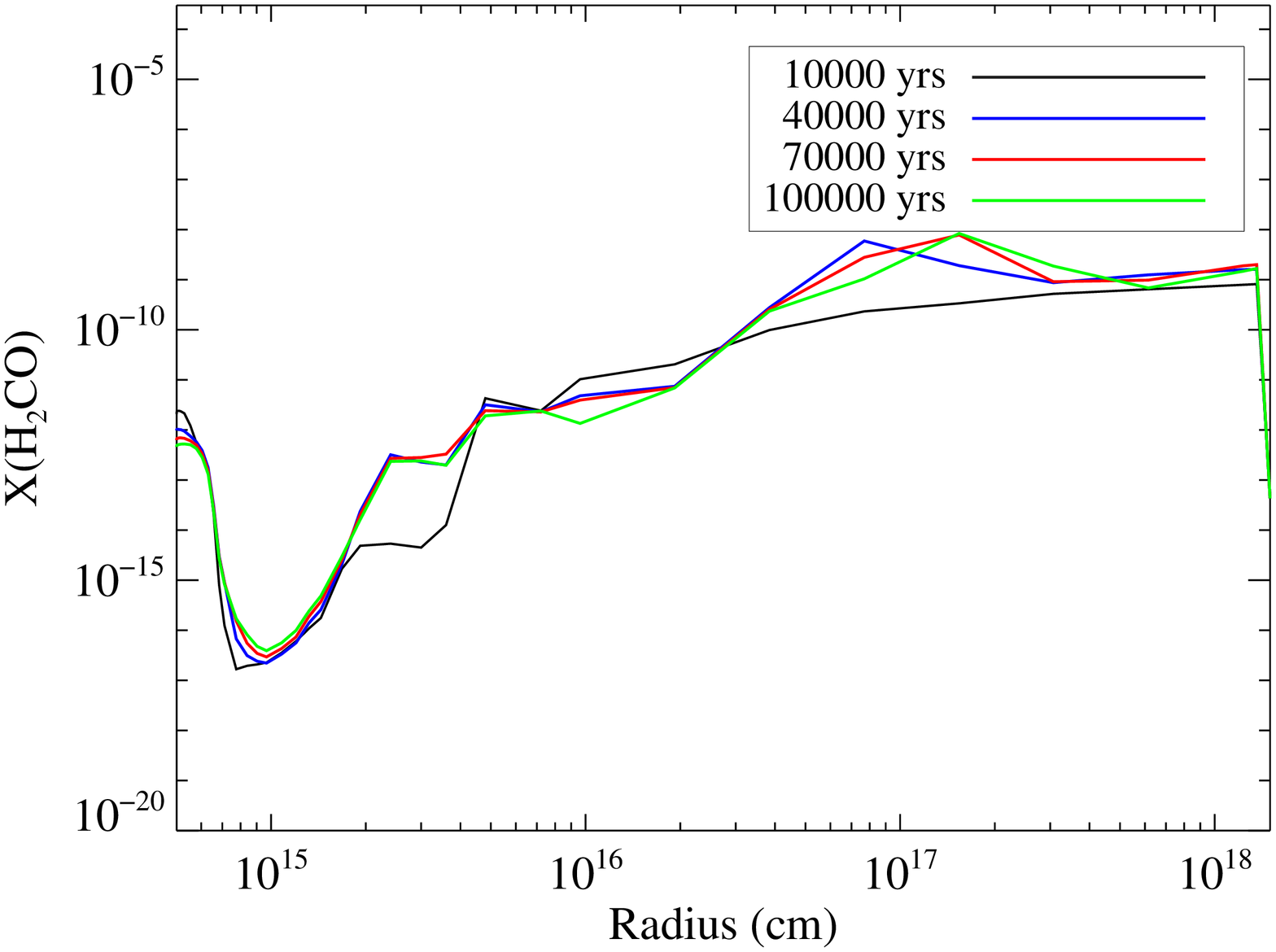} & \includegraphics[scale=0.32, angle=90]{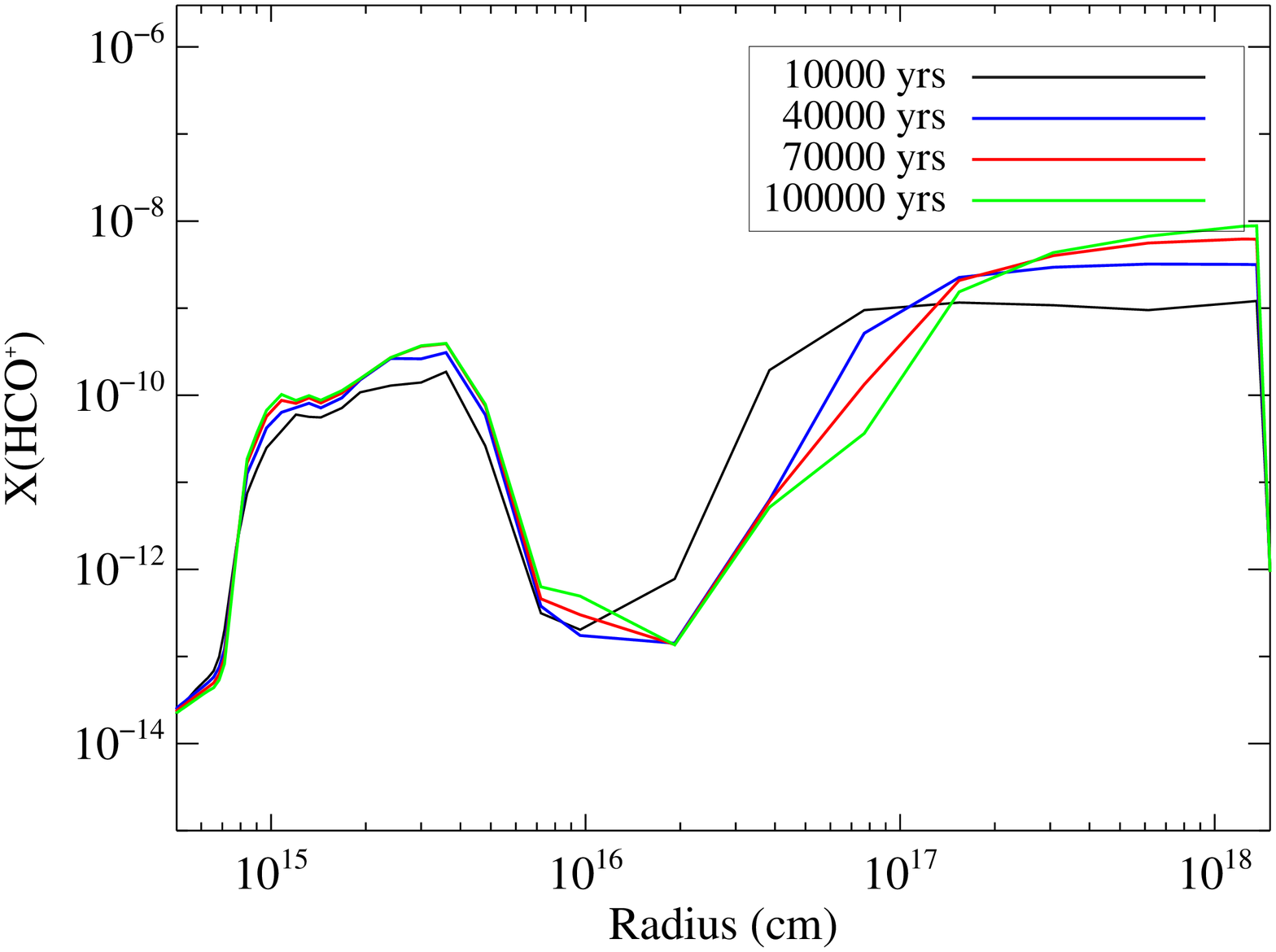} \\
\includegraphics[scale=0.32, angle=90]{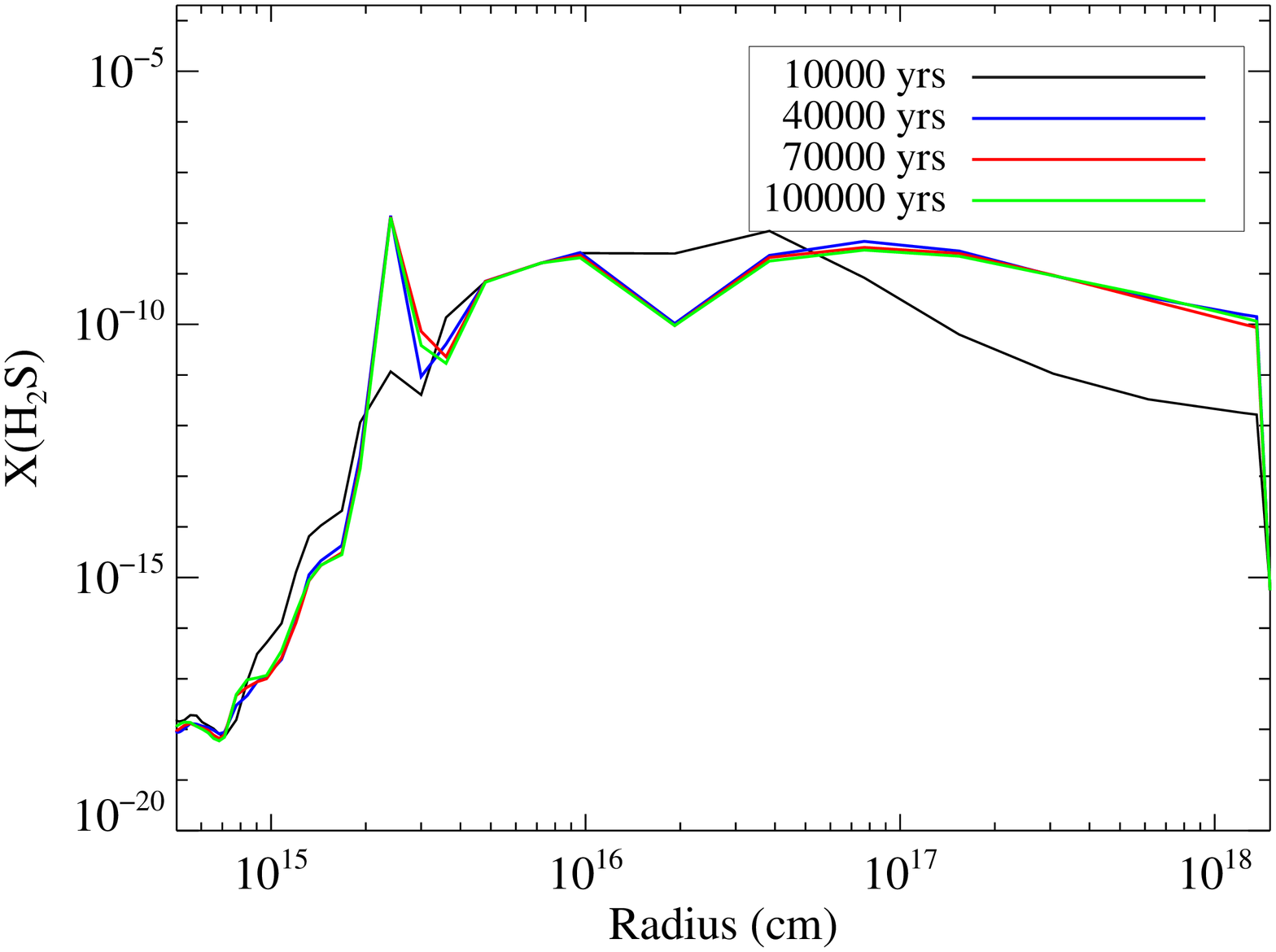} & \includegraphics[scale=0.32, angle=90]{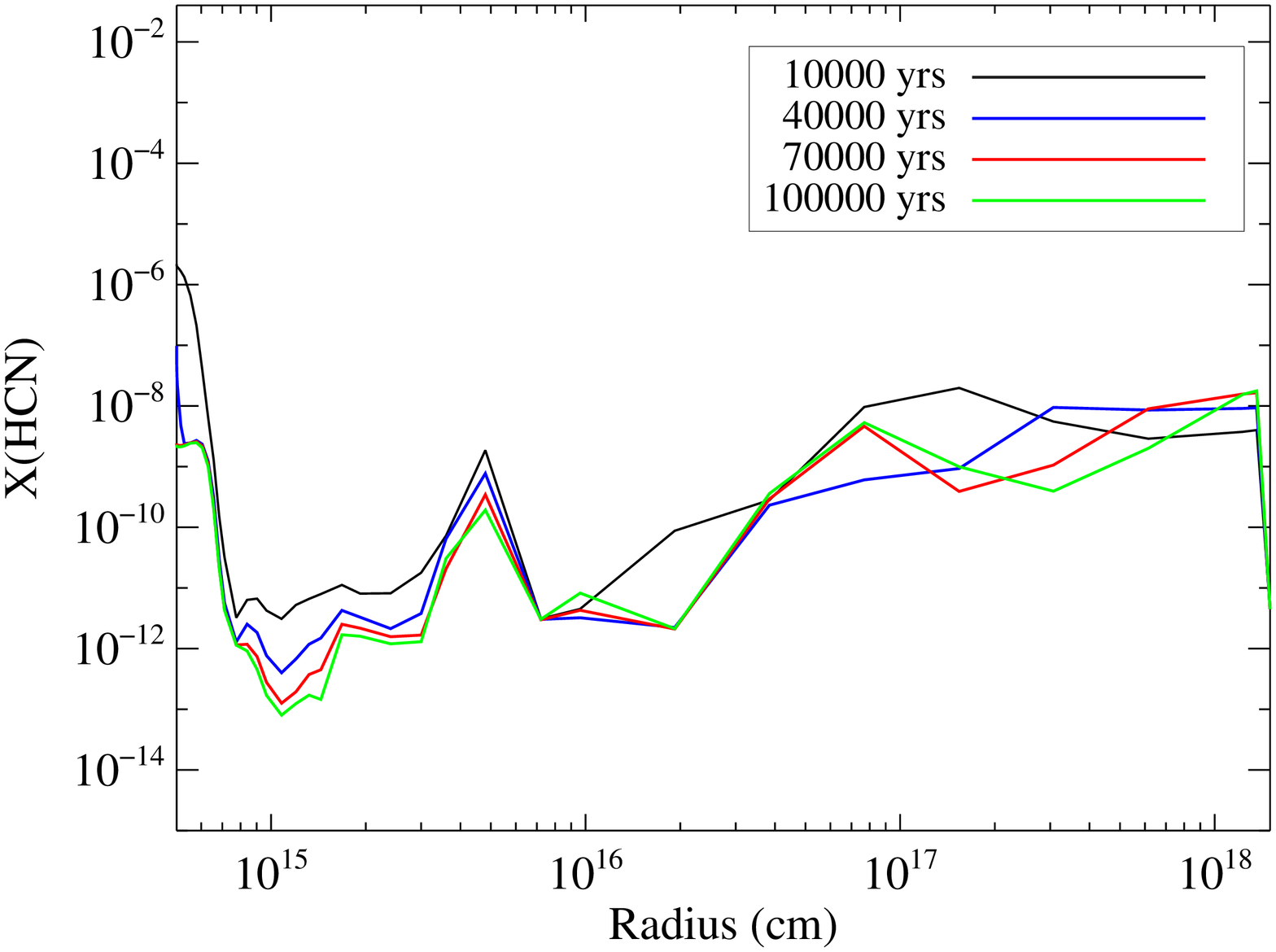} \\
\end{array}$
\end{center}
\caption{Time-dependent 1D chemical models for timescales 10$^{4}$ -- 10$^{5}$ yrs, which is 
the predicted lifespan of Class~0 objects, such as NGC~1333~IRAS~4A. The observed variations are not 
significant and thus we adopt the 10$^{4}$ models to compare with our empirical models.}
\label{fig:chem}
\end{figure*}

\begin{figure*}[h]
\begin{center}$ 
\begin{array}{cc}
\includegraphics[scale=0.32, angle=90]{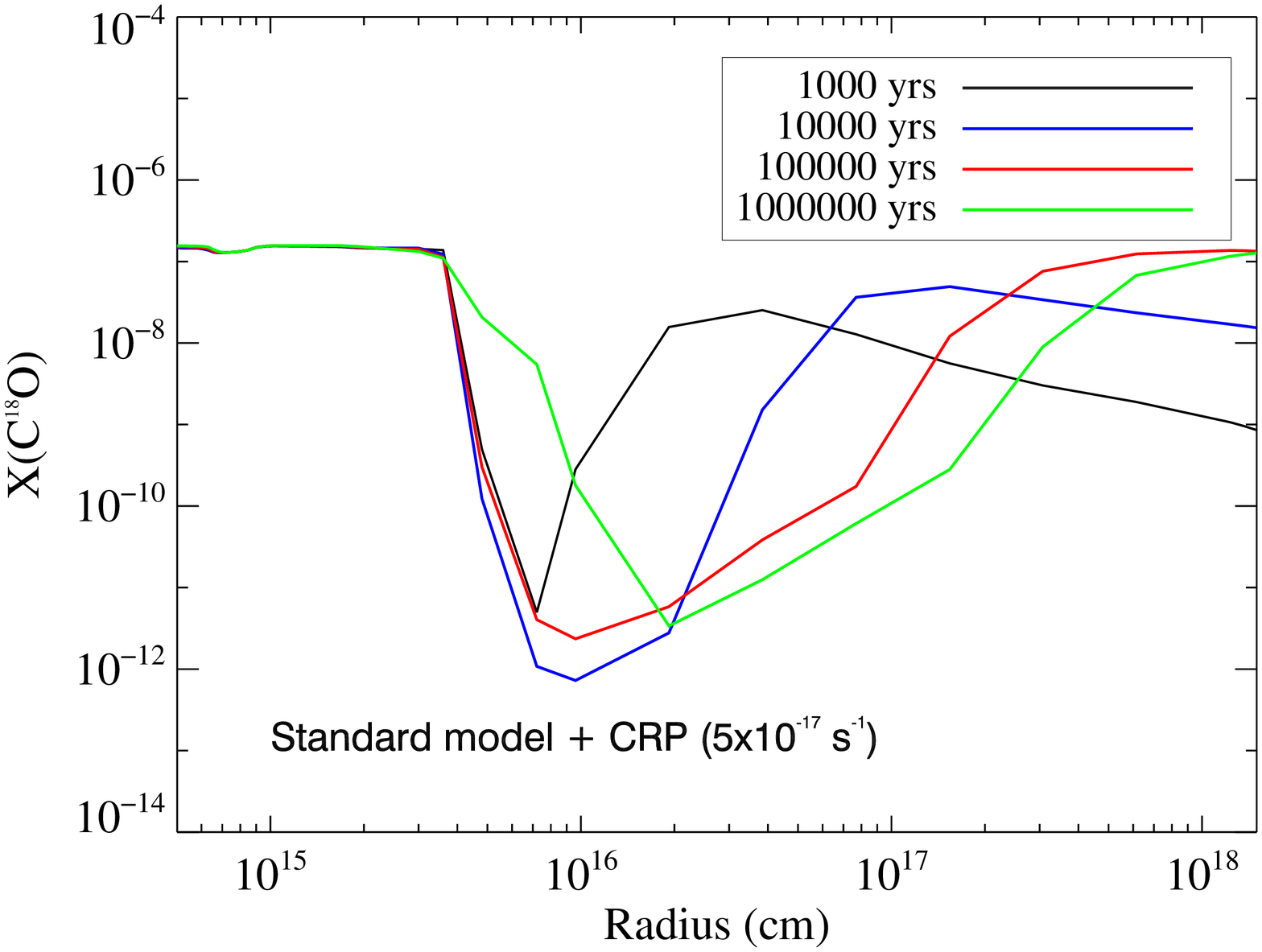} & 
\includegraphics[scale=0.32, angle=90]{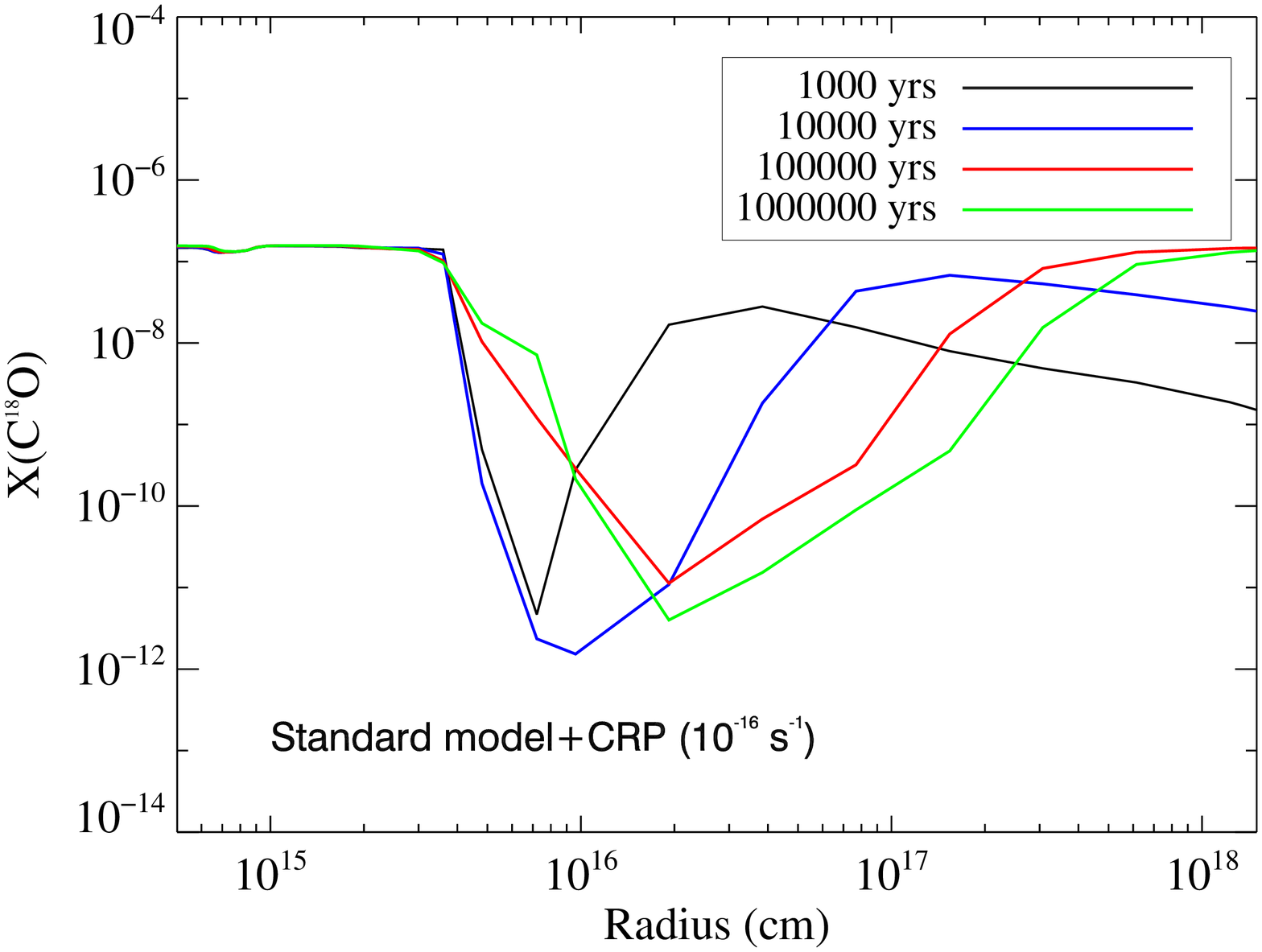} \\
\includegraphics[scale=0.32, angle=90]{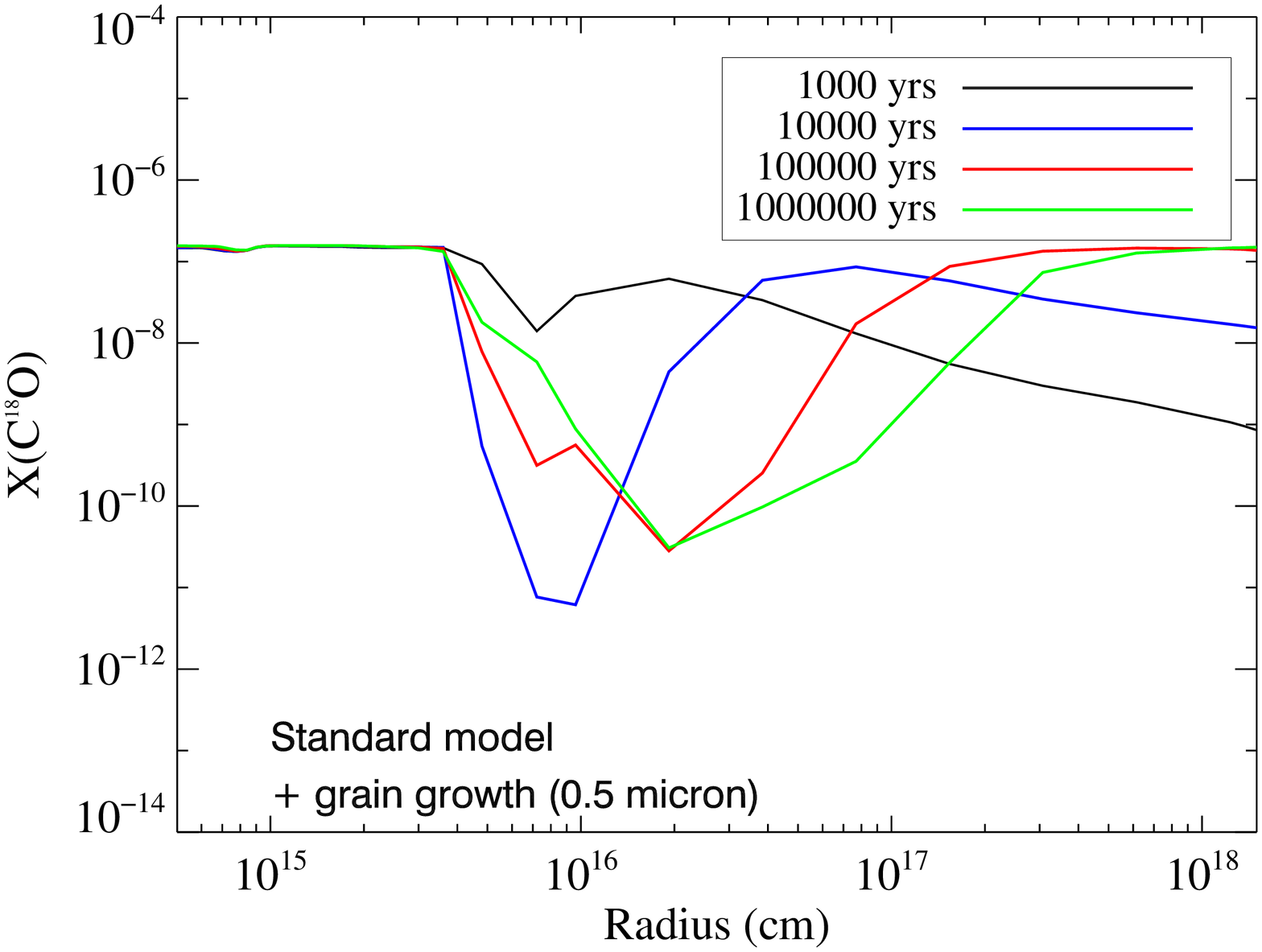} & \includegraphics[scale=0.32, angle=90]{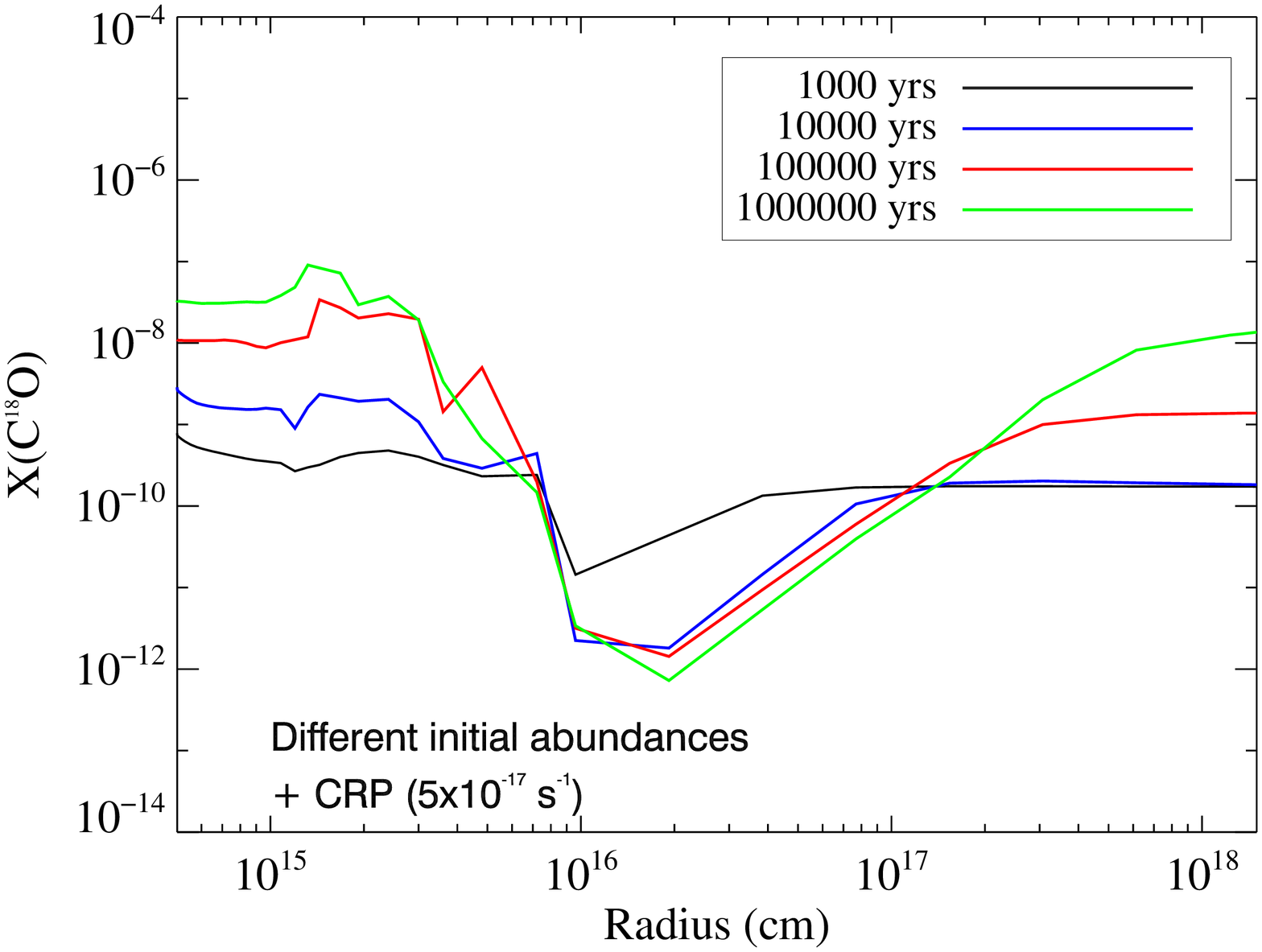} \\
\includegraphics[scale=0.32, angle=90]{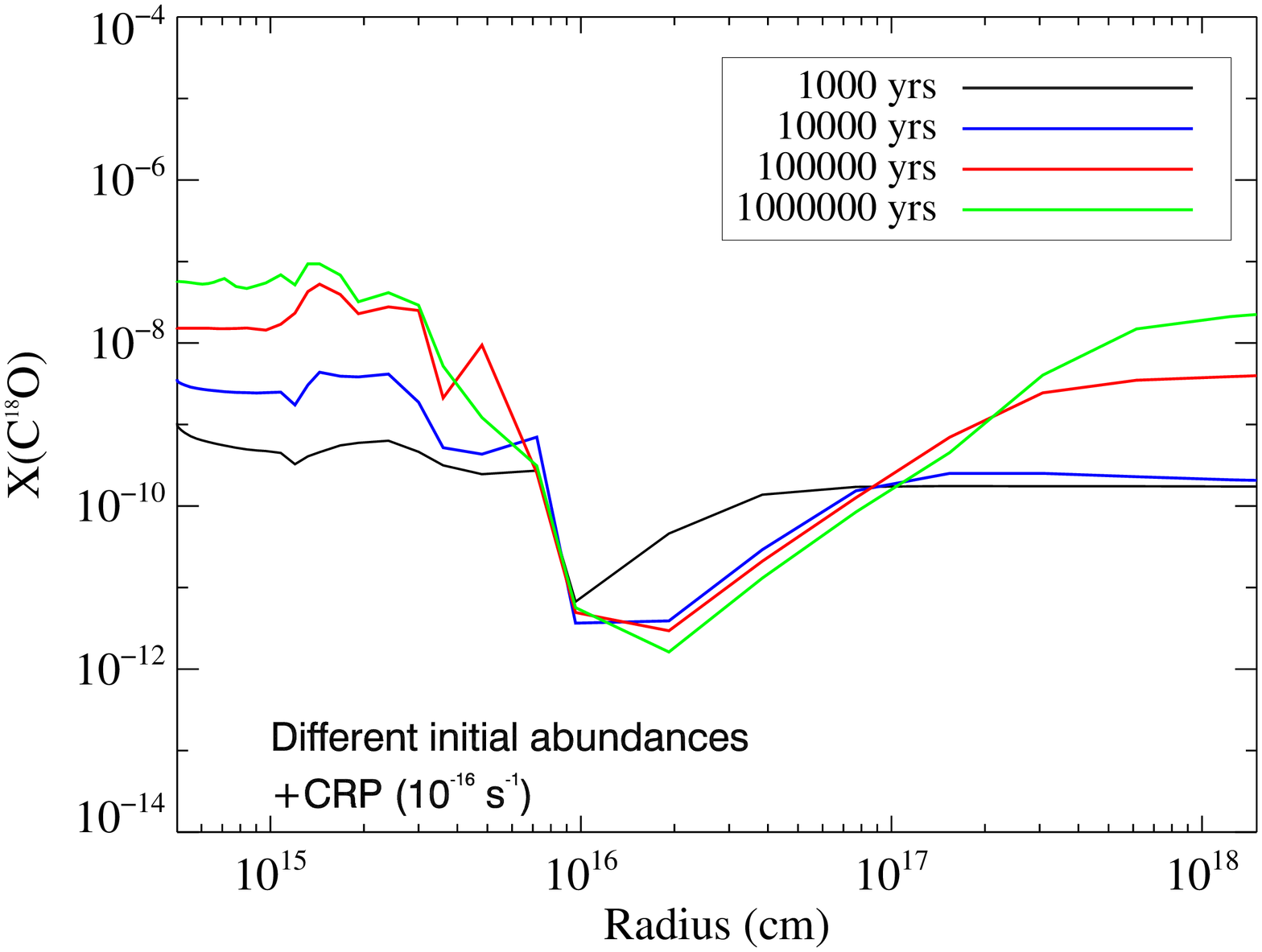} & \includegraphics[scale=0.32, angle=90]{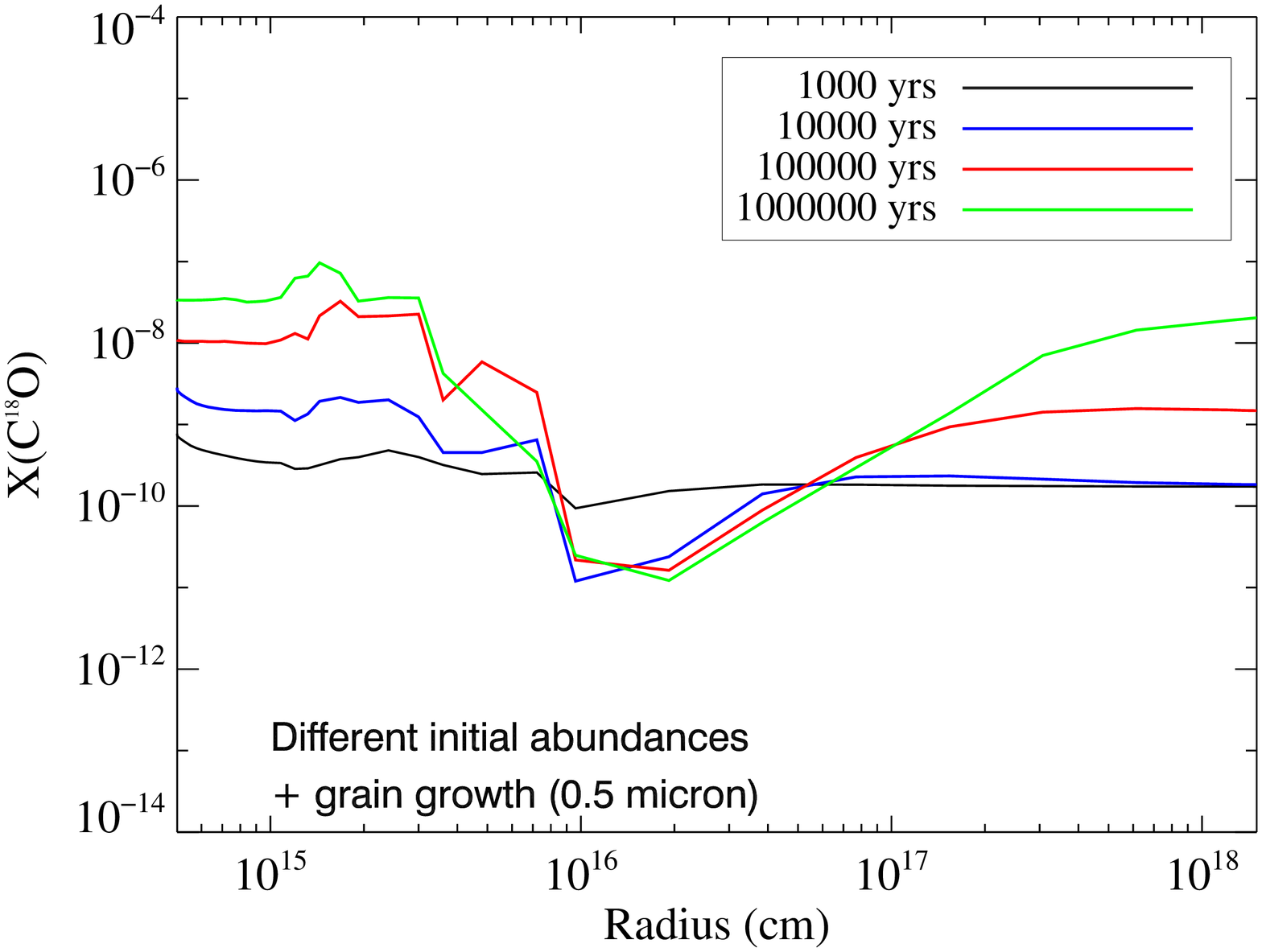} \\
\end{array}$
\end{center}
\caption{Time-dependent 1D chemical models for timescales 10$^{3}$ -- 10$^{6}$ yrs adopting different ``extreme'' input parameters.  
These models include twice higher cosmic ray ionization rate, a grain growth up to 0.5~$\mu$m and different initial abundances.}
\label{fig:chem2}
\end{figure*}

\end{appendix}
\end{document}